# Statistical Characteristics of the Heliospheric Plasma and Magnetic Field at the Earth's Orbit during Four Solar Cycles 20-23

*A.V. Dmitriev[1, 2], A.V. Suvorova[2], I.S. Veselovsky[2, 3]*

[1]Institute of Space Science, National Central University, Taiwan
[2]Skobeltsyn Institute of Nuclear Physics, Moscow State University, Russian Federation
[3]Space Research Institute (IKI), Russian Academy of Sciences,
Moscow, Russian Federation

## Abstract

The review presents analysis and physical interpretation of available statistical data about solar wind plasma and interplanetary magnetic field (IMF) properties as measured in-situ at 1 A.U. by numerous space experiments during time period from 1964 to 2007. The experimental information have been collected in the OMNI Web/NSSDC data set of hourly averaged heliospheric parameters for last four solar cycles from 20[th] to 23[rd]. We studied statistical characteristics of such key heliospheric parameters as solar wind proton number density, temperature, bulk velocity, and IMF vector as well as dimensionless parameters. From harmonic analysis of the variations of key parameters we found basic periods of 13.5 days, 27 days, 1 year, and ~11 years, which correspond to rotation of the Sun, Earth and to the solar cycle. We also revealed other periodicities such as specific five-year plasma density and temperature variations, which origin is a subject of discussion. We have found that the distribution of solar wind proton density, temperature and IMF is very close to a log-normal function, while the solar wind velocity is characterized by a very broad statistical distribution. Detailed study of the variability of statistical distributions with solar activity was performed using a method of running histograms. In general, the distributions of heliospheric parameters are wider during maximum and declining phase of the solar cycle. More complicated behavior was revealed for the solar wind velocity and temperature, which distribution is characterized by two- or even tree-peak structure in dependence on the phase of solar cycle. Our findings support the concepts of solar wind sources in the open, closed and intermittent magnetic regions on the Sun.

# 1. Introduction

Solar wind (SW) plasma and interplanetary magnetic field (IMF) parameters are measured in situ during space era nearly continuously onboard many spacecraft and satellites. The physical processes on the Sun and in the heliosphere leading to observed SW and IMF parameters and their variations are now rather well investigated and understood, though some unsolved problems still remaining [*Schwenn and Marsch*, 1990; *Burlaga*, 2005]. Namely, heating of the solar corona and generation of the solar wind and heliospheric magnetic field are still unresolved subjects of very intense investigations during last decades.

The SW and IMF data are processed and compiled in data bases, which contain hundreds of thousand hourly averages of the solar wind plasma and IMF parameters measured near Earth's orbit in 1964-2007 by the IMP, HEOS, VELA, OGO, ISEE, Prognoz, Wind and ACE satellites [*King*, 1981, *King and Papitashvili*, 2005]. The estimation of errors in those data is difficult because direct measurements were made with different instruments on different satellites and at different orbits. The data obtained are rather nonuniform in both spacing and relative and absolute accuracy. The procedure of relative intercalibration of detectors and introduction of correction is not complete [*King*, 1977; *Couzens and King*, 1986; *Freeman et al.*, 1993; *Russell and Petrinec*, 1993; *Zwickl*, 1993; *Dmitriev et al.*, 2005a; *King and Papitashvili*, 2005]. Detailed description of the data intercalibrations and corrections are presented at web site http://omniweb.gsfc.nasa.gov/html/omni2_doc.html.

The average values of SW and IMF parameters at the Earth orbit were calculated in a number of papers based on the analysis of these growing data sets [*Veselovsky et al.,* 1998a; 1999; 2000a, 2001]. In mentioned papers, long-term variations of the averaged density and other parameters of the solar wind and interplanetary magnetic field were analyzed using the data obtained from direct measurements at the Earth orbit from 1964 to 1996. A general trend was revealed for the entire period along with quite pronounced but comparatively small variations during solar cycles 20, 21, and 22. The results obtained highlighted the important role of different sources of the solar wind. At different phases of the solar cycle, open, closed, and intermittent magnetic-field configurations are typical of these sources [*Veselovsky et al.,* 1998a].

The variability and the periodicity of heliospheric parameters are of interest from the point of view of plasma dynamics on the Sun and in the interplanetary space as well as for the solar-terrestrial physics. The long-term and large-scale variations are described in numerous studies [e.g. *Crooker*, 1983; *Veselovsky*, 1984; 2004; *Schwenn*, 1990; *Zhang and Xu*, 1993; *Gazis*, 1996; *El-Borie et al.*, 1997]. The variability of the Sun as a star was traced in its integrated solar-wind mass and energy fluxes [*Veselovsky et al.,* 1999]. Direct plasma measurements in the heliosphere over more than the last thirty years indicated that these quantities have experienced relative variations by factors of 1.5-2, approximately in antiphase with the last three eleven-year solar cycles. A rising trend was noted over this time, with a similar relative-variation scale. This trend may be a manifestation of a "secular" cycle with duration of 60-70 years or longer.

The methods of the Fourier transform, spectrum-time analysis, and wavelet analysis were used to study the structure and dynamics of rhythmic and non-rythmic variations of the main SW and IMF parameters at time scales from days to tens of years. A large variety of the observed regular and irregular variations in the near-Earth heliosphere is explained by a number of reasons: (1) the variability of unstable processes in the region of solar wind formation, (2) the rotation of Sun and the associated inhomogeneities in its corona, (3) and the Earth's orbital motion. Irregular, a-periodic variations are present for all parameters and at all time scales. The most prominent regular variations are related to cyclic processes on the Sun and its rotation [*Veselovsky et al.,* 2000b; *Dmitriev et al.,* 2000].

The time-epoch analysis and hysteresis curves of the heliospheric parameters show some general and specific properties of the cycles [*Veselovsky et al.,* 2000b; 2001; *Dmitriev et al.,* 2002a; 2002b; 2005b]. Based on these results and using measured heliospheric parameters during the rising phase of the 23-rd solar cycle we were able to present some semi-quantitative estimations of the expected solar wind energy flux and the induced electric field for the time period after the solar maximum. The similarity between the rising phases of the 23-rd and 20-th solar cycles presented additional grounds for correct expectations of the lower maximum of the 23-rd solar cycle and the geomagnetic activity as compared with the 21-st and 22-nd solar cycles [*Dmitriev et al.,* 2002b].

The purpose of this paper is an extension of our statistical studies of SW and IMF parameters based on growing amount of direct in-situ measurements near the Earth orbit during space era. Common statistical properties are considered in a form of statistical distributions in Section 2. "Basic Statistical Properties". Some characteristic periods in variations of heliospheric parameters derived by a method of Fourier transform for unequally-spaced data are discussed in Section 3. "Characteristic periodicities". Variations of the parameters with solar cycle are studied by a method of running histograms in Section 4. "Solar cycle variations". Section 5. is "Summary and Conclusions".

## 2. Basic Statistical Properties

### Sunspot Number

In the first turn we consider sunspot numbers represented by the Wolf number ($W$) as a key heliospheric parameter related to the variations of solar activity. The Wolf number is measured continuously with 1-day step for many decades. It is a simple and robust parameter for comparisons and ordering of data in its regular and irregular behavior. Figure 1 shows 27-day running annual Wolf number for the time interval from 1963 to 2007. The smoothed time profile has four prominent maxima of solar cycles 20[th] to 23[rd], respectively, in 1969, 1980, 1989 and 2000. We can also indicate five solar minima in 1964, 1976, 1986, 1996 and about 2007, which allow easy estimation of the cycle duration: 12 years for the 20[th] cycle, 10 years for the cycles 21[st] and 22[nd], and ~11 to 12 years for the 23[rd] cycle. Those time intervals are in good agreement with empirical fact that large solar cycles (21[st] and 22[nd]) have shorter duration of ~10 years than small cycles (20[th] and 23[rd]) having duration of ~12 years. It is a manifestation of the well-known Waldmeier's rule [*Veselovsky and Tarsina,* 2002a]. Hence the last four solar cycles exhibit two periods of 10 years and of 12 years, i.e. around 11-year cycle.

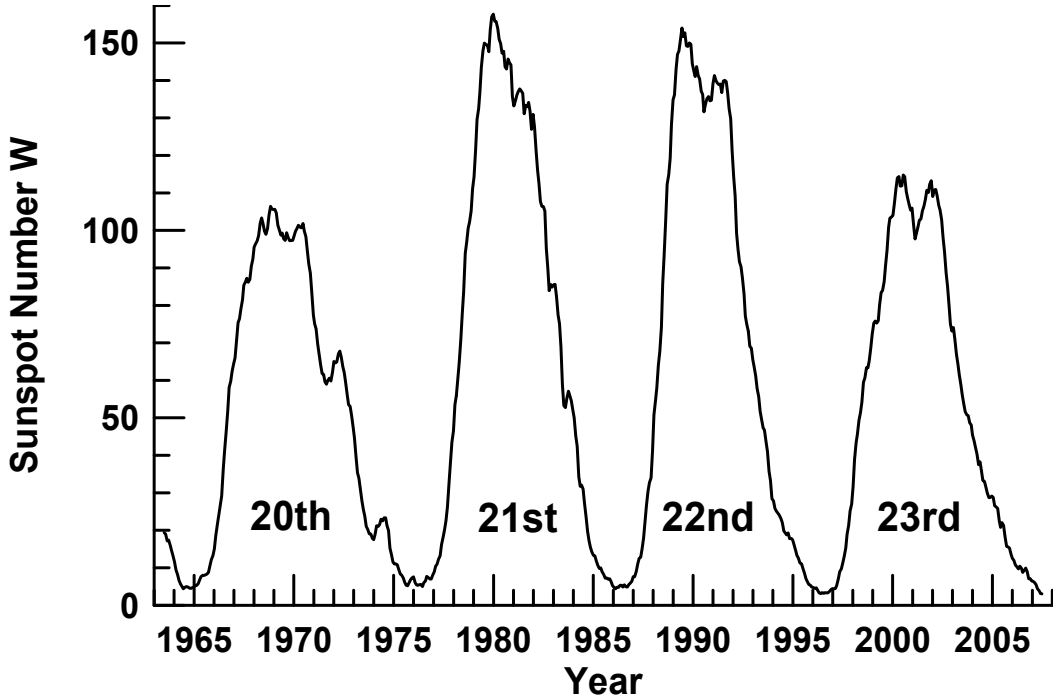

Figure 1. Time profile of the 27-day running annual Wolf number W from 1963 to 2007.

Further analysis of the dynamics of sunspot number requires a study of its statistical properties. Statistical distributions of the *W* are presented in Figure 2. The daily value of sunspot number varies in very wide range from 0 to 302. In the linear scale, the distribution is smooth and decreases fast from maximum at *W*=0. There is a little plateau at *W* of ~100. Hence in the linear representation the dominant contribution to the distribution function is produced by relatively small sunspot number of *W*<100.

The situation changes dramatically when we represent the sunspot number in logarithmic scale. Namely, instead of values *W* we consider their decimal logarithms *lg*(*W*). The statistical distribution of *lg*(*W*) has two maxima. The first one is formed by small values of *W*<10 and the second peak with amplitude of ~1400 at *W*~100 is wide and most prominent. Apparently the values of *W*=0 are not accounted in this distribution. From 1963 to 2007 the number of days with *W*=0 is 1300. That is much smaller than the number of days forming the wide second maximum. As we can see in Figure 1, that peak is mainly contributed by long-lasting solar maxima of the moderate 20[th] and 23[rd] cycles. Hence the logarithmic representation of sunspot number allows grouping the vast majority of solar active days in a wide peak around *W*=100. As a result, considering the *lg*(*W*) we get ability to study in detail the sunspot number enhancements, rather than sunspot number decreases in vicinities of the solar minima, which are mostly prominent in the linear scale.

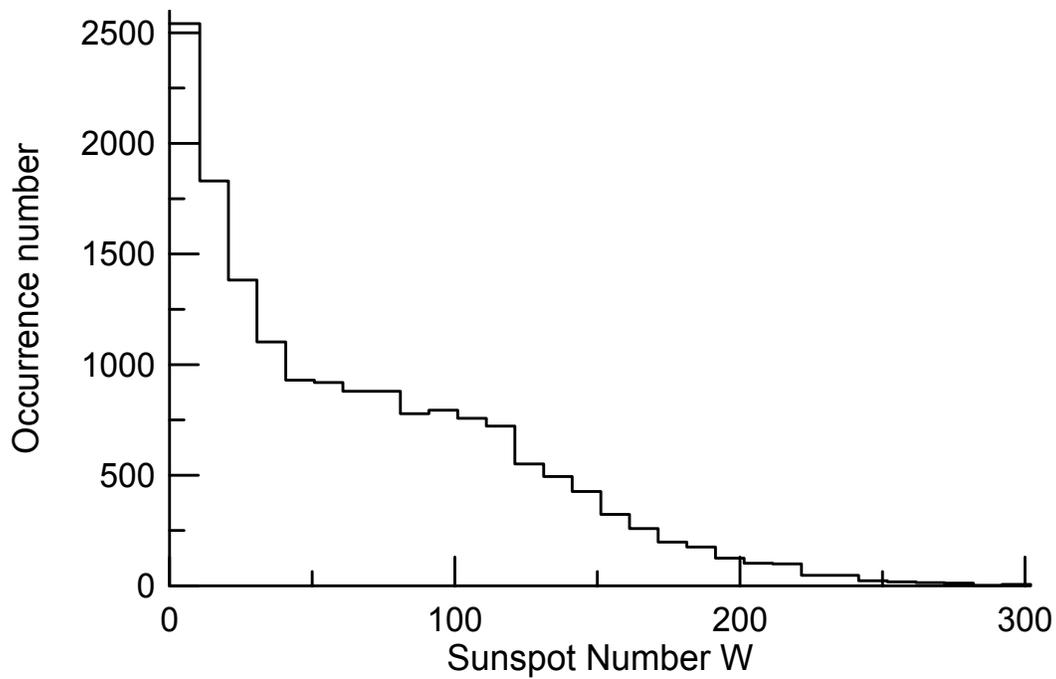

a)

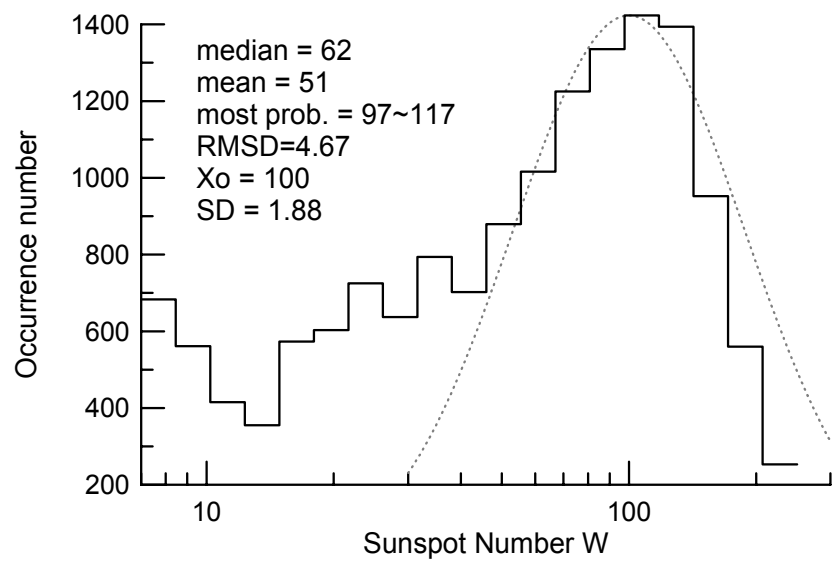

b)

Figure 2. Statistical distribution of the daily Wolf numbers for time interval 1963 to 2007: (a) linear and (b) logarithmic scale. In the latter case, the distribution contains two maxima: at small $W$ (from 0 to 7) and at large $W \sim 100$. The dotted line is a lognormal distribution with mode $X_0$=100 and standard deviation $\sigma$=1.88 (see Formulae 6).

## Data Coverage

   Hourly averaged data on heliospheric parameters for time period from 1963 to 2007 contains about 400,000 consequent hours of measurements. However, about one third of the data is occupied by gaps. Figure 3 shows the data coverage for SW and IMF data at each year from 1963 to 2007. The poor coverage (below 50%) of data takes place during early years of interplanetary measurements in 1963 to 1972 and during interval of 1983 to 1993, when single satellite IMP-8 was operating in the interplanetary medium. Though there are many data gaps occur during the other years, especially for the plasma parameters.

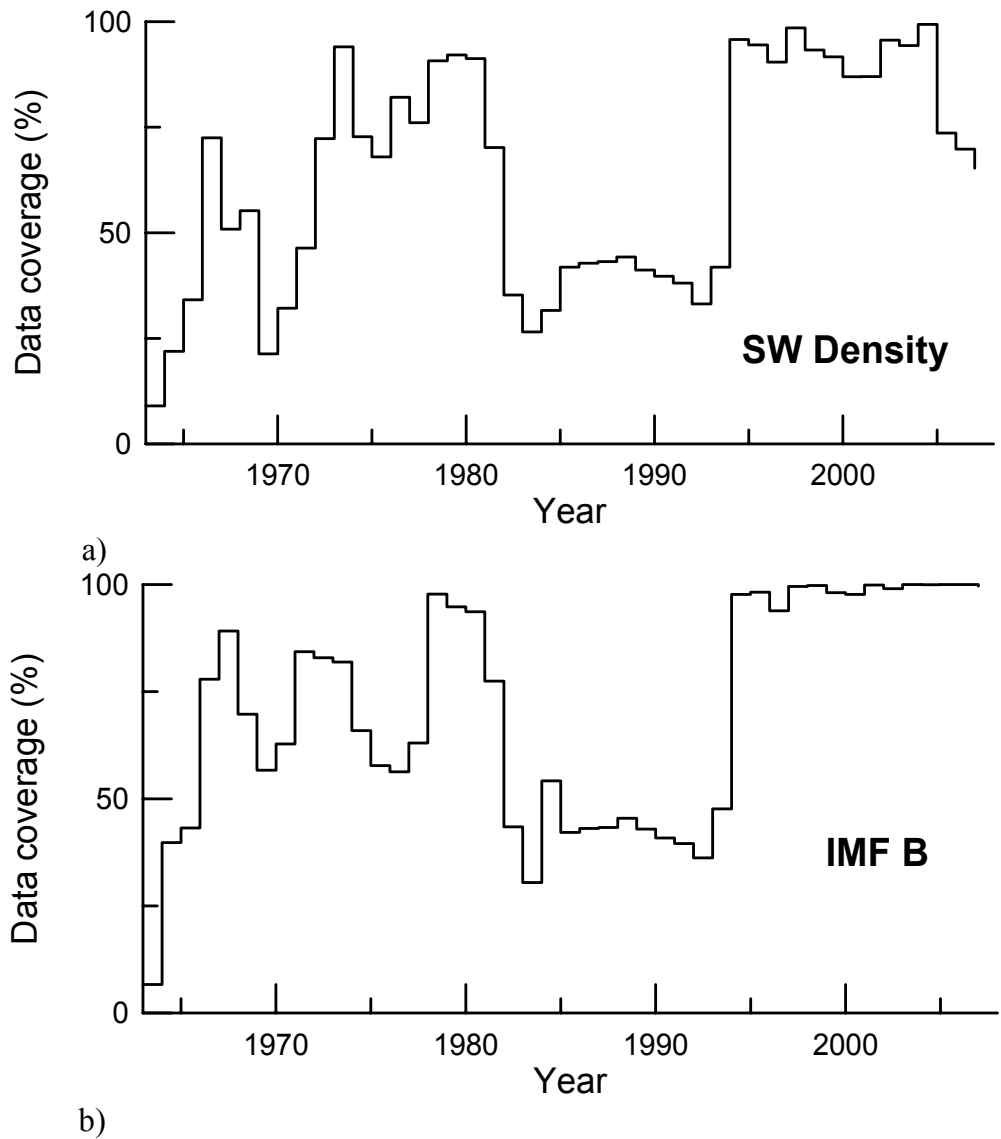

a)

b)

Figure 3. (Continued)

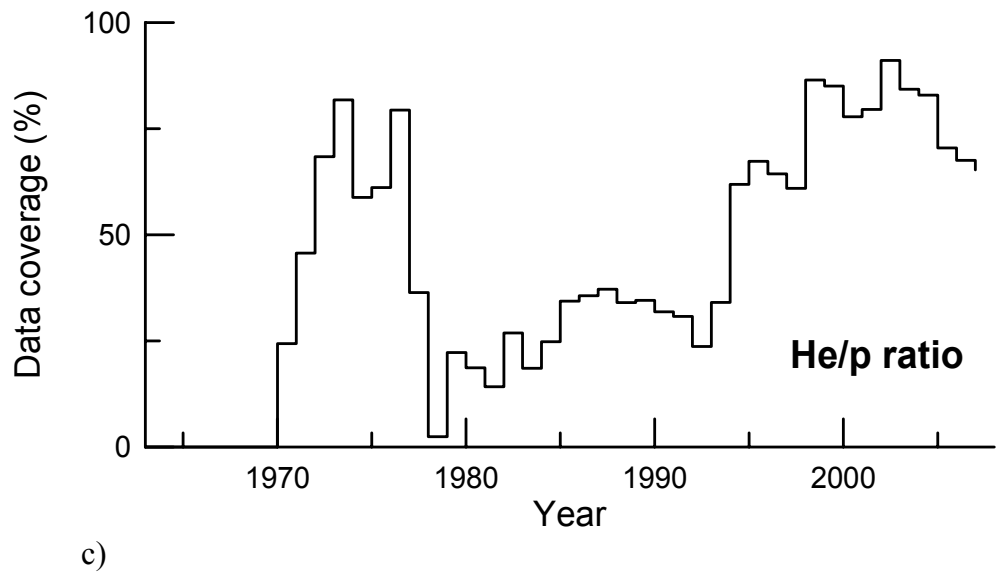

c)

Figure 3. Annual data coverage for the datasets of:
(a) solar wind density, (b) IMF strength, and (c) helium to proton ratio.

Duration of the data gaps varies in a wide range from 1 hour to more than 10 days as shown in Figure 4. There are various reasons causing the gaps. Before 1996, the near-earth interplanetary measurements were conducted by high apogee satellites, which spent a large portion of time inside the magnetosphere and magnetosheath. Data gaps were caused also by restricted volume of onboard data storages and infrequent sessions of data transmission. Measurements of modern interplanetary monitors such as Wind and ACE have many data gaps due to malfunctions of onboard equipment.

Most serious and long-lasting malfunctions are caused by radiation damage during intense solar proton events, which duration can achieve up to several days [*e.g. Dmitriev et al.*, 2005b]. Because of specific of experimental methodic, the solar energetic particles impinging upon the plasma detectors affect mostly the data on density and temperature. The measurements of IMF and plasma velocity are less sensitive to the radiation effects. Apparently, studying of such "full of holes" data sets requires using specific methods of statistical analysis.

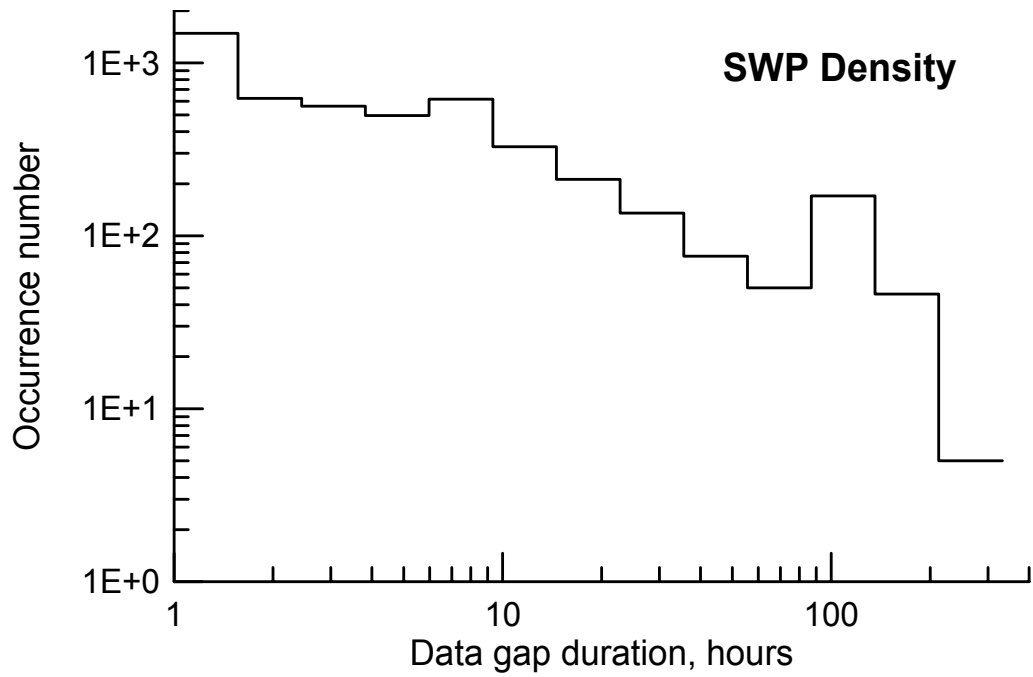

a)

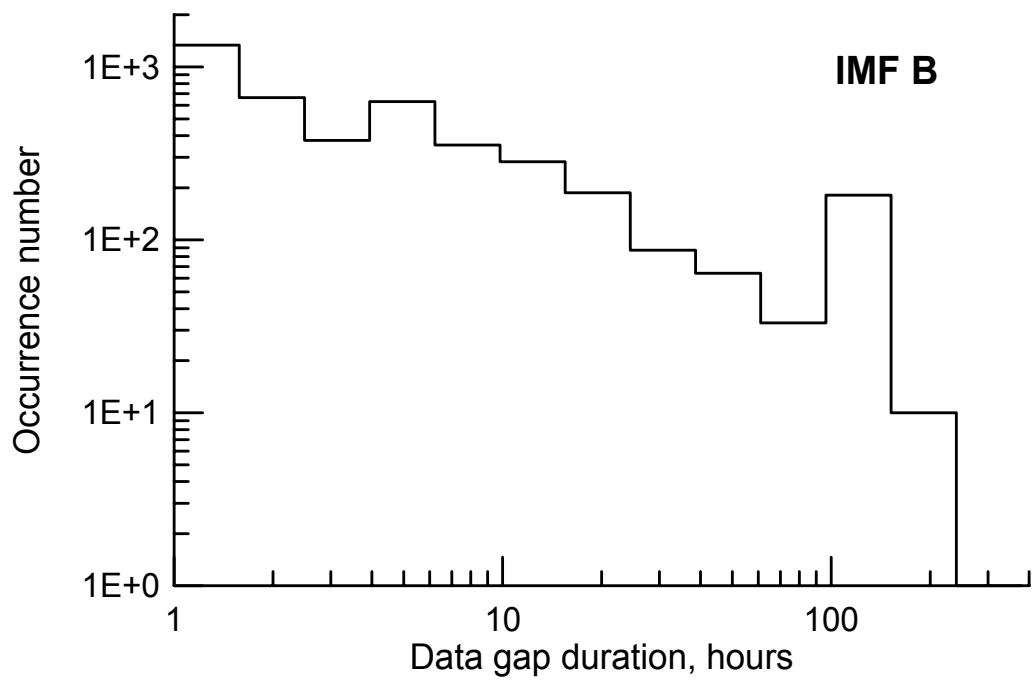

b)

Figure 4. Occurrence number distribution for the duration of data gaps in the data sets of:
(a) solar wind density and (b) IMF strength.

**Probability Distribution Functions**

Statistical study of a heliospheric parameter is aimed determination of two basic statistical moments: most probable value, or mode, and dispersion, i.e. variability of the parameter around its mode. One of mostly common approaches for the statistical distributions is a normal probability distribution function (PDF). An assumption that the measured parameter is distributed normally is prevalent in many studies of the space physics.

The normal PDF of a random variable $x$ is expressed by a Gaussian function:

$$P(x) = A \exp\left[-\frac{(x - X_0)^2}{2\sigma^2}\right] \qquad (1)$$

where $A$ is amplitude of the Gaussian, $X_0$ is mode or most probable value, and $\sigma$ is standard deviation ($SD$), a measure of the dispersion of distribution. An interval within one standard deviation around the mode accounts for ~68% of the dataset, while two and three standard deviations account 95% and 99.7%, respectively. A half width at middle height ($hwmh$) of the normal distribution is simply related to the $SD$ as $hwmh = \sigma\sqrt{2\ln(2)}$. Here $ln$ means a function of natural logarithm.

In experiment, statistical distributions of measured parameters are usually characterized by median and first four moments: mean, root mean square deviation ($RMSD$), skewness and kurtosis. The median is a middle number separating the higher half of statistical distribution from the lower half. The Gaussian function is symmetric relative to the $X_0$. Hence its mode and median are equal. The mean of $N$ independent measurements of the random variable $x_i$ defined as follow:

$$\langle X \rangle = \frac{1}{N} \sum_{i=1}^{N} x_i \qquad (2)$$

Very important property of the normal distribution is that the mean is equal to mode and median. Apparently, the number $N$ should be as more as possible. The proof of equality between the $X_0$ and $<X>$ as well as proofs of other important properties of the normal distribution can be found in many books devoted to statistics [e.g. *Mood et al.*, 1974].

The RMSD is calculated from the following expression:

$$RMSD = \sqrt{\frac{1}{N-1} \sum_{i=1}^{N} (x_i - \langle X \rangle)^2} \qquad (3)$$

Because the set of measurements $x_i$ has been already used for determination of the average $<X>$, the number of independent measurements in calculation of the *RMSD* becomes *N-1* as represented in the denominator. The *RMSD* of normal distribution is equal to one standard deviation $\sigma$. That is another important property of the normal PDF.

The third moment of a statistical distribution is skewness:

$$k_3 = \frac{1}{N} \sum_{i=1}^{N} \left( \frac{x_i - \langle X \rangle}{RMSD} \right)^3 . \tag{4}$$

The skewness is a measure of the lack of symmetry of the distribution. A positive (negative) number indicates a higher number of large (small) values of the parameter than would be expected for the normal PDF, which has zero skewness.

The fourth statistical moment, kurtosis, is defined as follow:

$$k_4 = \frac{1}{N} \sum_{i=1}^{N} \left( \frac{x_i - \langle X \rangle}{RMSD} \right)^4 - 3 \tag{5}$$

The kurtosis is a measure of the flatness (negative value of $k_4$) or peakedness (positive value of $k_4$) of the distribution relative to the normal PDF with the same mean and dispersion. Note that normal distribution has zero kurtosis.

Thus, under the assumption of normal distribution the determination of mode and standard deviation can be simply substituted by calculation of mean and *RMSD*. Actually those moments can be calculated for any variable, which statistical distribution might differ from the normal, i.e. have non-zero skeweness and/or kurtosis. However in such a case the first two statistical moment have lost their important statistical sense. Namely, the mean might be different from the mode, i.e. we lost information about the most probable value of measured parameter. The *RMSD* is not equal to the standard deviation $\sigma$, i.e. the variation of parameter can not be determined in standard manner. In this case it might be possible to find such a represnetation of the parameter that allows approaching its statistical distribution to the normal PDF.

The logarithmic scale is one of very useful representations. Here we can introduce a lognormal PDF as follow [e.g. *Hartlep et al.*, 2000]:

$$P(x) = A \exp \left[ -\frac{1}{2} \left( \frac{\ln(x) - \ln(X_0)}{\ln(\sigma)} \right)^2 \right] \equiv A \exp \left[ -\frac{1}{2} \left( \frac{\ln(x / X_0)}{\ln(\sigma)} \right)^2 \right] \tag{6}$$

The lognormal PDF has the same properties as normal distribution but in logarithmic scale. Indeed, replacing $ln(x)$ by $z$, $ln(X_0)$ by $Z_0$ and $ln(\sigma)$ by $\delta$ we can rewrite Equation 4 as:

$$P(z) = A \exp \left[ -\frac{1}{2} \left( \frac{z - Z_0}{\delta} \right)^2 \right] \tag{6a}$$

This Equation is exactly the same as Equation 1. Taking the exponent we can simply convert numbers from logarithm to linear scale. The exponent of most probable of lognormal distribution is equal to mode $X_0$. In other words, the lognormal distribution reaches the maximum at $X_0$. However, the dispersion of lognormal PDF has a different meaning such that the *hwmh* should be redefined as a ratio of $x/X_0$ at middle height (*rmh*), which is equal to $rmh = \sigma^{\sqrt{2\ln(2)}}$. Therefore, the standard deviation $\sigma$ in the logarithmic scale has a meaning of standard relative deviation, i.e. a standard ratio relative to mode. In the lognormal distribution the parameter $\sigma$ is dimensionless. In linear scale, the lognormal distribution is asymmetric and its dispersion is characterized by upper $SD_{up}$ and lower $SD_{lo}$ standard deviations:

$$SD_{up} = X_0 \cdot \sigma - X_0 = X_0(\sigma - 1) \tag{7a}$$

$$SD_{lo} = X_0 - X_0/\sigma = X_0(1 - 1/\sigma) \tag{7b},$$

which are not equal, because of $\sigma$ is always more than 1. Indeed, the logarithmic dispersion $\delta = ln(\sigma)$ should be always positive.

In the logarithmic scale we can also introduce the first four statistical moments of the distribution of random variable. Namely, the mean value $<X>_{\ln}$ can be defined as follow:

$$\left\langle X \right\rangle_{\ln} = \exp(\left\langle \ln(x) \right\rangle) \tag{8}$$

where

$$\left\langle \ln(x) \right\rangle = \frac{1}{N}\sum_{i=1}^{N}\ln(x_i) \tag{8a}$$

It is very easy to show that the mean in logarithmic scale is equal to the geometric mean:

$$\left\langle X \right\rangle_{\ln} = \left(\prod_{i=1}^{N} x_i\right)^{\frac{1}{N}} \tag{8b}$$

The lognormal distribution has the same achievement as the normal one: mean $<X>_{\ln}$ is equal to mode $X_0$ and to median of logarithms.

The second moment $RMSD_{\ln}$ is introduced as follow:

$$RMSD_{\ln} = \exp\left\{\sqrt{\frac{1}{N-1}\sum_{i=1}^{N}(\ln(x_i) - \ln\left\langle X \right\rangle_{\ln})^2}\right\} \tag{9}$$

For large number of independent measurements $N$ the $RMSD_{\ln}$ approaches to the standard deviation $\sigma$ of lognormal PDF. Note that the $RMSD_{\ln}$ is also dimensionless and has a meaning of relative deviation from the mean $<X>_{\ln}$. Hence in linear scale, the upper and lower deviations from the mean are calculated as $<X>_{\ln} \cdot RMSD_{\ln}$ and $<X>_{\ln}/RMSD_{\ln}$, respectively.

The third and fourth statistical parameters can be defined in logarithmic scale as follows:

skewness $k_3 = \dfrac{1}{N} \sum_{i=1}^{N} \left( \dfrac{\ln(x_i) - \ln\langle X \rangle_{/\ln}}{\ln(RMSD_{\ln})} \right)^3$  (10)

and

kurtosis $k_4 = \dfrac{1}{N} \sum_{i=1}^{N} \left( \dfrac{\ln(x_i) - \ln\langle X \rangle_{/\ln}}{\ln(RMSD_{\ln})} \right)^4 - 3$  (11)

The skewness and kurtosis calculated in the logarithmic scale have the same properties as those in the linear scale. Hence they can be used as a measure of similarity of the statistical distribution to lognormal PDF. For a log-normally distributed parameter the skewness and kurtosis are equal to zero, and the most probable and standard deviation can be easily calculated, respectively, as mean $<X>_{\ln}$ and $RMSD_{\ln}$.

In practice however, the number $N$ is not infinitely large and statistical distributions may have various shapes, which in general differ from the normal PDF. The mode, i.e. most probable value, is different from mean and median, and may be very different for strongly skewed distributions. As an example we demonstrate the statistical distribution of sunspot number (see Figure 2). Direct determination of the most probable of statistical distribution is not very accurate because of the following circumstances. The statistical distribution of empirical parameter is discrete with finite size of bins, because of limited number of measurements $N$. Very small size of the bins leads to appearance of many subsidiary peaks with low statistical significance, such that the most probable is hidden by noise. Hence the size of bins is selected in such a way as to provide a smooth shape of statistical distribution. As a result the accuracy of mode and dispersion are limited by the width of bin, which might be pretty wide and, thus, the accuracy becomes very poor. Another way of fitting the statistical distribution by standard, say normal, PDF is also unreliable, because in general the distributions are different from the normal.

In such situation, as a first step we compare the mode, mean and median in order to verify the normality of distribution. As a second step, the statistical distribution is approached, if possible, to the normal shape. Namely we choose such a representation (linear or logarithmic), which provides minimal difference between the mode, median and mean. Hence in the new representation the statistical distribution becomes more symmetric and its most probable is close to average. Our choice is based on a very important property of median: it is invariant in transformations between the linear and logarithmic scales. The quality of our approach is expressed numerically by the skewness and kurtosis.

In the following sections we will show that the logarithmic representation allows symmetrization of the statistical distributions for many heliospheric parameters. The number of bins is chosen to achieve a smooth shape of statistical distribution. For most of parameters that number is about 30. Using the statistical distribution, we estimate the most probable value of parameter. The median, mean and $RMSD$ are calculated in the most appropriate scale, where the distribution is mostly close to the normal PDF.

The distribution is fitted by normal or lognormal PDF in the following way. The mode $X_0$ of PDF is supposed to be equal to median. The standard deviation $SD$ is calculated by fitting the shape of statistical distribution by a PDF. The results are presented in Table 1. There we also indicate a type of PDF and percentage of data covered by that PDF.

**Table 1.  Main statistical numbers of key heliospheric parameters at the Earth orbit during 1963-2007.**

| | statistics | min | max | median | mean | mode | *RMSD* | *SD* | $k_3$ | $k_4$ | PDF | % |
|---|---|---|---|---|---|---|---|---|---|---|---|---|
| W | 16437* | 0 | 302 | 62 | 51 | 97~117 | 4.66 | 1.88 | | | two-peak | - |
| $V$, km/s | 265699 | 156 | 1200 | 420 | 430 | 370~390 | 1.25 | 1.26 | .42 | -.32 | flatten | - |
| $T$, $10^3$ K | 227883 | 2.9 | 7000 | 85 | 83 | 72~94 | 2.3 | 2.43 | -.14 | -.31 | lognormal | 94 |
| $n$, cm$^{-3}$ | 244625 | 0.1 | 118. | 5.3 | 5.4 | 4.3~5.3 | 2.0 | 2.0 | .008 | .59 | lognormal | 95 |
| $He/p$ | 163052 | 0.001 | 0.4 | 0.037 | 0.034 | 0.03~0.05 | 2.0 | 1.7 | -.79 | 1.4 | skewed | - |
| $P_d$, nPa | 244625 | 0.01 | 79 | 2. | 2. | 1.8~2.25 | 1.86 | 1.74 | -.01 | 1.7 | lognormal | 95 |
| $S_k$, erg/cm$^2$·s | 244625 | .0015 | 30 | 0.43 | 0.44 | 0.37~0.49 | 2.0 | 1.92 | .08 | 1.3 | lognormal | 98 |
| $B$, nT | 271938 | 0.4 | 55.8 | 5.9 | 6.0 | 5.5~6.5 | 1.55 | 1.48 | .12 | .85 | lognormal | 95 |
| $B_{xy}$, nT | 271938 | 0.14 | 52. | 4.6 | 4.5 | 4.3~4.7 | 1.74 | 1.55 | -.78 | 2.8 | lognormal | 82 |
| $B_x$, nT | 271938 | -40.1 | 29.4 | 0. | 0. | ±3 | 3.90 | - | - | - | two-peak | - |
| $B_y$, nT | 271938 | -38.8 | 46.1 | 0. | 0. | ±3 | 4.30 | - | - | - | two-peak | - |
| $B_z$, nT | 271938 | -46.3 | 36.8 | 0. | 0. | -0.6~1.5 | 3.1 | 2.09 | -.01 | 9.1 | normal | 88 |
| $E_y$, mV/m | 242155 | -34 | 30 | 0 | 0 | -0.4~0.2 | 1.55 | 1.0 | -0.1 | 23. | normal | 88 |
| $M_a$ | 228053 | 1.1 | 93. | 8.4 | 8.4 | 8.~9.4 | 1.54 | 1.44 | .025 | 1.7 | lognormal | 95 |
| $M_s$ | 227171 | 2.6 | 72. | 12.7 | 13. | 11.~13.8 | 1.35 | 1.30 | 0.57 | 1.3 | lognormal | 92 |
| $\beta$ | 212221 | 0.001 | 87. | 0.48 | 0.43 | 0.44~0.65 | 2.61 | 2.12 | -.92 | 2.5 | lognormal | 89 |

* Number of days

## Solar Wind Plasma

There are many indications that SW parameters have lognormal distribution. *Veselovsky et al.* [1998b] revealed that statistical distributions of hourly averages of SW proton density and temperature during three solar cycles from 20 to 22 are well fitted by the lognormal PDF. Statistical analysis of SW data acquired from the Wind satellite in 1995 to 1998 also reveal that hourly averages of solar wind velocity, density and temperature have lognormal distribution [*Burlaga and Szabo*, 1999, *Burlaga and Lazarus*, 2000]. *King and Papitashvili* [2005] studying the hourly averaged SW plasma data during the 23$^{rd}$ solar cycle, use also logarithmic representation of the proton density and temperature and mention that the distribution of density is close to the lognormal.

Statistical distributions of the SW parameters are shown in Figure 5. All the parameters are represented in the logarithmic scale. In Figures 5a and 5b we find that the distributions of proton density $n$ and temperature $T$ are very close to lognormal PDF with the mode, respectively, 5.3 cm$^{-3}$ and 85·$10^3$ K and standard relative deviation, respectively, 2 and 2.43. They have pretty small skewness and kurtosis. In contrast, the distributions of SW velocity and $He/p$ ratio are different from lognormal (Figure 5c and 5d). The statistical distributions of $n$ and $T$ are fitted very well by the lognormal PDF within 2-$\sigma$ interval that corresponds to ~95% of the total statistics, which contain 244625 and 227883 hourly averaged measurements, respectively, of the density and temperature.

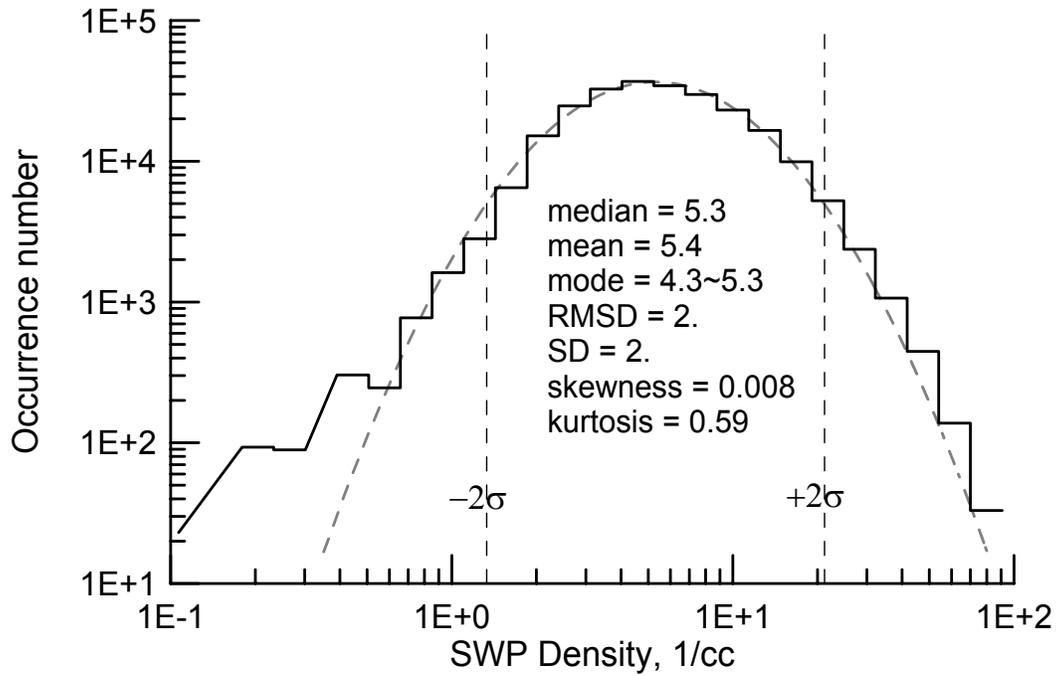

a)

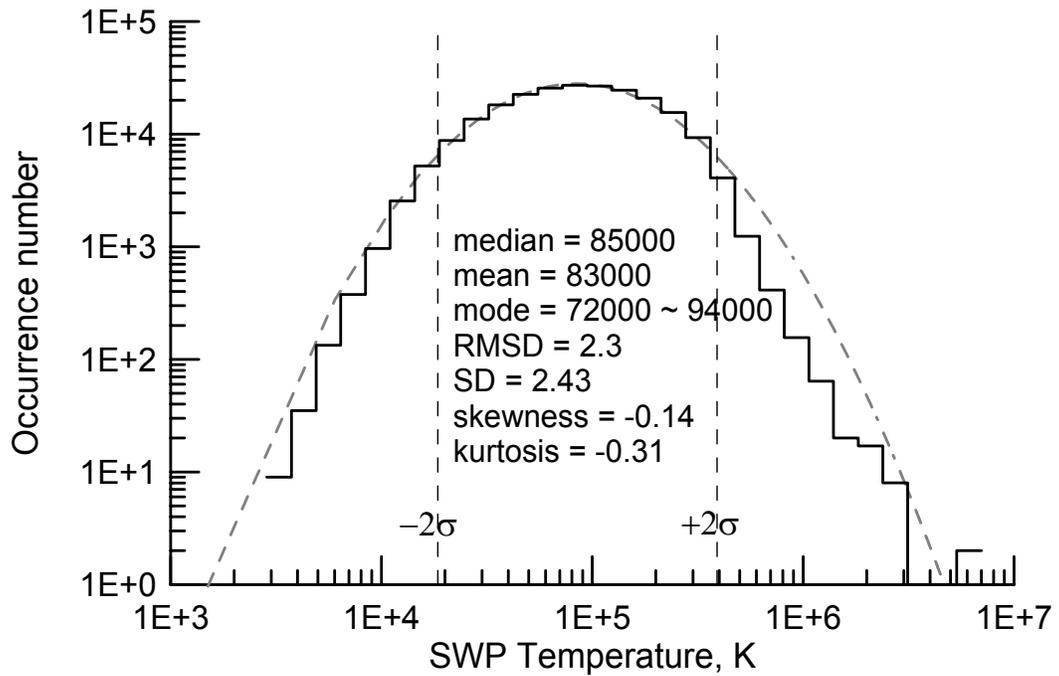

b)

Figure 5. (Continued)

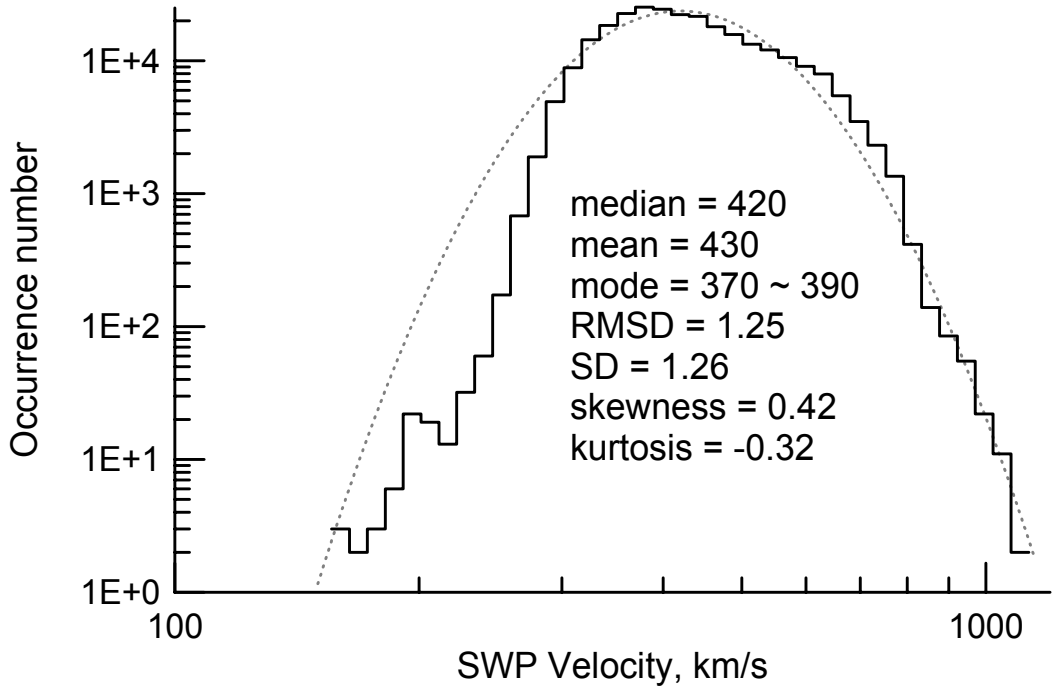

c)

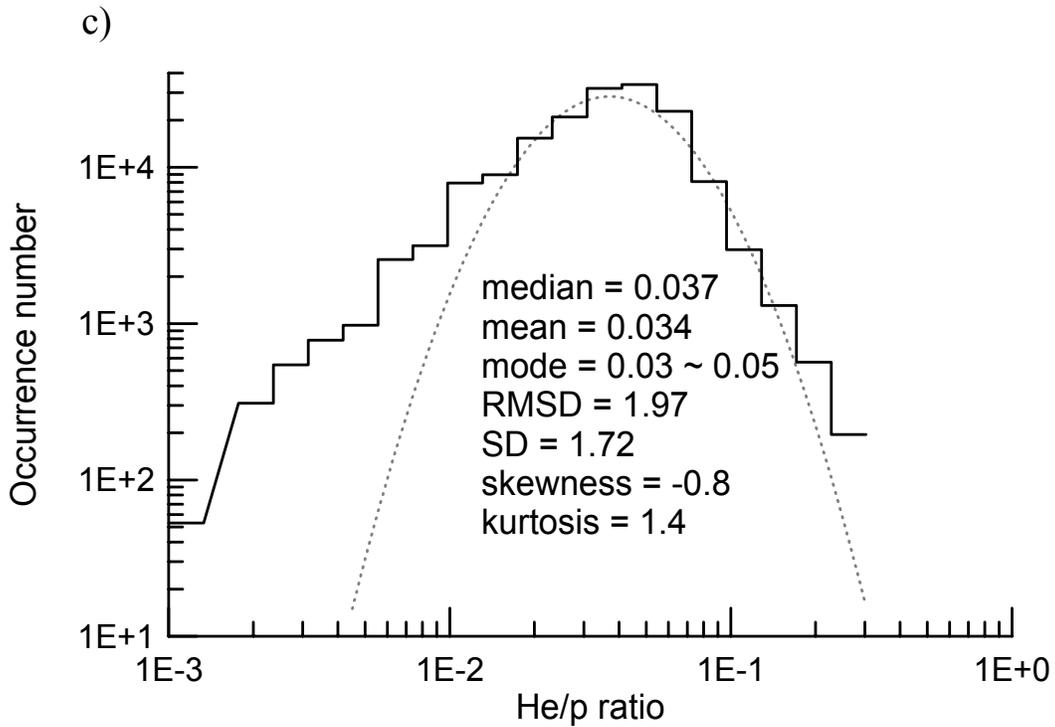

d)

Figure 5. (Continued)

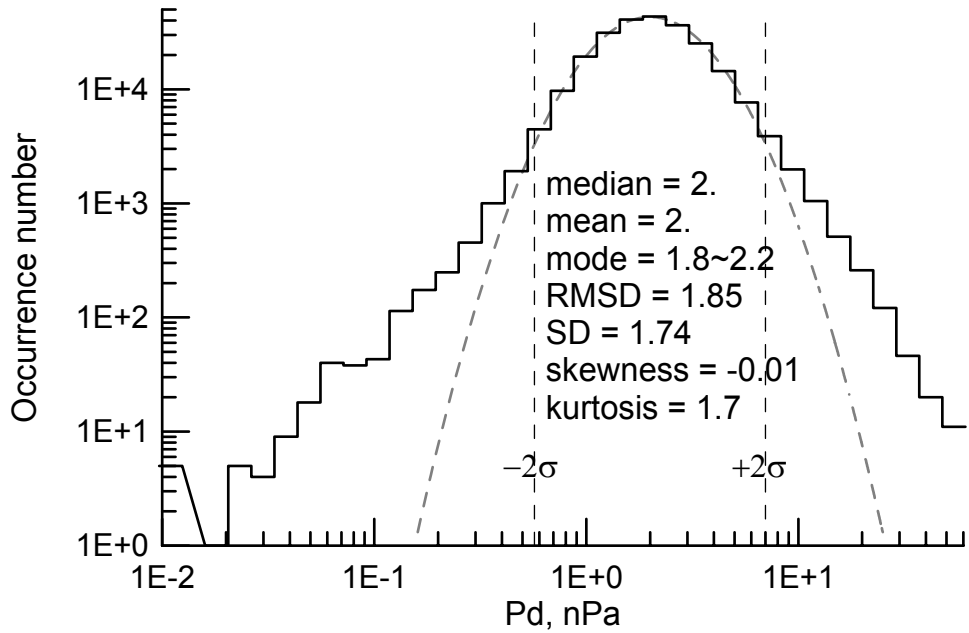

e)

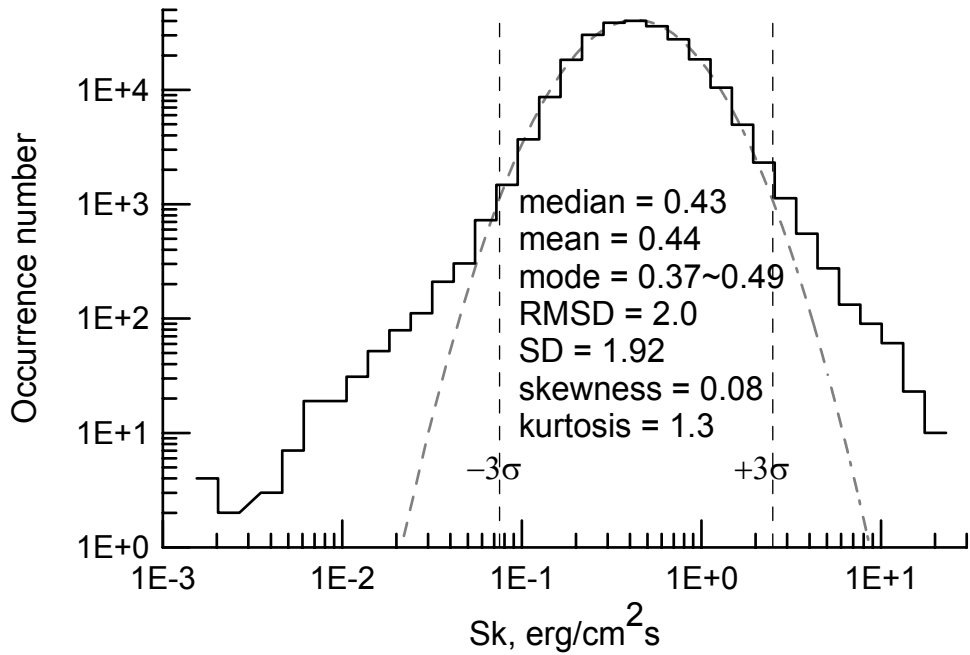

f)

Figure 5. Statistical distributions of the solar wind parameters represented in logarithmic scale for time interval from 1963 to 2007: (a) proton density, (b) proton temperature, (c) SW velocity, (d) He/p ratio, (e) dynamic pressure $P_d$, and (f) kinetic energy flux density $S_k$. Fitting by lognormal distribution function is shown by dashed line for good fit or by dotted line for poor fit. Vertical thin dashed lines restrict the best-fit intervals.

A characteristic feature of the solar wind density within the 2-$\sigma$ interval is an anticorrelation with the SW velocity such that fast streams originating mainly in coronal holes in average have lower density than dense slow SW originating around coronal streamers [*Ipavich et al.*, 1998]. Beyond of the 2-$\sigma$ interval, the proton density distribution deviates from the lognormal PDF such that the wings of distribution become less steep. At small values, the 2-$\sigma$ level restricts densities smaller than 1 cm$^{-3}$. Interplanetary structures with very small proton density are associated with transient SW structures, in particular, expanding ejecta and post-shock flows [*Richardson et al.*, 2000].

Most prominent deviation from the lognormal PDF is revealed at extremely small densities of $n$<0.4 cm$^{-3}$. Note that the proton density can be less than 0.1 cm$^{-3}$ for several hours [*Usmanov et al.*, 2005], but the OMNI dataset has no information about such events, because the density is recorded only with one position after decimal point. Time intervals of extremely low densities, so-called SW disappearance events [*Lazarus*, 2000], are accompanied with stable unipolar magnetic field and highly nonradial SW flow, which is not associated with any transient structures [*Crooker et al.*, 2000]. These density anomalies are caused by a rarefaction at the trailing edge of relatively fast flow that formed as a result of suppression of coronal outflow from a region that earlier provided fast wind flow [*Usmanov et al.*, 2005]. The solar sources of the density anomalies are found as either active region open fields or small coronal hole boundaries embedded in or near large active region located close to central meridian [*Janardhan et al.*, 2008].

The proton density distribution is slightly deviated from the PDF at densities above 20 cm$^{-3}$. High densities are usually formed in the strongly compressed leading regions associated with CIRs and sheaths of CMEs, and in particular with interplanetary shocks [e.g. *Borrini, et al.*, 1982a; *Crooker et al.*, 2000]. Strongly compressed regions of high proton density can be also formed as a result of interaction between the leading or rear edges of ICME with the ambient SW [*Bothmer and Schwenn*, 1995; *Dal Lago et al.*, 2001]. Extremely high densities (up to $n$~100 cm$^{-3}$) are observed in solar eruptive filaments/prominences, which have characteristics of the chromosphere, i.e. consist of dense, cold material with abundance of He [*Burlaga et al.*, 1998]. Note that the SW structures with extremely high densities have relatively short duration and, thus, the hourly averaged density reaches only maximum of 118 cm$^{-3}$.

The distribution of SW proton temperature is fitted very well by the lognormal PDF in the range from ~10000 K to ~400000 K. Inside that range the temperature correlates with the SW velocity [*Lopez*, 1987] such that fast SW is much hotter than the slow one. Outside the 2-$\sigma$ range, the wings of distribution turn down and go lower than the lognormal PDF, demonstrating a deficiency of extremely low and high temperatures. The deficiency can be also revealed from the comparison of the temperature dispersions: the calculated *RMSD* of 2.3 is less than standard deviation of 2.43 derived from the fitting.

Previous studies showed that the intervals of cold SW with proton temperature $T$<15000 K are often sustained for substantial periods, up to several days, accompanying a very slow SW propagating with speeds of 200 to 350 km/s [*Freeman and Lopez*, 1985]. It is claimed that the cold SW is not a separate component of the SW, but rather part of the continuum of the SW below 500 km/s, which satisfies a continuous linear relationship between the temperature and velocity [*Lopez and Freeman*, 1986]. In addition, plasma structures with very low proton temperature occur inside the ICME and in solar eruptive filaments/prominences [*Richardson and Cane*, 1995; *Burlaga et al.*, 1998]. Hence there should be no natural reasons for the deficiency of low-temperature events. To explain this discrepancy we can assume that the contribution of cold and slow SW to the total statistics might be reduced due to interaction with hot and fast SW streams, which sweep-away the slow solar wind.

In Figure 5b more prominent deficiency is found at very high temperatures of >400000 K. Those temperatures are associated mainly with the fast SW streams expanding from coronal holes at middle and high heliographic latitudes as revealed by Ulysses [*McComas et al.*, 2003]. Because of curved heliospheric streamer belt, a portion of fast and hot SW streams can penetrate to the ecliptic plane. The deficiency of very high temperatures might indicate to restricted contribution of that portion of fast SW to the total statistics of temperature.

It is interesting to note a far tail of the statistical distribution at extremely high temperatures of $>2 \cdot 10^6$ K. Such high temperatures are observed downstream of very strong interplanetary shocks [*Skoug et al.*, 2004]. They are far exceed those predicted from empirical relationship between the temperature and speed [*Lopez and Freeman*, 1986]. The extremely high temperature is a result of conversion of the kinetic energy of the interplanetary disturbance into thermal energy of the shocked gas. Hence they are generated by such extremely fast SW transients as ICMEs.

From the above we can see that the thermodynamic properties of SW plasma are related with the SW velocity $V$. The statistical distribution of the SW velocity is different from the lognormal PDFs (see Figure 5c). The distribution has a relatively large positive skewness. The most probable of ~380 km/s is shifted toward lower velocities relative to the median of 420 km/s. In the linear scale (not shown), the distribution is even far from the normal PDF, because in that representation the mean value of 443 km/s differs very much from the median of 420 km/s and from the most probable of 360 ~ 380 km/s.

In the range from 300 km/s to ~700 km/s the statistical distribution has a very wide peak, which contains more than 95% of the total statistics of 265699 hourly averages. That wide peak is formed by various kinds of the solar wind from slow plasma streams in the heliospheric current sheet to fast streams from the coronal holes [e.g. *Smith*, 2001; *McComas et al.*, 2003]. Note that the speed of fast streams correlates well with the size of coronal holes located near the central solar meridian [*Veselovsky et al.*, 2006; *Vrsnak et al.*, 2007].

It is interesting that the distribution extends smoothly from the peak of most probable to very high speeds of >800 km/s. Those fast SW streams are related to fast interplanetary transients, such as ICME and other eruptive events. Note that the highest SW speed of more than 2000 km/s was observed in the interplanetary sheath region leading by extremely fast ICME [*Skoug et al.*, 2004].

The number of events with slow SW decreases abruptly at velocities below 300 km/s. That deficiency of very slow SW might be due the "sweep-away" effect, which was discussed above in regard to deficiency of very low SW temperatures. It is difficult to interpret the extremely slow SW speeds of ~200 km/s. Some of those events correspond to intervals of extremely low SW density. Others might be due to unaccounted encounters to the magnetosheath or comet tails [*Baker et al.*, 1986; *Oyama et al.*, 1986]. Further studies of that subject are required.

**Table 2. Average values of the fluxes for the solar wind
at the Earth orbit during 1963-2007.**

| Physical quantity | Formula | Mean value |
|---|---|---|
| Mass flux density | $J = m_p D V$ | $4.0 \cdot 10^{-16}$ g/cm²·s |
| Momentum flux density (dynamic pressure) | $P_d = m_p D V^2$ | $2.0 \cdot 10^{-8}$ erg/cm³ (nPa) |
| Kinetic energy flux density | $S_k = \frac{1}{2} m_p D V^3$ | 0.44 erg/cm²·s |
| Potential energy flux density | $S_p = \frac{1}{2} m_p V_g^2 D V$, where $V_g = 618\ km/s$ | 1.7 erg/cm²·s |
| Enthalpy flux density for protons | $S_t = \frac{5}{2} n T V$ | $6.8 \cdot 10^{-1}$ erg/cm²·s |
| Magnetic energy flux density | $S_m = V \dfrac{B^2}{8\pi}$ | $6.1 \cdot 10^{-3}$ erg/cm²·s |

Based on the key plasma parameters, we calculate average physical numbers characterizing the SW plasma flow at the Earth orbit (see Table 2). As we have found, most of the SW parameters have better representation in logarithmic scale, where their statistical distributions are very close to the lognormal PDF. The average values of physical numbers from the Table 2 have been also calculated in the logarithmic scale, i.e. those are geometric mean. In calculation we take into account that the SW density $D$ is contributed by both protons with atom mass $A=1$ and helium ions with atom mass $A=4$.

$$D = n(1 + 4 \cdot He/p),\qquad (12)$$

where $n$ is concentration of protons per cubic centimeter and $He/p$ is a ratio of helium content to proton concentration. Here we suppose that the speed of helium ions is equal to the speed of protons. That assumption is reasonable because the average difference between the He and proton speeds was found in the SW observations of only about 10 km/s [*Ogilvie et al.*, 1982].

Statistical distribution of the He to proton ratio is presented in Figure 5d. The shape of distribution is different from both lognormal and normal PDF. The skewness and kurtosis are relatively large. Helium to proton ratio is averaged about 0.04 and vary from $10^{-3}$ to ~0.3. In Figure 3c we can see that the data coverage for the *He/p* is poor. There is no data before 1970. More-less regular measurements started only in 1998. As a result, the *He/p* is known only for less than 50% of measurements in 1963 to 2007 and, thus, the statistics is not very representative. Usually the data gaps are filled by the average value of 0.04. This assumption is very rough. The statistical distribution of *He/p* ratio has a long tail toward very small values of ~$10^{-3}$. Note that the ratio can be even smaller but in the OMNI dataset, only three positions after decimal point were used to record that value. In average, the *He/p* ratio correlates with the SW velocity [*Aellig et al.*, 2001]. Smaller *He/p* ratio is characteristic of the slow SW in the heliospheric current sheet and the ratio of >0.04 corresponds to fast streams from coronal holes [*Borrini et al.*, 1981; McComas e al., 2003]. Very high ratios of >0.1 occur within the ICMEs and eruptive filaments [*Borrini et al.*, 1982b; *Burlaga et al.*, 1998; *Skoug et al.*, 2004].

Interaction of the SW plasma with planetary atmospheres and magnetospheres as well as with interstellar gas is controlled mainly by such important parameter as SW dynamic pressure $P_d$. The dynamic pressure is contributed by both proton and helium ion population, and, hence it is calculated as follow

$$P_d = 1.67 \cdot 10^{-5} D V^2 \tag{13},$$

The pressure $P_d$ and velocity $V$ are expressed, respectively, in nPa and km/s.

Figure 5e shows the statistical distribution of SW dynamic pressure. The distribution is fitted very well by the lognormal PDF within 2-$\sigma$ vicinity of the median of 2 nPa, i.e. about 95% of statistics satisfy to lognormal distribution. The same properties of the dynamic pressure were revealed in previous studies of ~1 min averages of SW parameters measured by the ACE and Wind satellites, and of hourly averages from the OMNI data base [*Dmitriev et al.*, 2002a; 2004; 2005c].

The distribution of $P_d$ has zero skewness. However, pretty large positive kurtosis indicates to excess peakedness due to long tails of the distribution at small and high pressures. The tail at low pressures of <0.5 nPa is contributed mainly by low-density plasma structures. Small amount of slow SW streams is characterized by relatively high density, and hence does not contribute to the low-pressure tail. In contrast, the redundant statistics at high pressures of >7 nPa is mostly associated with fast SW transient events, which are often accompanied with dense regions of plasma compression. We should emphasize that SW structures with very low and very high dynamic pressure occupy only less than 5% of total statistics.

Statistical distribution of the kinetic energy flux density $S_k$ presented in Figure 5f is very similar to the distribution of dynamic pressure. The distribution is fitted by the lognormal PDF within 3-$\sigma$ vicinity of the median of 0.43 erg/cm$^2$s, i.e. more than 98% of statistics are distributed lognormally. The skewness of distribution is zero and kurtosis is relatively small. The moderate peakedness is due to excess of statistics at very small and very large magnitudes, which contribute only to less than 2% of statistics.

## Interplanetary Magnetic Field

Numerous studies are devoted to statistical properties of the IMF at various heliocentric distances and in various time scales. Most of studies note that statistical distribution of IMF intensity $B$ is different from the normal PDF. Some authors consider the lognormal PDF as the best fit for the distribution of $B$ [*Burlaga and King*, 1979; *Burlaga and Ness*, 1998; *Veselovsky et al.*, 1998b; 2000a; *Burlaga*, 2001b; *Dmitriev et al.*, 2005c]. However, *Feynman and Ruzmaikin* [1994] considering 3-hour averages, found that the statistical distribution of $B$ is different from the log-normal, because of non-zero skewness toward the small values, and relatively large positive kurtosis, corresponding to larger peakedness relative to a normal distribution. *Hartlep et al.* [2000] propose another approach of the IMF distribution in a form of fixed mean and normally distributed components. In particular, they reveal very good correspondence with the observed IMF statistical distribution when the normal component magnitude distribution is axisymmetric about the mean field (which is mainly aligned with the Archimedean spiral) but admits a high degree of variance anisotropy, with parallel variance much less than perpendicular variance. On the other hand, *Bieber et al.* [1993] revealed that the amplitudes of the spectra of variations are comparable for the IMF components, respectively, in the north-south direction, perpendicular to the Archimedean spiral in the ecliptic plane, and parallel to the Archimede spiral.

Figure 6a shows statistical distribution of the IMF intensity $B$ represented in logarithmic scale. The distribution is very close to lognormal PDF within 2-$\sigma$ and 1.5-$\sigma$ deviations, respectively, toward low and high intensities relative to the average of 6 nT. So more than 95% of the statistics satisfy to the lognormal distribution. There are no any well-defined correlations of the magnetic field with other heliospheric parameters in that range. The skewness of distribution is relatively small. The positive non-zero kurtosis is due to pretty prominent wings, which exceed the lognormal PDF at very low and very high magnitudes and contribute less than 5% of total statistics.

The range of very weak IMF intensities of <2 nT contains less that 1% of the total statistics. *Zurbuchen et al.* [2001] studying the magnetic field depletions, or so-called magnetic holes, reveal that they can last for up to several hours. The magnetic holes are associated with increases in the SW density and temperature and large magnetic field rotations. However, they are not associated with large-scale magnetic field polarity changes. From analysis of the chemical composition, the authors conclude that the magnetic holes very likely develop in the heliosphere and are not of direct solar origin.

Very strong IMF of >13 nT is observed in 4% of cases, which are characterized by wide variability of the SW parameters from very small to very large values. An excess of strong fields relative to the lognormal distribution was reported in several studies [e.g. *Burlaga and Szabo*, 1999]. There are various physical processes contributing to the long tail at large IMF magnitudes. It is well known that the magnetic field can be enhanced significantly inside the ICMEs [*Burlaga et al.,* 1987; 2001; *Owens et al.*, 2005]. It was found that the IMF intensity correlates well with the speed of ICME [*Owens and Cargill*, 2002]. The IMF is also enforced in the interplanetary sheath and other compressed regions formed due to interaction of high speed structures with the ambient SW [*Burlaga and King*, 1979; *Borrini et al.*, 1982a; *Bothmer and Schwenn*, 1995; *Dal Lago et al.*, 2001; *Owens et al.*, 2005].

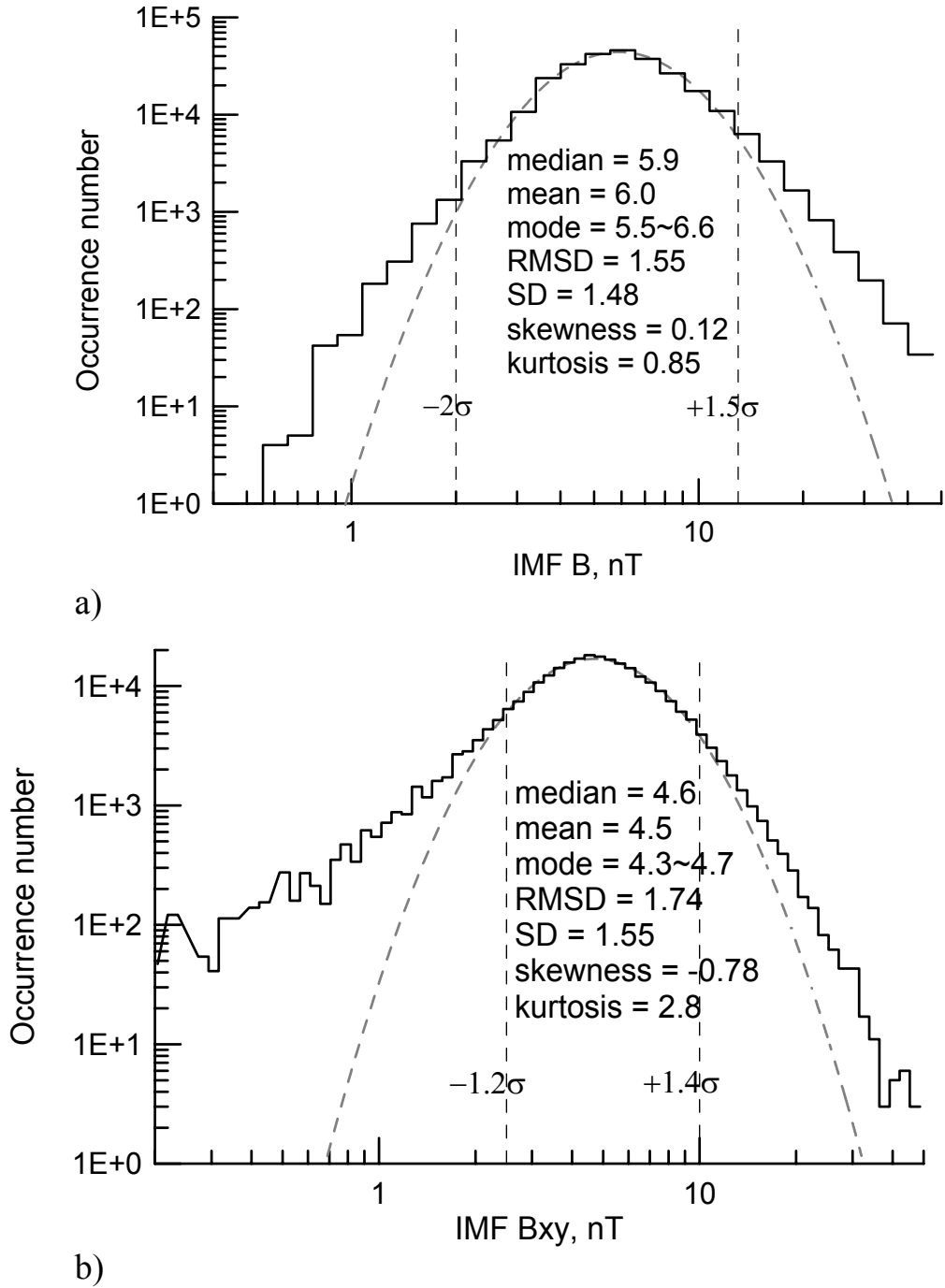

a)

b)

Figure 6. Statistical distributions of the IMF for time interval from 1963 to 2007 in logarithmic scale: (a) strength $B$, (b) projection of the IMF vector onto the ecliptic plane $B_{xy}$. Dashed curves depict the best fit of the $B$ and $B_{xy}$ by the lognormal PDF. Vertical thin dashed lines restrict the best-fit intervals.

The vector of IMF can be presented as a sum of three orthogonal components $B_x$, $B_y$, and $B_z$. In the GSE coordinate system, X-axis is pointed to the Sun, Y-axis lies in the ecliptic plane and directed duskward, and Z-axis is perpendicular to the ecliptic plane and directed northward. Projection of the IMF vector to the ecliptic plane $B_{xy}$ is simply presented as the vector sum of $B_x$ and $B_y$ components: $\vec{B}_{xy} = \vec{B}_x + \vec{B}_y$, and the magnitude of $B_{xy}$ is equal to $\left| \vec{B}_{xy} \right| = \sqrt{B_x^2 + B_y^2}$ .

Statistical distribution of the $B_{xy}$ magnitude is presented in Figure 6b. It is close to the lognormal PDF with average of 4.6 nT within a narrow interval from $-1.2$ to 1.4 standard deviations, i.e. ~80% of the total statistics of $B_{xy}$ satisfy to lognormal distribution. The distribution is slightly skewed toward small values. That skewness is mainly due to very prominent tail extending to small magnitudes. The tail together with excess of large values of $B_{xy}$ leads to a pretty large positive kurtosis and large *RMSD*=1.74 relative to the standard deviation of 1.55. The tail at large magnitudes of $B_{xy}$ (>10 nT) has apparently the same nature as the tail of high IMF intensities. The abundant statistics at small values of $B_{xy}$ (<2 nT) is mainly contributed by the variations of IMF orientation in Alfvén waves and will be discussed later.

We find that the problem of lognormal distribution of IMF intensity comes to a problem of lognormal distribution of the component $B_{xy}$ in the ecliptic plane. The statistical distribution of $B_{xy}$ has been studied in detail by *Luhmann et al.* [1993] on the base of an ideal Archimedean spiral model of IMF. In that model the solar magnetic field is stretched out to a spiral by the expanding SW plasma from the rotating Sun. As a result, the radial $B_x$ component decreases as a square of distance from the Sun and the tangential $B_y$ component is formed due to rotation of the Sun. $B_y$ and $B_z$ components decrease inversely proportional with distance. In this model the changes in SW velocity should anticorrelate with changes in the magnitude of $B_{xy}$. *Luhmann et al.* [1993] found that the IMF vector is indeed oriented along the Archimedean spiral but they did not find distinct anticorrelation between the velocity and $B_{xy}$ magnitude. Following to *King et al.* [1981] the authors concluded that the solar source field variation must play an important role in the observed variability of IMF at the Earth orbit.

In Figure 7 we show statistical distributions of the IMF components $B_x$ and $B_y$ as well as two-dimensional probability distribution P($B_x$, $B_y$). The distributions of $B_x$ and $B_y$ have a typical two-peak shape with long tails as reported in previous studies [*Luhmann et al.*, 1993; *Dmitriev et al.*, 2004]. The peaks are situated at about 3 nT. Apparently, those peaks correspond to the average value of $B_{xy}$ =4.5 nT. The long tails correspond to the excess of large values in the $B_{xy}$ statistical distribution.

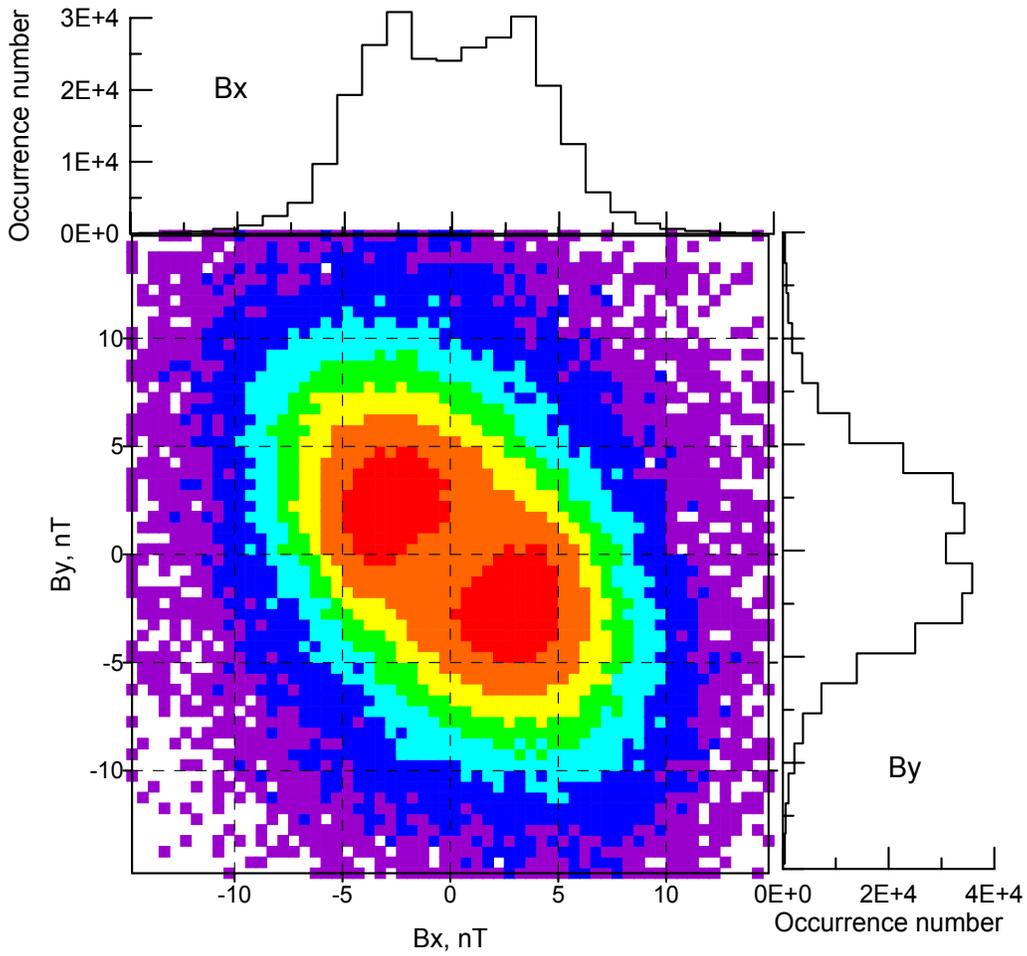

Figure 7. Statistical distribution of the IMF $B_x$ and $B_y$ components in GSE coordinate system. Two-dimensional distribution is presented in rainbow color scale: from small occurrence numbers of a few counts (violet) to maximum statistics of >500 counts (red). The top and right histograms represent statistical distributions of the $B_x$ and $B_y$ component, respectively. The distributions of the $B_x$ and $B_y$ components correspond to predominant orientation of the IMF vector along the Archimedean spiral, which has an angle of ~135° relative to the sunward direction (pointed by positive $B_x$).

The two-dimensional distribution of the components $B_x$ and $B_y$ has a long ridge corresponding to a predominant orientation of the $B_{xy}$ vector along a line, which is inclined on ~-45° relative the X-axis. The same predominant orientation of the IMF was reported earlier [e.g. *Luhmann et al.*, 1993; *Veselovsky and Tarsina,* 2001]. This orientation is very close to the Archimedean spiral. In the solar equatorial plane the angle $\alpha$ between the Archimedean spiral and X-axis is calculated as follow:

$$\alpha = \tan^{-1}\left(\frac{2\pi r}{\tau_s V}\right), \qquad (14)$$

where $r$ is heliocentric distance, $V$ is the SW velocity and $\tau_s$=25.4 days is the sidereal rotation period of the Sun at the equator. Taking distance $r$=1.5 $10^8$ km and median SW velocity of 420 km/s directed antiparallel to the X-axis, we obtain $\alpha$=-45° at the Earth orbit.

We should emphasize that the predominant orientation along the Archimedean spiral is revealed even at very large $B_{xy}$ magnitudes of >15 nT. Such strong magnetic field in the ecliptic plane can not be only the result of field line stretching. The ecliptic magnetic field described by the ideal Parker spiral model is given by

$$\left|B_{xy}\right| = B_0 a^2 r^{-1} \sqrt{r^{-2} + \Omega^2 / V^2} \tag{15}$$

were $a$ is the radius of the solar source surface. Using this equation we simply estimate that variations of the solar wind velocity with average of 420 km/s and dispersion of ~1.25 (see Table 1) cause the relative variation of ~1.24 in the $B_{xy}$ magnitude. That variation is apparently smaller than logarithmic $RMSD$ of 1.74 and standard deviation of 1.55, which we find from the statistical distribution of the $B_{xy}$ (Figure 6b).

Hence it is rather possible that significant part of the $B_{xy}$ variations is originated from solar sources. Variations of the Archimedean spiral angle from $\alpha$~-55° in slow SW ($V$=300 km/s) to $\alpha$~-31° in the fast SW ($V$=700 km/s) also contribute to a pretty large dispersion of the ridge of most probable values in the two-dimensional distribution P($B_x$,$B_y$) presented in Figure 7.

Another important heliospheric phenomenon contributing to the dispersion of magnetic field components is large-amplitude Alfvén waves propagating outward from the Sun [*Belcher and Davis*, 1971; *Tsurutani and Gonzalez*, 1987; *Tsurutani et al.*, 1995]. They have a broad wavelength range up to 5·$10^6$ km and beyond, which corresponds to period of hours. Most Alfvén waves in the interplanetary medium are likely the undamped remnants of waves generated at the Sun. They occur mainly in high-speed SW streams and on their trailing edges where the velocity slowly decreases. In the Alfvén waves, the magnetic field orientation varies such that one IMF component increases and another decreases. As a result, those variations can contribute a lot to the statistics at both very small and large values of $B_x$ and $B_y$. The largest amplitude Alfvénic fluctuations of ~10 nT in the IMF component are observed in the compression regions at the leading edges of high-velocity streams, i.e. in the CIR region.

The Alfvén waves are one of the sources of IMF fluctuations perpendicular to the ecliptic plane, i.e. variations of the $B_z$ component, which statistical distribution is presented in Figure 8a. The distribution has a zero skewness and can be fitted well by a normal PDF within 2-$\sigma$ interval (i.e. ±4.18 nT) around the average of 0 nT. However the $RMSD$ of 3.1 nT is much larger than the standard deviation $SD$ of 2.1 nT, because of presence of very prominent tails, which contain about 13% of the total statistics. Because of those tails the kurtosis is positive and very large. The large peakedness of the $B_z$ distribution is also reported by [*Feynman and Ruzmaikin*, 1994]. The excess of large $B_z$ magnitudes can be explained by the Alfvén waves only partially. Large $B_z$ is generated due to compression in the CIR regions and in the interplanetary sheaths. In the latter case the $B_z$ can achieve extremely high values but for a short time. The long-duration large and extremely large $B_z$ occurs in the ICME and as a result of interaction between the ICME and other SW structures [*Bothmer and Schwenn*, 1995; *Dal Lago et al.*, 2001].

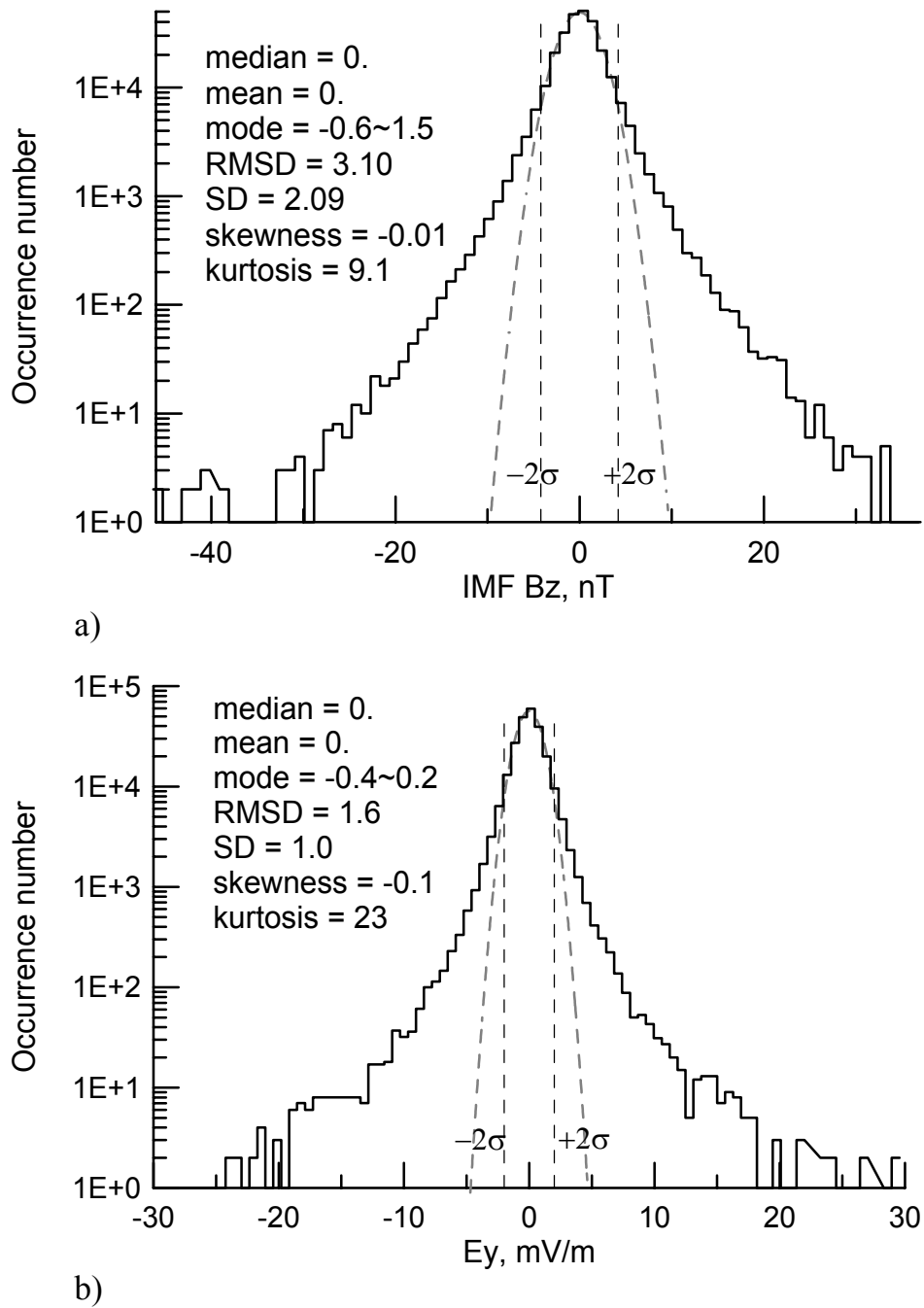

a)

b)

Figure 8. Statistical distributions of (a) IMF $B_z$ component and (b) Y-component of induced interplanetary electric field $E_y$ for time interval 1963 to 2007. Dashed curve depicts the fitting by normal PDF. Vertical thin dashed lines restrict the best-fit intervals.

The IMF $B_z$ and induced interplanetary electric field $V \times B$ play a key role in transmission of the SW energy into the Earth's magnetosphere and hence control the system of magnetospheric currents [*Burton et al.*, 1975; *Akasofu*, 1979; *Iijima and Potemra*, 1982; *Tsyganenko*, 2002a,b; *Tsurutani et al.*, 2004]. Geoeffective $E_y$ component of the induced electric field is defined as follow:

$$E_y = 10^{-3} V_x \cdot B_z \qquad (16)$$

Here the radial component of the velocity $V$ is negative and expressed in km/s, $B_z$ in nT and $E_y$ in mV/m. Note that in this case the IMF component $B_z$ is represented in the GSM coordinate system, which is related to the orientation of the Earth dipole axis projection to YZ plane. In the GSM system the Z component is contributed mainly by the IMF $B_z$ and partially by the $B_y$ components represented in the GSE coordinate system.

Statistical distribution of the induced electric field $E_y$ is presented in Figure 8b. The distribution is similar to one of the $B_z$ (Figure 8a). It is well fitted by the normal PDF within 2-$\sigma$ interval from –2 to 2 mV/m around the zero average. The distribution is symmetrical relative to the mode (skewness $k_3$=0). The kurtosis is large because of great excess of large $E_y$ magnitudes, which form prominent wings extending up to extremely high values of ~30 mV/m. As a result only 88% of the statistics at relatively low $E_y$ magnitudes are distributed normally. The wings are contributed by both the excess of large intensities of the IMF $B_z$ (see Figure 8a) and the abundant statistics of the fast SW streams (see Figure 5c). Note that the strongest magnitudes of $E_y$ occur in extremely fast interplanetary transients (ICME and related sheath regions), which often contain very strong IMF $B_z$.

*Burton et al.* [1975] found a criterion for the onset of geomagnetic storms: a storm starts when the $E_y$ is larger than 0.5 mV/m. Using the statistical distribution of $E_y$ we can find that the criterion is satisfied in 30% of cases. So about one third of time the magnetosphere stays under magnetic storm conditions.

## Relevant Physical Quantities

Using measured parameters of the SW and IMF we calculated various quantities characterizing average physical properties of the interplanetary medium at the Earth orbit and listed them in Tables 2, 3 and 4. In the previous sections we found that statistical distributions of all measured parameters of the solar wind and IMF intensity are very close to lognormal PDF. It is easy to show that the lognormal distribution is multiplicative, i.e. multiplication/division of two random variables distributed log-normally has also lognormal distribution. As an example we can indicate the log-normally distributed SW dynamic pressure $P_d$. Hence physical quantities being a multiplication of log-normally distributed measured parameters should be represented in the logarithm scale. So we calculate average logarithms of the physical quantities (Equation 8).

**Table 3. Mean heliospheric plasma conditions at the Earth orbit during 1963-2007.**

| Physical quantity | Formula | Mean value |
|---|---|---|
| *Alfvèn velocity* | $V_A = \dfrac{B}{\sqrt{4\pi\, nm_p}}$ | 56.8 km/s |
| Sonic velocity for protons | $c_S = \sqrt{\dfrac{5T}{3m_p}}$ | 33.7 km/s |
| Alfvèn-Mach number | $M_A = \dfrac{V}{V_A}$ | 7.7 |
| Sonic Mach number for protons | $M_S = \dfrac{V}{c_S}$ | 13. |
| Gas-kinetic proton pressure (Thermal proton pressure) | $P_t = nT$ | $.62\cdot10^{-10}$ erg/cm$^3$ ($10^{-2}$ nPa) |
| Magnetic pressure | $P_m = \dfrac{B^2}{8\pi}$ | $1.4\cdot10^{-10}$ erg/cm$^3$ ($10^{-2}$ nPa) |
| Proton gas-kinetic to magnetic pressure ratio | $\beta_p = \dfrac{8\pi\, nT}{B^2}$ | 0.43 |
| Coulomb collision time for electrons | $\tau_e \cong 10^{-2} T_e^{3/2} n^{-1}$ | $\sim 9.9\cdot10^4$ s |
| Coulomb collision time for protons | $\tau_p \cong 0.6 T_p^{3/2} n^{-1}$ | $\sim 2.9\cdot10^6$ s |

**Table 4. Main plasma characteristics calculated for mean heliospheric parameters at the Earth orbit during 1964 -1996.**

| Physical quantity | Formula | Mean value |
|---|---|---|
| *Plasma frequency* | $\omega_0 = \sqrt{\dfrac{4\pi n e^2}{m_e}}$ | $1.30 \cdot 10^5$ s$^{-1}$ |
| Electron cyclotron frequency | $\omega_{ce} = \dfrac{eB}{m_e c}$ | $1.04 \cdot 10^3$ s$^{-1}$ |
| Proton cyclotron frequency | $\omega_{cp} = \dfrac{eB}{m_p c}$ | $0.56$ s$^{-1}$ |
| Upper hybrid frequency | $\omega_{h1} = \left( \omega_0{}^2 + \omega_{ce}{}^2 \right)^{1/2} \approx \omega_0$ | $1.30 \cdot 10^5$ s$^{-1}$ |
| Lower hybrid frequency | $\omega_{h2} = \left( \omega_{cp} \omega_{ce} \right)^{1/2}$ | $24.1$ s$^{-1}$ |
| Mean thermal speed of protons | $V_p = \sqrt{\dfrac{3T}{m_p}}$ | $45.8$ km/s |
| Larmor radius for protons | $r_p = \dfrac{V_p}{\omega_{cp}}$ | $81.8$ km |

Average values of the energy, momentum and mass fluxes for the SW are shown in Table 2. Note that here we consider SW density $D$ which accounts the helium contribution (Equation 12). One can see that the largest energy flux density is carried in the shape of potential and kinetic energies of the solar wind. The density of enthalpy (thermal) and magnetic energy fluxes are smaller on about 2 orders of magnitude. The total energy flux density of ~2.3 erg cm$^{-2}$ s$^{-1}$ amounts only a small portion of the total energy flux density emitted by the Sun in the form of electromagnetic radiation [*Veselovsky et al.*, 1999].

Mean heliospheric plasma and IMF quantities are listed in Table 3. In the present case we use only proton concentration $n$ and temperature $T$ in the solar wind. The neglect of helium contribution leads to ~10% overestimation of the Alfvèn speed, while the Alfvèn Mach number is underestimated on ~10%. Other plasma quantities such as the sonic speed $c_s$, sonic Mach number $M_s$, gas-kinetic pressure $P_t$ and plasma $\beta$ depend strongly on electron temperature, which is not available in most cases.

The electron temperature $T_e$ varies in very wide range from ~5·10$^4$ to ~10$^6$ K, and it has only very weak correlation with the proton temperature [*Newbury et al.*, 1998; *Salem et al.*, 2003]. It was found that in average the electron to proton ratio $T_e$/T varies from 0.5 in the fast wind with velocities of ~700 km/s to ~4 in the slow solar wind. The average electron temperature is ~1.4·10$^5$ K, i.e. almost two-time higher than the average proton temperature of 8.5·$10^4$ K (see Table 1). Hence neglect of the electron temperature may lead to ~50% underestimation of the sonic speed $c_s$ and 50% overestimation of the sonic Mach number $M_s$. The magnitudes of gas-kinetic pressure $P_t$ and plasma $\beta$ might be about two-time underestimated. That is only rough estimation, because of the absence of electron temperature data.

From Table 3 we can conclude that in average the SW flow at the Earth orbit is supersonic ($M_s>1$) and superalfvènic ($M_a>1$). Interaction of such SW with the magnetosphere obstacle causes generation of fast magnetosonic wave enveloping the magnetosphere, or so-called bow shock [e.g. *Spreiter et al.,* 1966]. The Alfvènic and sonic Mach numbers and plasma $\beta$ are the key parameters controlling the bow shock formation, i.e. conditions for SW flow about the magnetosphere [e.g. *Dmitriev et al.*, 2003]. Statistical distributions of dimensionless quantities $M_a$, $M_s$ and $\beta$ are presented in Figure 9. As one can see, those distributions can be well fitted by a lognormal PDF as reported before by *Dmitriev et al.* [2003]. Similar behavior was found with 1 min data [*Mullan and Smith,* 2006].

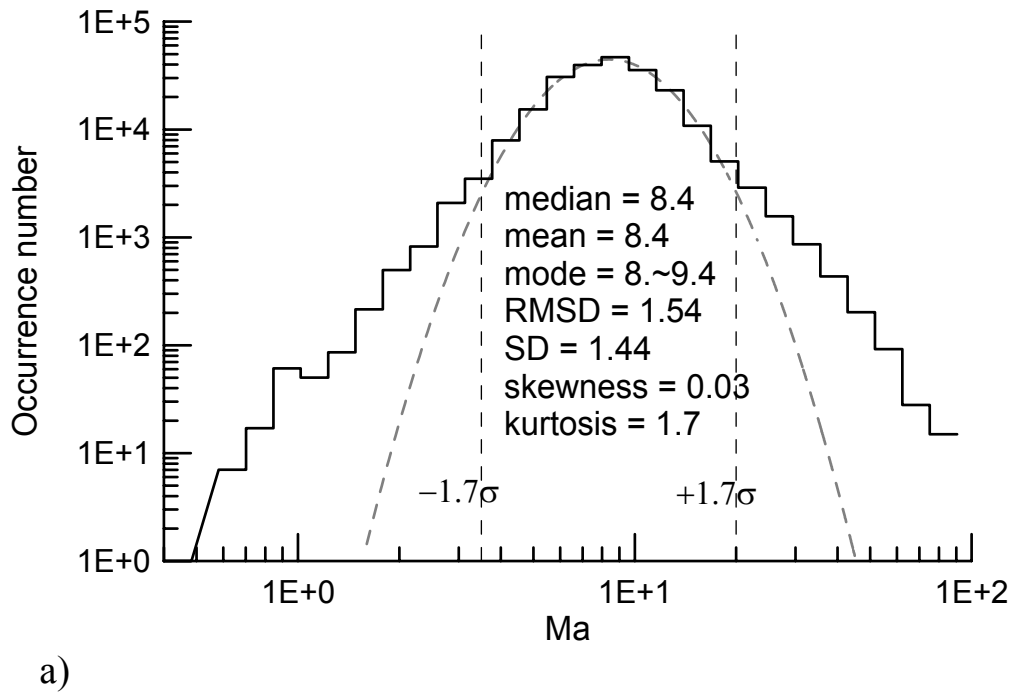

a)

Figure 9. (Continued)

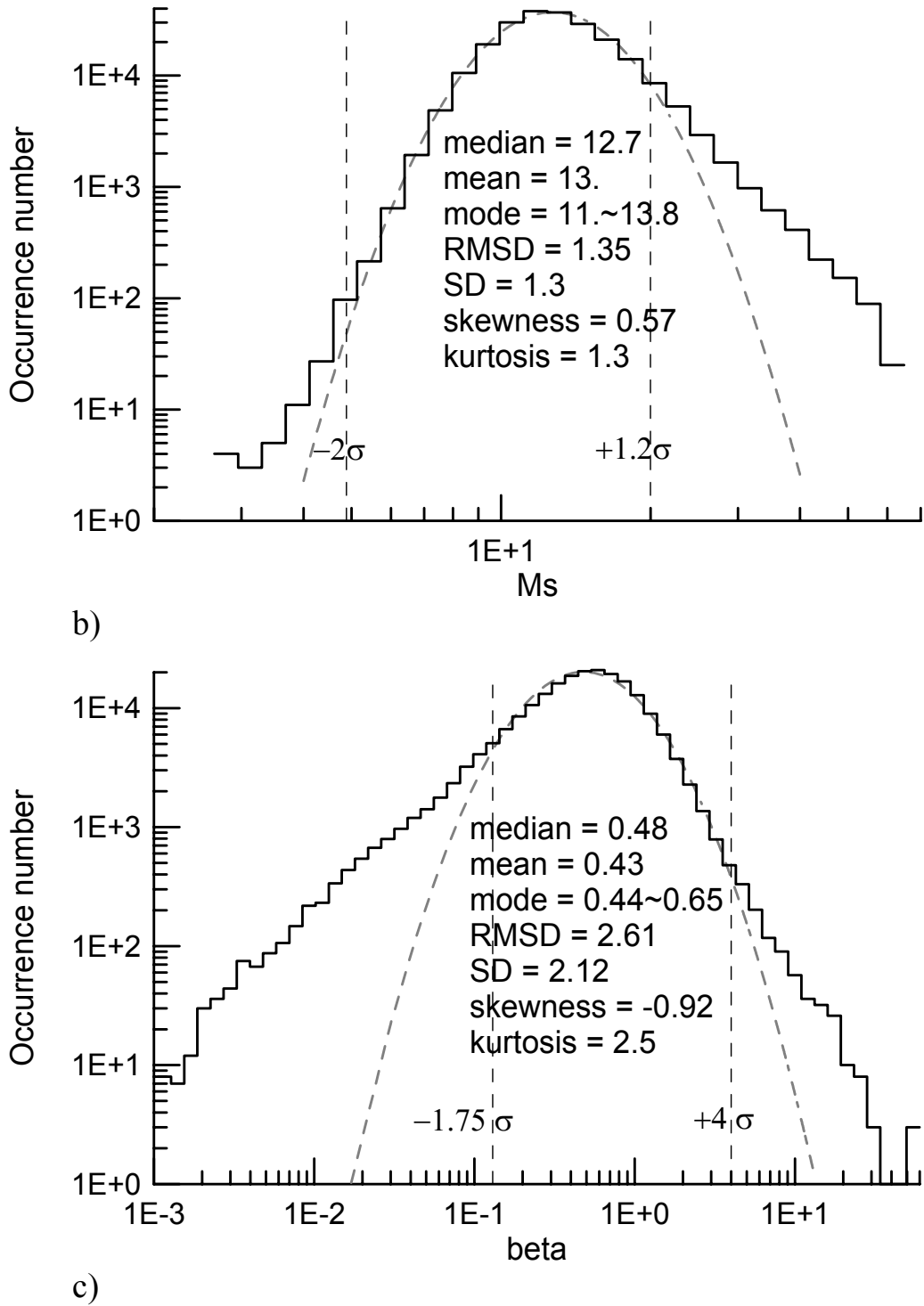

Figure 9. Statistical distributions of the dimensionless quantities in logarithmic scale for time interval 1963 to 2007: (a) Alfvén-Mach number $M_a$, (b) Sonic Mach number $M_s$, and (c) thermal to magnetic pressure ratio $\beta$. The distributions are fitted well by the lognormal PDF shown by dashed line. Vertical thin dashed lines restrict the best-fit intervals.

The distribution of Alfvèn Mach number (Figure 9a) is very close to the lognormal PDF within 1.7-$\sigma$ interval around the average of 8.4 that contains ~95% of total statistics. The distribution has zero skewness and relatively large positive kurtosis, which is caused by abundant statistics at small and large $M_a$ values. From analysis of statistical distributions of the SW velocity, density and IMF intensity, we can find that the tail of relatively small values of $M_a$ can be contributed by relatively small SW densities and/or strong IMF. Such conditions often occur inside ICMEs. It is important to note the excess of extremely small values of $M_a$, which can be sometimes less than 1. Correction to the He abundance helps a little. Hence in some very rare cases the SW can be subalfvènic. *Usmanov et al.* [2005] have studied this problem and find that events of very low Alfvèn Mach number at the Earth orbit are associated mainly with extremely low SW density (<0.3 cm$^{-3}$) and only few events are due to very high IMF intensity of >10 nT. On the other hand, the extremely high $M_a$ can occur inside the magnetic holes [*Zurbuchen et al.*, 2001].

Contrary to the $M_a$, the sonic Mach number $M_s$ is always larger than 1, as we can see from its statistical distribution in Figure 9b. About 92% of the $M$s statistics is fitted well by the lognormal PDF in the range from 5 to 20 with average of ~13. The events with very low $M_s$ of <4 are very rare. The positive skewness and relatively large positive kurtosis are due to the excess of large values of $M_s$. The tail of large $M_s$ can be contributed by fast and cold SW structures, such as fast ICMEs.

Figure 9c shows statistical distribution of the thermal to magnetic pressure ratio (plasma $\beta$). The distribution is close to the lognormal PDF with mode of 0.48 and standard deviation of $\sigma$=2.12 within interval restricted by lower limit of -1.75$\sigma$ ($\beta$~0.1) and upper limit of 4$\sigma$ ($\beta$~4). About 89% of the statistics are covered by the lognormal distribution. The skewness of statistical distribution is large negative and the kurtosis is relatively large positive because of significant excess of small values, which form a long tail extending to very small $\beta$ of <10$^{-3}$. In that region the magnetic pressure $P_m$ is dominant and exceeds the gas-kinetic pressure $P_t$ on orders of magnitude. This tail is contributed by cold plasma structures of SW with strong magnetic field, such as ICMEs and heliospheric current sheet. It is important to note that accounting the electron temperature leads to increase of the average value of $\beta$ up to 1 and even more and can modify the tail of low $\beta$ because the slow SW streams in the heliospheric current sheet are characterized by very high electron temperature contribution ($T_e/T$~4). Hence the problem of very low $\beta$ is a subject of future studies.

Considering statistical distribution of $\beta$ we find that the thermal energy of the SW plasma $P_t$ is comparable or even less than the energy of interplanetary magnetic field, $P_m$. Both of them amount of about 10$^{-10}$ erg/cm$^3$ (see Table 3). That is at least two orders of magnitude less than the SW plasma momentum flux density (or dynamic pressure, $P_d$), which amounts of 2•10$^{-8}$ erg/cm$^3$ at the Earth orbit (see Table 2). Note that under some circumstances inside the ICMEs the dynamic pressure $P_d$ can be reduced by several times because of low density, and becomes comparable with significantly enhanced pressure of very intense magnetic field.

# 3. Characteristic Periods

Study of periodicities in dynamics of the heliospheric parameters is difficult because of very large amount (~30%) of data gaps having a wide range of durations (see Figures 3 and 4).

This problem is solved using various methods for sparse data processing from simple averaging and smoothing to interpolation by polynomials and splines. Also various techniques are applied for studying of periodicities in long time profiles of the heliospheric parameters containing hundreds of thousand data points such as widely used fast Fourier transform (FFT), wavelet techniques, spectral wave analysis (SWAN diagrams) and sophisticated methods of maximum entropy and singular spectrum analysis [e.g. *Szabo et al.*, 1995; *Rangarajan, and Barreto*, 2000; *Veselovsky and Tarsina,* 2002b; *Mursula and Vilppola*, 2004; *Kane*, 2005].

In the present study we use a method of Fourier transform for unequally-spaced data proposed by *Deeming* [1975]. That method does not need any specific data processing for elimination of the data gaps. Only linear detrending of the data is desirable. Obviously, the spectrum derived by that method suffers from frequency interference, which is caused by both the finite data length and data spacing. The interference produces numerous subsidiary peaks. In this situation it is rather difficult to distinguish between the meaningful and spurious peaks. One of the ways to resolve this problem is a method of reducing statistics [*Dmitriev et al.*, 2000]. Namely, the initial data set is reduced chaotically by increasing the number of data gaps with a random duration and spacing. The meaningful peaks survive even after the statistics reduction on several tens of percent (up to 85%). The subsidiary peaks change fast with the reduction, i.e. their amplitudes and locations vary and new peaks occur.

We should note that 'meaningful' in this context does not mean any quantitative, but only some qualitative characterization and indication for future more rigorous and complete studies. The method of integral Fourier transforms and all its modifications as applied for finite, non-complete non-stationary data sets obtained with a finite time resolution and accuracy of noisy data has its own limitations and restricts correct interpretation of obtained 'periodicities' with non-defined reservations from the side of high (Nyquist sampling) and low (finite data set length) frequencies as well as spurious harmonics. We caution the reader against straight applications of results without notification of accuracy and stress their illustrative and very preliminary nature only. The same can be said also about subsequent periodicities described below.

One can easily reveal by using all these methods the robust signatures of the presence of the solar cycle, the Earth rotation and the Sun rotation signals in SW and IMF data obtained near the Earth. They are seen as a group of peaks in periodograms around corresponding characteristic durations of ten years, a year and a month, correspondingly. The accuracy and the reliability of numbers characterizing the positions and amplitudes of individual peaks is highly limited and not analyzed. Because of this, there is no big physical sense in all these details, which are not quantified in the present study, as well as in many similar works. Nevertheless, they are traditionally marked also in our study only for the purpose of further studies. As for eigenfrequecies of the solar oscillations manifested in the SW and IMF parameters [*Thomson et al.*, 2002], there is no unique interpretation as yet in our opinion.

**Solar Periodicities**

Periodogram of daily sunspot number variations in 1963 to 2007 is presented in Figure 10. The periods are changed consequently from 3 days to 16437 days (45 years) with a step of 1-day. In calculation of spectral amplitudes we use logarithm of the sunspot number $W$, because the lognormal scale is more representative, as we have demonstrated in the previous section (see Figure 2). The zero magnitudes of $W$ are considered as data gaps. Hence this periodogram contains mainly information about the rising, maximum and declining phases of the four solar cycles 20 to 23.

Strongest amplitude is revealed for the solar cycle periodicity of 10.6 years. This period is a result of superposition of two strong 10-year cycles with two relatively weak 12-year cycles. As we can see in Figure 10, the four last solar cycles do not demonstrate 22-year period in sunspot number. This is because of weak 23rd solar cycle [e.g. *Dmitriev et al.*, 2005c]. That cycle breaks a so-called "Gnevyshev-Ohl" empirical rule that the odd cycles are higher than the previous even cycles [*Gnevyshev and Ohl*, 1948]. Such violations happened also in more distant epochs. Tentatively they could be associated with longer secular cycles and trends. Physical origins of solar cycles are not known. They can be the internal property of the Sun as a star and its dynamo action as mostly believed now. Otherwise, planetary influences and interstellar causes could be involved. For example, it is believed sometimes that orbital rotation of giant planets (Jupiter, Saturn, Uranus and Neptune) is a natural source of the solar activity north-south asymmetries, and decadal and secular variations in the range of periods from ~11 years to ~165 years [e.g. *Juckett*, 2000].

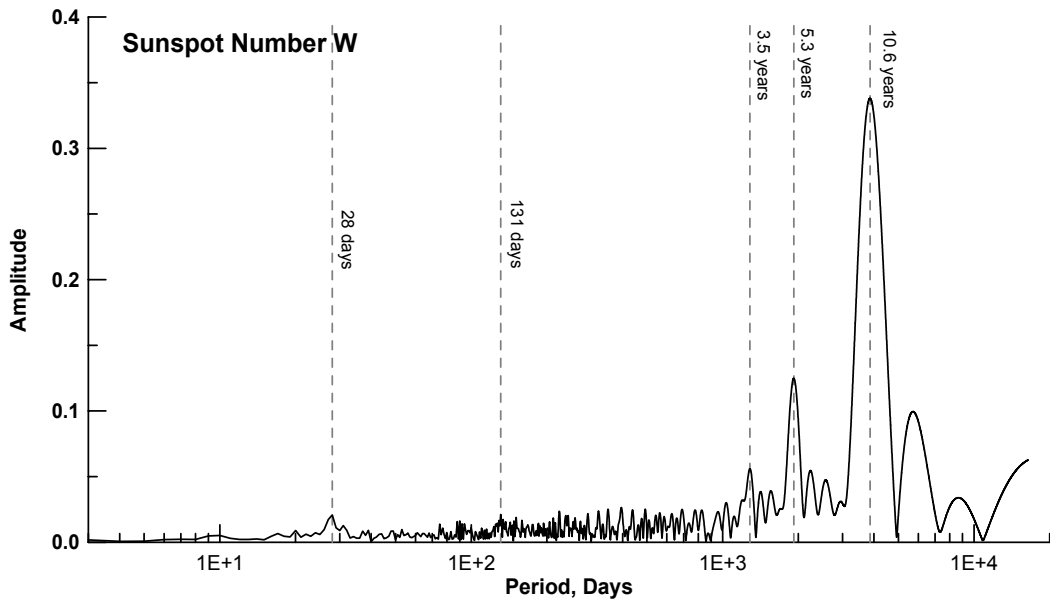

Figure 10. Periodogram of the sunspot number $W$ for time interval 1963 to 2007. Some characteristic periods are indicated by vertical dashed lines.

Periods of 5.3 years and 3.5 years may correspond, respectively, to 1/2 and 1/3 of the 10.6-year solar cycle. Similar oscillations with 5.2-year and 3.2-year periods as well as ~2.5-year and ~1.9-year periodicities have been also found being meaningful from analysis of the sunspot number variations in 1975-2001 by a method of maximum entropy [*Kane*, 2005]. Recently different authors report those periodicities for various solar indices such as solar magnetic field, coronal index etc. [*Kane*, 2005; *Mavromichalaki et al.*, 2005]. Using a basic wavelet technique for analysis of the solar magnetic flux, *Valdés-Galicia et al.* [2005] reveal periodicities of ~5, ~3, ~1.7, and ~1.3 years and demonstrate their alternating importance during consecutive odd 21st and even 22nd solar cycles. *Mavromichalaki et al.* [2005] also mention about variability of amplitudes and periods of the solar periodicities in the range from 0.5 year to 3 years. Note that the periods less than 3 years have relatively small amplitudes.

It is rather difficult to judge whether the periods of several years are sub-harmonics of the ~11-year solar cycle, i.e. correspond to double and triple etc. frequencies, or they are manifestations of physical processes driving the solar activity. In Figure 1 we can find that the smoothed sunspot number demonstrates various periodicities, which vary from cycle to cycle. For example, during maximum and declining phase of the 20th cycle we clearly see variations with ~2-year period. However during the 21st cycle a 1-year period prevail. The rising and declining phases of cycles 22nd and 23rd are smooth and have no any preferable periods. Prominent quasi-biennial variations during solar maxima are associated with so-called Gnevyshev gaps and the solar magnetic field reversal [*Gnevyshev and Ohl*, 1948]. Recently *Charvátová* [2007] has claimed that the period of 1.6, 2.13 and 6.4 years are associated with the solar interior motion due to disturbing effects from the rotation of terrestrial planets Mercury, Venus, Earths and Mars. Numerical estimation of the power and significance of the planetary effect to the solar variations is a subject for further investigations.

A periodicity around of 28 days in Figure 10 is apparently associated with the synodic period of the Sun rotation as seen from the Earth (27.3 days). The synodic period is also found in various solar indices [e.g. *Mursula and Zieger*, 1996; *Mavromichalaki et al.*, 2005]. In Figure 10 we find a periodicity of ~131 days, which is related neither to the solar rotation nor to the solar cycle. The 131-day period is revealed in hard X-ray flare activity during declining phase of solar cycle 23 [*Jain et al.*, 2008]. The nature of that period is still unclear.

## SW Periodicities

The periodicities of heliospheric parameters having 1-hour time resolution are studied using quasi-logarithmic fragmentation of the periods. Namely, in the range of periods from 1 to 100 days we use 6-hour step, and further in ranges of 100 days ~ 1.1 year, 1.1 ~ 5.5 years, 5.5 ~ 11 years, and > 11 years we use respectively, 12-hour, 1-day, 10-day and 1-month steps. The fragmentation helps to reduce the calculation time. Note that changing the fragmentation does not affect significantly on the resultant periodicities.

Figure 11 demonstrates periodograms of the SW parameters. We analyse the logarithms of magnitudes of parameters. We do not consider periods longer than solar cycle because they are affected by interference with the finite length of data sets. As one can see in Figure 11a, the periodicities of solar wind proton velocity and temperature are very close.

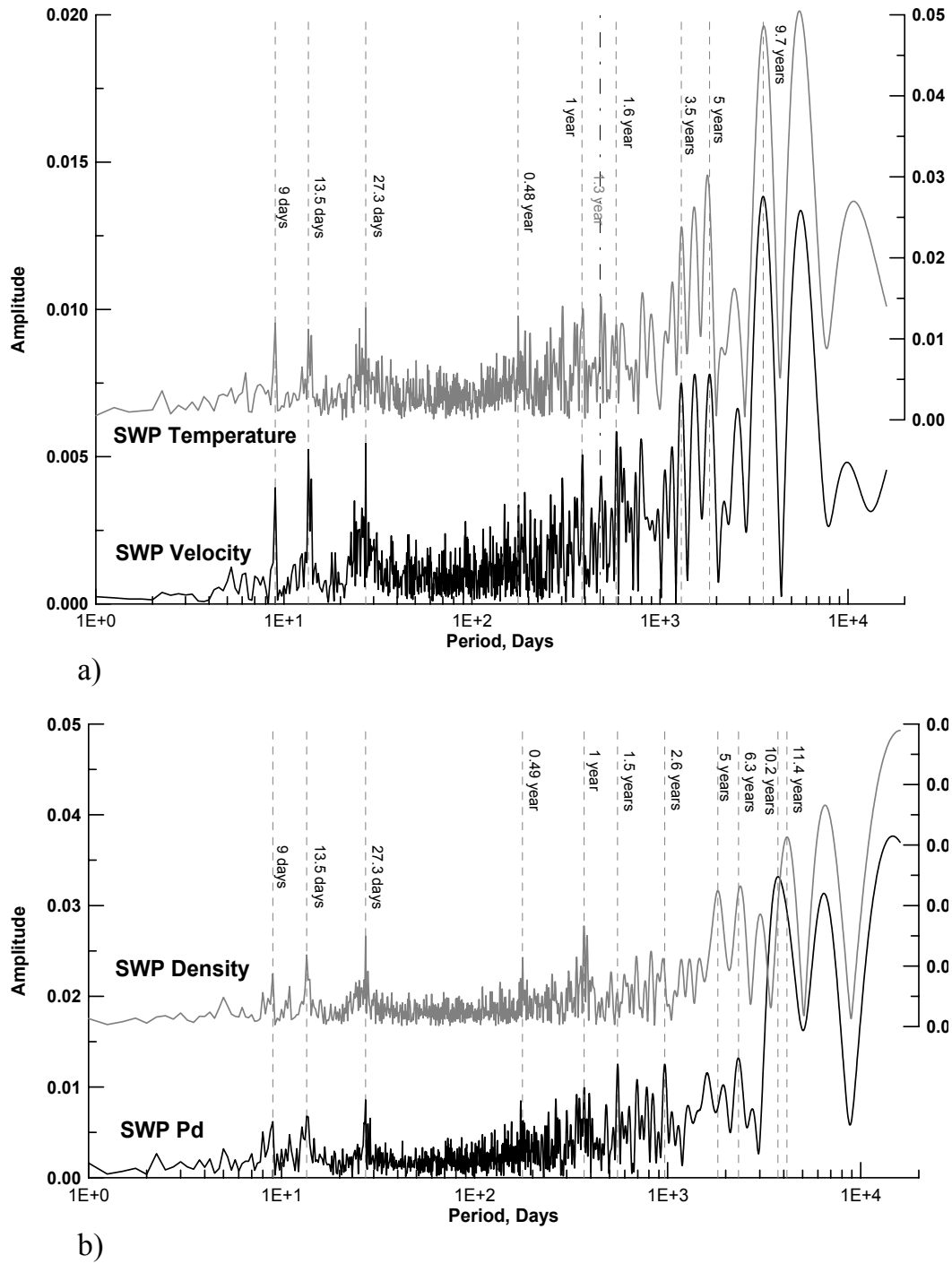

Figure 11. Periodogram of the solar wind plasma parameters for time interval 1963 to 2007 (a) proton velocity (black curve) and temperature (grey curve, right axis); (b) density (grey curve, right axis) and dynamic pressure $P_d$ (black curve). Characteristic periods are indicated by vertical dashed lines. The alternating period of 1.3 year is indicated by vertical dashed dotted line.

The periods of 27.3 days, 3.5 years and 5 years practically coincide with the characteristic periods in sunspot number. We can also find periods of 13.5 days and 9 days, which are observed for practically all heliospheric parameters and correspond, respectively, to 1/2 and 1/3 of the solar synodic period of 27.3 days [*Mursula and Zieger*, 1996; *Dmitriev et al.*, 2000; *Neugebauer et al.*, 2000; *Burlaga and Forman*, 2002; *El-Borie*, 2002; *Bolzan et al.*, 2005]. *Mursula and Zieger* [1996] demonstrate that the ~9-day and 13.5-day sub-periods might be spurious due to the effect of interference with data gaps. On the other hand the periodicity of 13.5 days can be a manifestation of four-sector solar wind structure. Note, that this question needs more investigation in future with a more complete data set.

The period of 9.7 years is shorter than the 10.6-year solar cycle in sunspot number. The relatively short cycle in the solar wind velocity and temperature is also found in previous studies [*Dmitriev et al.*, 2000; *Neugebauer et al.*, 2000; *Rangarajan and Barreto*, 2000; *El-Borie*, 2002; *Kane*, 2005]. This period might be related to ~9.6 year periodicity of solar coronal-hole area [*McIntosh et al.*, 1992]. As we know, the solar wind velocity correlates with the size of coronal holes [*Veselovsky et al.*, 2006; *Vrsnak et al.*, 2007].

In Figure 11a we find relatively strong variations with period of 1.3 years in the velocity and 1.6 years in the temperature. The latter one can be attributed to the effect of terrestrial planet [*Charvátová*, 2007]. Numerous studies regard the 1.3-year periodicity [*Richardson et al.*, 1994; *Gazis et al.*, 1995; *Paularena et al.*, 1995]. A method of dynamic power spectrum [*Szabo et al.*, 1995] and wavelet transform [*Mursula and Vilppola*, 2004] demonstrate the alternating importance of 1.3-year period, which is dominant during the 22[nd] cycle and vanishes in the 21[st] solar cycle. *Rangarajan and Barreto* [2000] show that the 1.3-year period alternates with 3.5-year and 5-year periodicities of the SW velocity. It is reasonable to assume that the 1.3-year period might be an 1/8 harmonic of the solar cycle.

In Figure 11 we also find periods of 1 year and ~0.5 year, which are apparently associated with the Earth's orbital rotation. Note that the annual periodicity of solar wind velocity is revealed only in the near-Earth's experiments, while the Voyager 2 and Pioneer 10 missions do not observe that periodicity in the outer heliosphere [*Mursula and Vilppola*, 2004]. The annual periodicity is originated from two geometric effects: 7° tilt of ecliptic plane relative to the solar equator, and nonzero eccentricity of the earth orbit with perihelion of $146 \cdot 10^6$ km and aphelion of $152 \cdot 10^6$. As a result, the Earth orbital rotation is characterized by annual variation of heliographic latitude within the range from -7° to 7°, and ~4% variation of the heliocentric distance. The north-south asymmetry of the Sun can be involved for a tentative explanation of the annual variations. The annual variations have been found in the solar wind plasma velocity, density and temperature [*Dmitriev et al.*, 2000; *Neugebauer et al.*, 2000].

Periodicities of the SW density and dynamic pressure are presented in Figure 11b. They have similar solar synodic and annual periods. The SW density demonstrates the periods of 5 years, 6.3 years and 11.4 years. The 5-year period corresponds to the sunspot number periodicity of 5.1 years, while the 6.3-year period might be associated with the effect of terrestrial planets [*Charvátová*, 2007]. Note that the sunspot period of 3.5 years is absent in the solar wind density.

The amplitudes of 5-year and ~6.3-year variations in the density are comparable with the 11.4-year peak. Unexpected and high significance of the five-year wave in the proton density variations was reported for solar cycles 20[th] to 22[nd] [*Dmitriev et al.*, 2000] and also partially for the 23[rd] solar cycle [*Kane*, 2005].

The solar wind dynamic pressure (Figure 11b) is characterized by periods of 1.5 years, 2.6 years, 6.3 years and 10.2 years. Note that the pressure is a multiplicative parameter of the solar wind velocity and density. In this context the 1.5-year and 6.3-year periods are inherited, respectively, from the velocity and density. The period of 2.6 years might be a meaningful result of interference between the 5-year periodicity in the density and 1.6-year periodicity in the velocity. By the same way the period of 10.2 years might be a superposition of the velocity and density cycles.

## IMF Periodicities

Periodograms of the IMF intensity $B$ and magnitude of $B_{xy}$ component are presented in Figure 12a. As one can see, their profiles are practically identical. Hence the variation of magnetic field in the ecliptic plane is the major source of the IMF variations. They are represented by solar synodic harmonics and annual variations. Their cycle periodicities are very close to the sunspot number characteristic periods of 3.5 years, 5.3 years and 10.6 years. There are a number of reports about pretty strong 1.7-year periodicity observed in the IMF intensity at the Earth's orbit during 20 to 23 solar cycles [*Rouillard and Lockwood*, 2004; *Mursula and Vilppola*, 2004]. Indeed, in Figure 10a one can distinguish a moderate intensification around the period of ~1.7 years (~600 days). This period can be probably attributed to the 1/6 harmonic of the IMF cycle, i.e. $10.3 / 6 \approx 1.7$ years.

The basic solar synodic period of 27.3 days is absent in the variations of magnetic field intensities. This fact is revealed in several studies but its origin is not clear [*Mursula and Zieger*, 1996; *Dmitriev et al.*, 2000; *Rouillard and Lockwood*, 2004]. It is pretty possible that diminishing of the amplitude of the 27.3-day periodicity is related to substantial enhancements of the IMF intensity in the corotating interaction regions. Because of tilted and curved sector boundary, the CIRs pass the Earth at least two times per solar rotation, i.e. the period of $27.3 / 2 = 13.6$ days should be dominant for the IMF $B$ and $B_{xy}$ intensities.

Figure 12b shows periodograms of the IMF $B_x$ and $B_y$ components. They have very similar periodicities, which however are different from those in IMF intensity. Note that the variable-polarity components are analyzed in linear scale, while the intensities are converted to logarithms. Variations of the $B_x$ and $B_y$ components are characterized by dominant periods of 28 days and 1 year. *Neugebauer et al.* [2000] obtain the same result for the $B_x$ component. It is suggested that the 27-day period is most prominent because the Earth encounters two IMF sectors per solar rotation. The annual variation of the Earth's heliographic latitude modulates the duration of encounters to the north or south IMF sector.

It is important to emphasize that the solar cycle period and its harmonics vanish in the periodograms of variable-polarity components as mentioned by *Neugebauer et al.* [2000]. Figure 12b illustrates this fact clearly. The amplitudes of ~3-year, ~5.9-year and 9.8-year periodicities are diminished and only ~20-year cycle is still prominent. Note that the solar cycle in the $B_x$ and $B_y$ components is represented by the period of ~9.8 years. This period is close to the 9.7-year cycle in the solar wind plasma velocity and temperature that is attributed to variations of the size of unipolar solar coronal holes [*McIntosh et al.*, 1992].

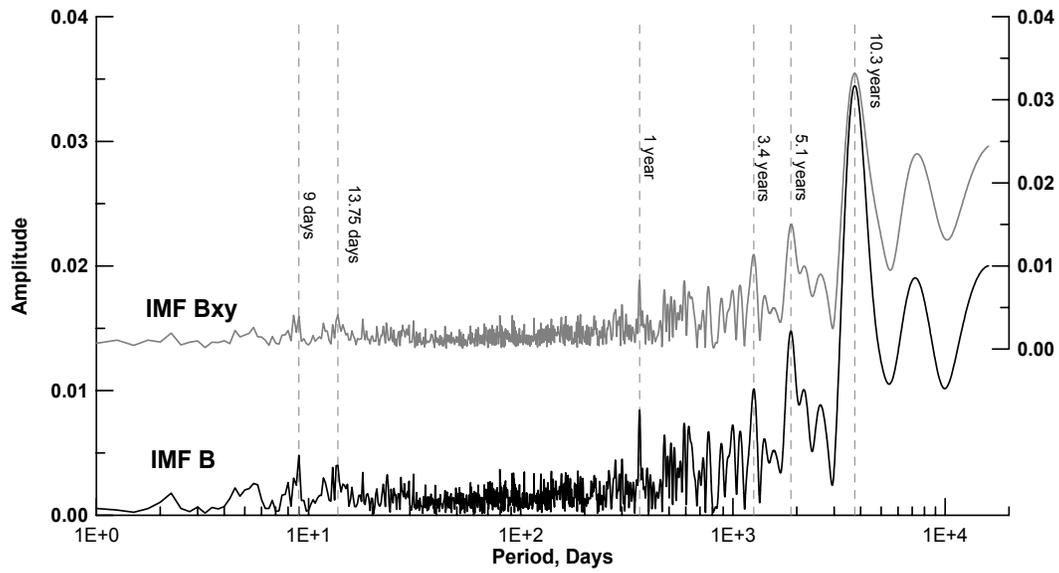

a)

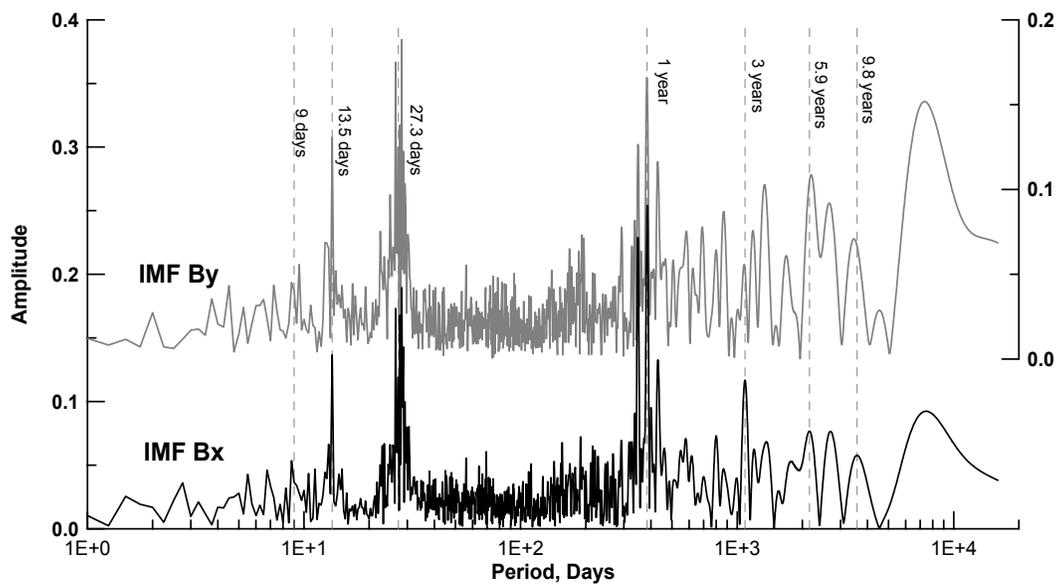

b)

Figure 12. (Continued)

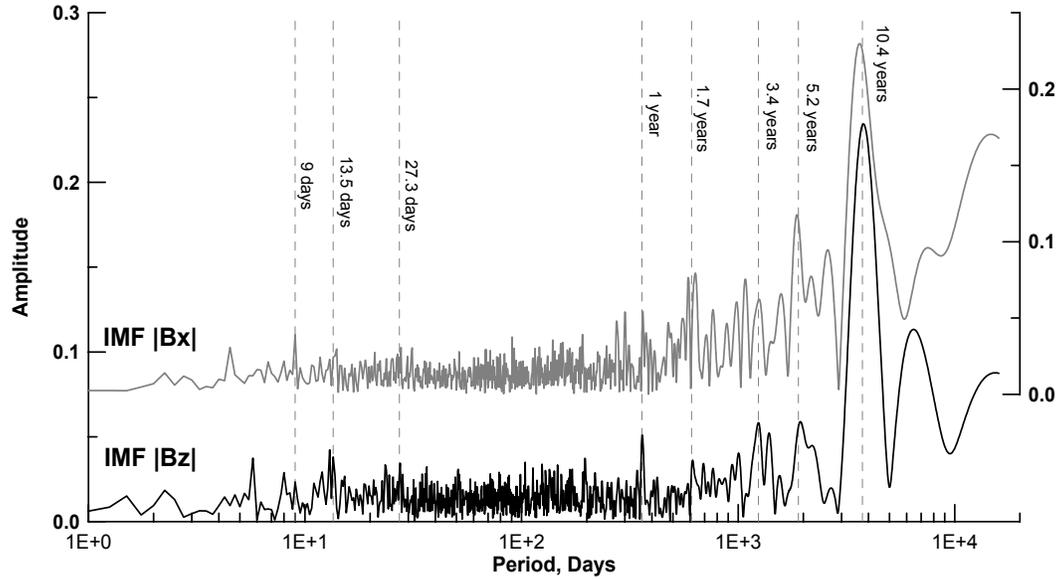

c)

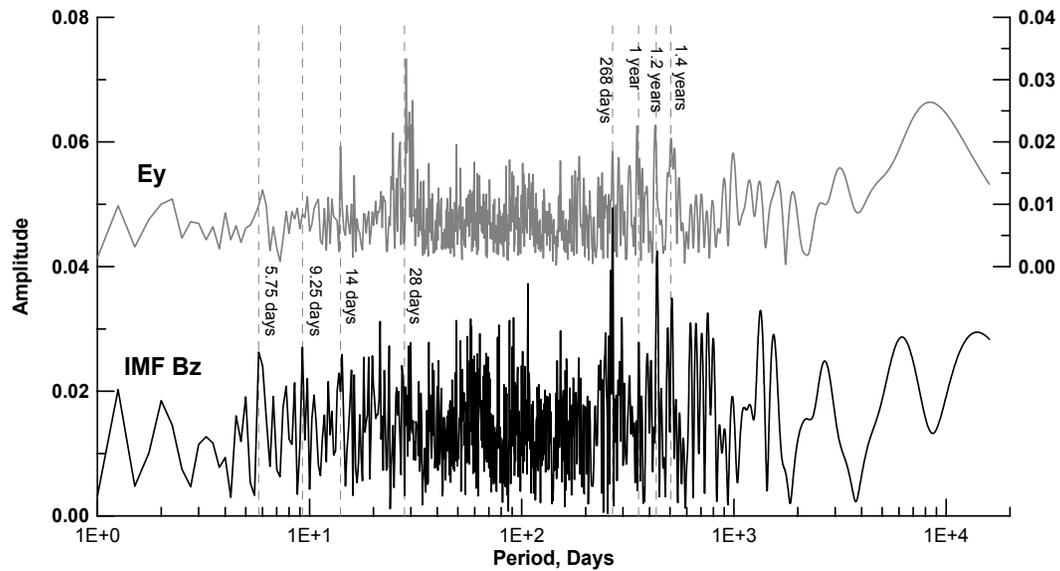

d)

Figure 12. Periodogram of the IMF parameters for time interval 1963 to 2007: (a) strength (black curve) and $B_{xy}$ component (grey curve, right axis); (b) $B_x$ (black curve) and $B_y$ component (grey curve, right axis); (c) magnitudes of IMF components $B_z$ (black curve) and $B_x$ (grey curve, right axis), (d) $B_z$ component (black curve) and $E_y$ component of induced electric field (grey curve, right axis). Characteristic periods are indicated by vertical dashed lines.

The above effects disappear when we consider intensities (i.e. absolute values) of the IMF components. Figure 12c shows periodograms of the absolute values of IMF $B_x$ and $B_z$. They are pretty similar one to other and they both very close to the periodograms of IMF $B$ and $B_{xy}$ intensities (see Figure 12a). Namely, the periodicities are represented by the cycle of 10.4 years, which is very close to the solar cycle of 10.6 years, and by harmonics of 10.4 / 2 = 5.2 years, 10.4 / 3 ≈ 3.4 years, 10.4 / 6 ≈ 1.7 years. There are two strong periodicities of 1 year and 13.5 days, while the solar synodic period of 27.3 days practically vanishes.

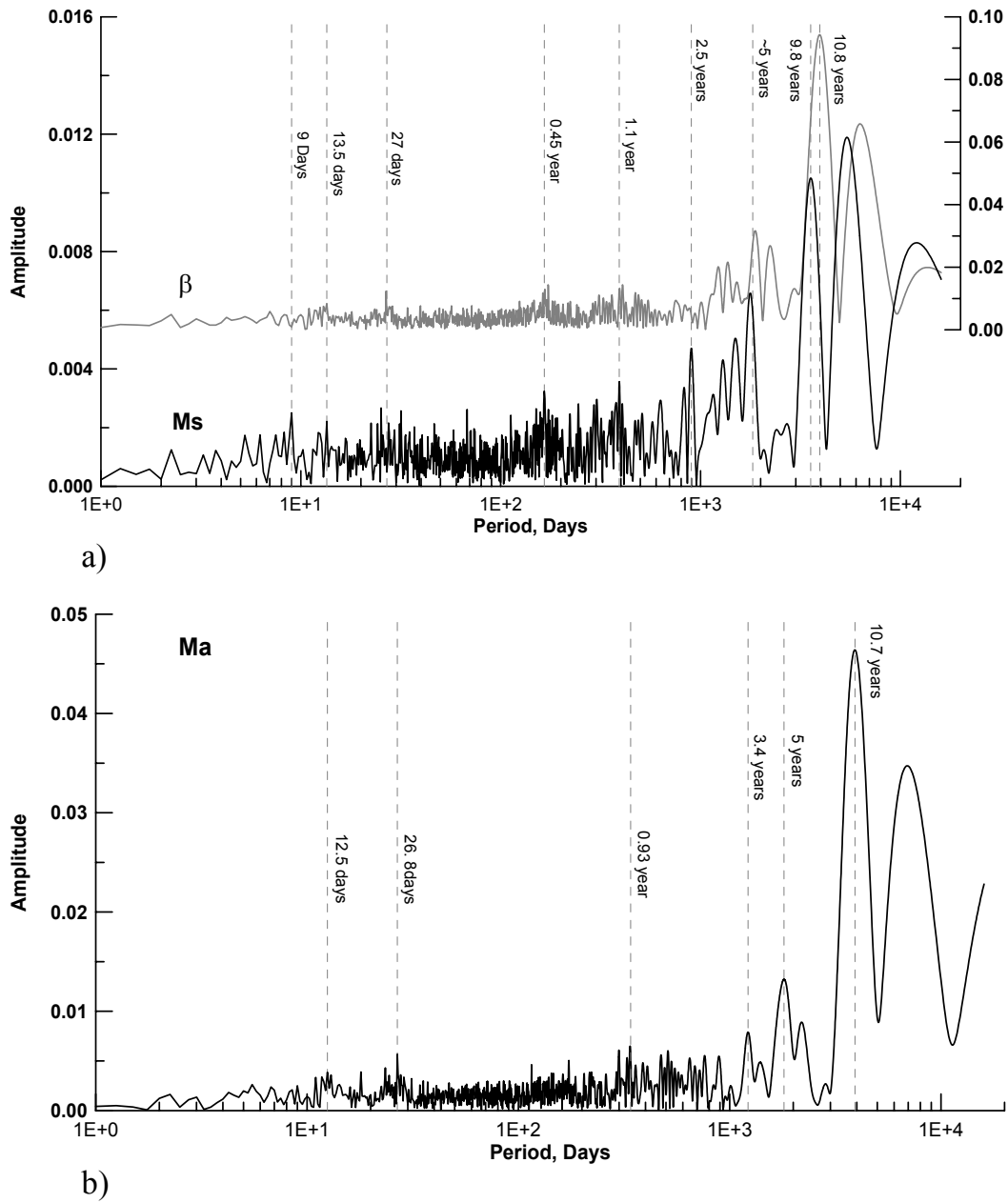

Figure 13. Periodogram of the dimensionless parameters for time interval 1963 to 2007: (a) Sonic Mach number $M_s$ (black curve) and thermal to magnetic pressure ratio $\beta$ (grey curve, right axis); (b) Alfvén Mach number $M_a$. Characteristic periods are indicated by vertical dashed lines.

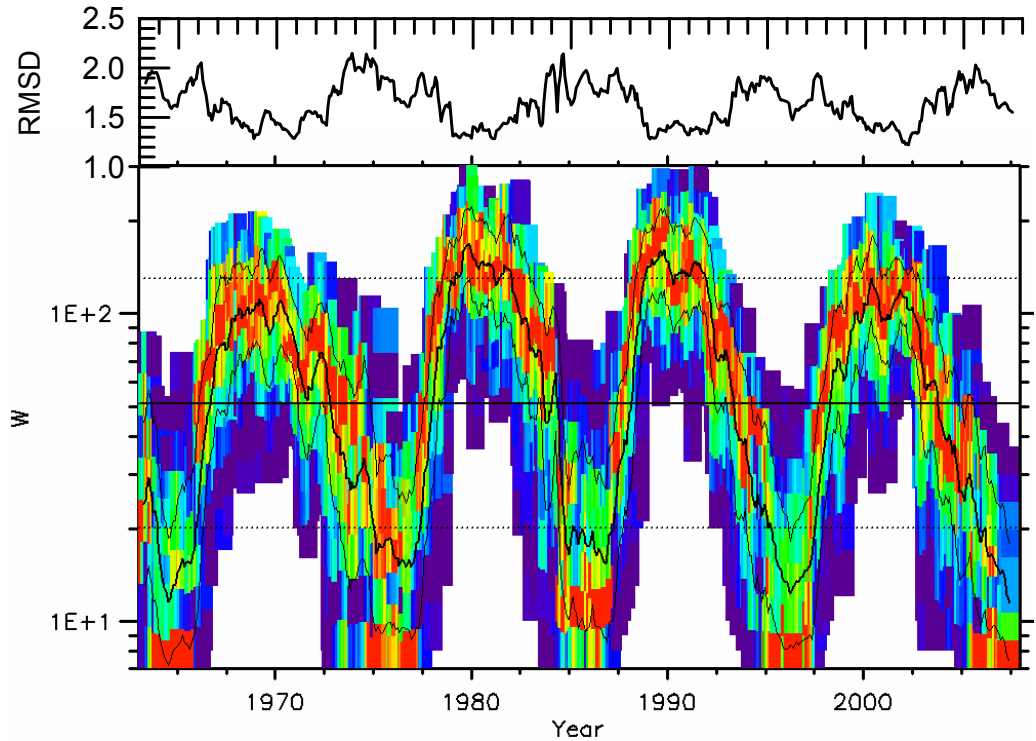

Figure 14. Running histogram of statistical distribution of sunspot number *W* for time interval from 1963 to 2007. Occurrence number in the running histograms is indicated in rainbow palette from violet (minimum) to red (maximum). Top panel shows the running dispersion of distribution. Running mean and running 1-σ deviation are indicated, respectively, by thick and thin black curves. The average is indicated by horizontal black solid line. The 1-σ and 3-σ corridors are restricted, respectively, by horizontal black dotted lines and white dotted lines.

Periodograms of the IMF $B_z$ component and of the $E_y$ component of induced interplanetary electric field (see Equation 16) are shown in Figure 15d. Similarly to the $B_x$ and $B_y$ components the periodograms of variable-polarity components $B_z$ and $E_y$ do not demonstrate prominent solar cycle periods of >2 years. In contrast to the $B_x$ and $B_y$, the 27.3-day and 13.5-day periodicities in $B_z$ component are diminished and only harmonics of 9.25 days and 5.75 days are represented [*Mursula and Zieger*, 1996]. That is a manifestation of very high variability of the IMF $B_z$ within even half of the solar rotation.

The dominant periodicities of IMF $B_z$ are 268 days (0.73 year) and 1.2 years. A broad peak near 250-285 days was revealed from power spectral analysis of cosmic-ray intensity during the period 1964-1995 [*El-Borie and Al-Thoyaib*, 2002]. The origin of 0.73-year periodicity is still unclear and hence that should be a subject of future investigations. The period of 1.2~1.4 years was revealed in the dynamics of $B_z$ by different spectral methods [*Paularena et al.*, 1995; *Szabo et al.*, 1995]. This variable period was discussed above in the context of the solar wind velocity variations.

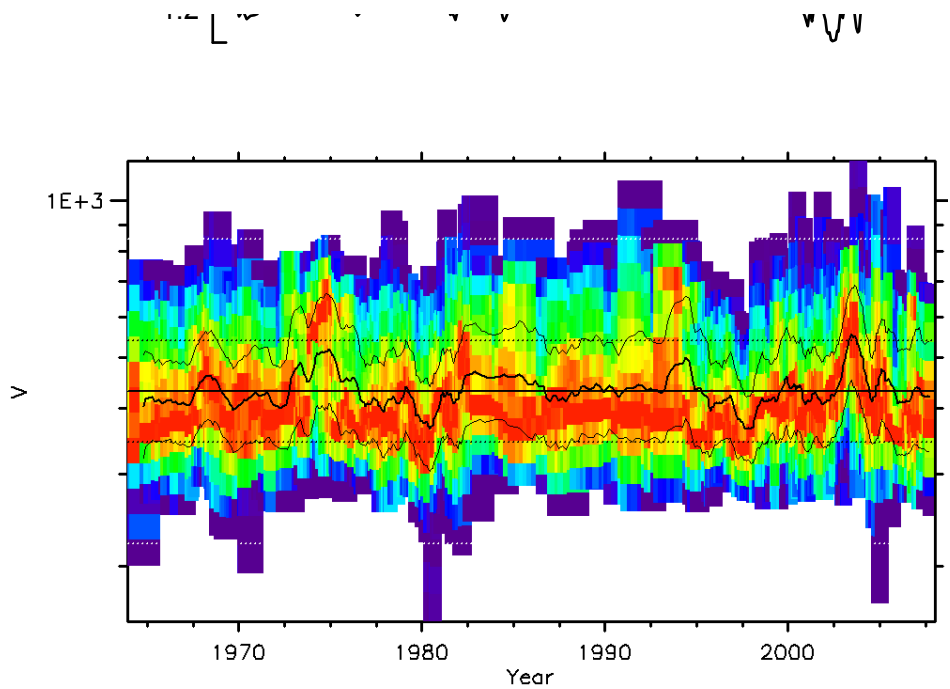

a)

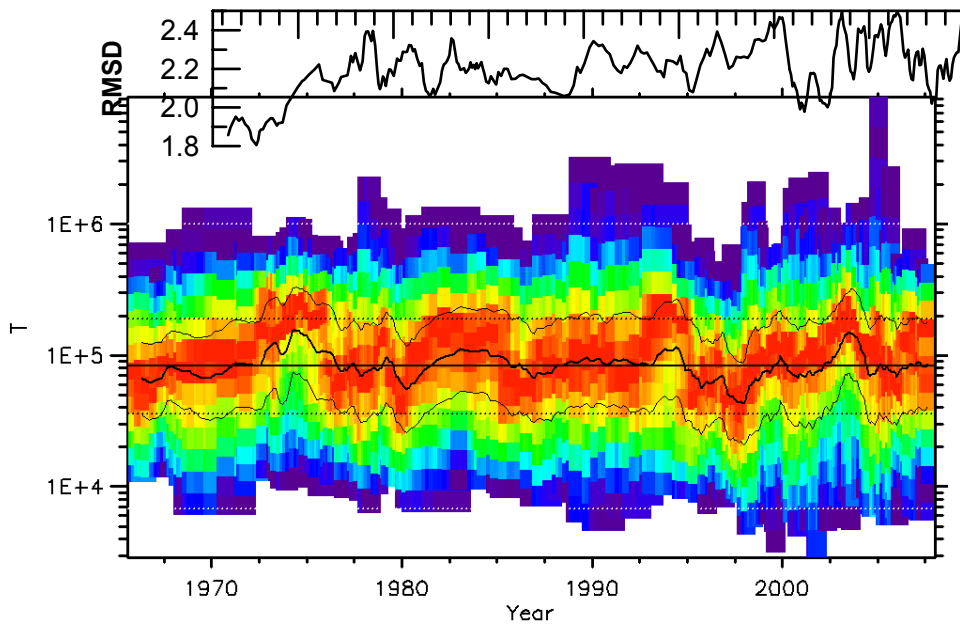

b)

Figure 15. (Continued)

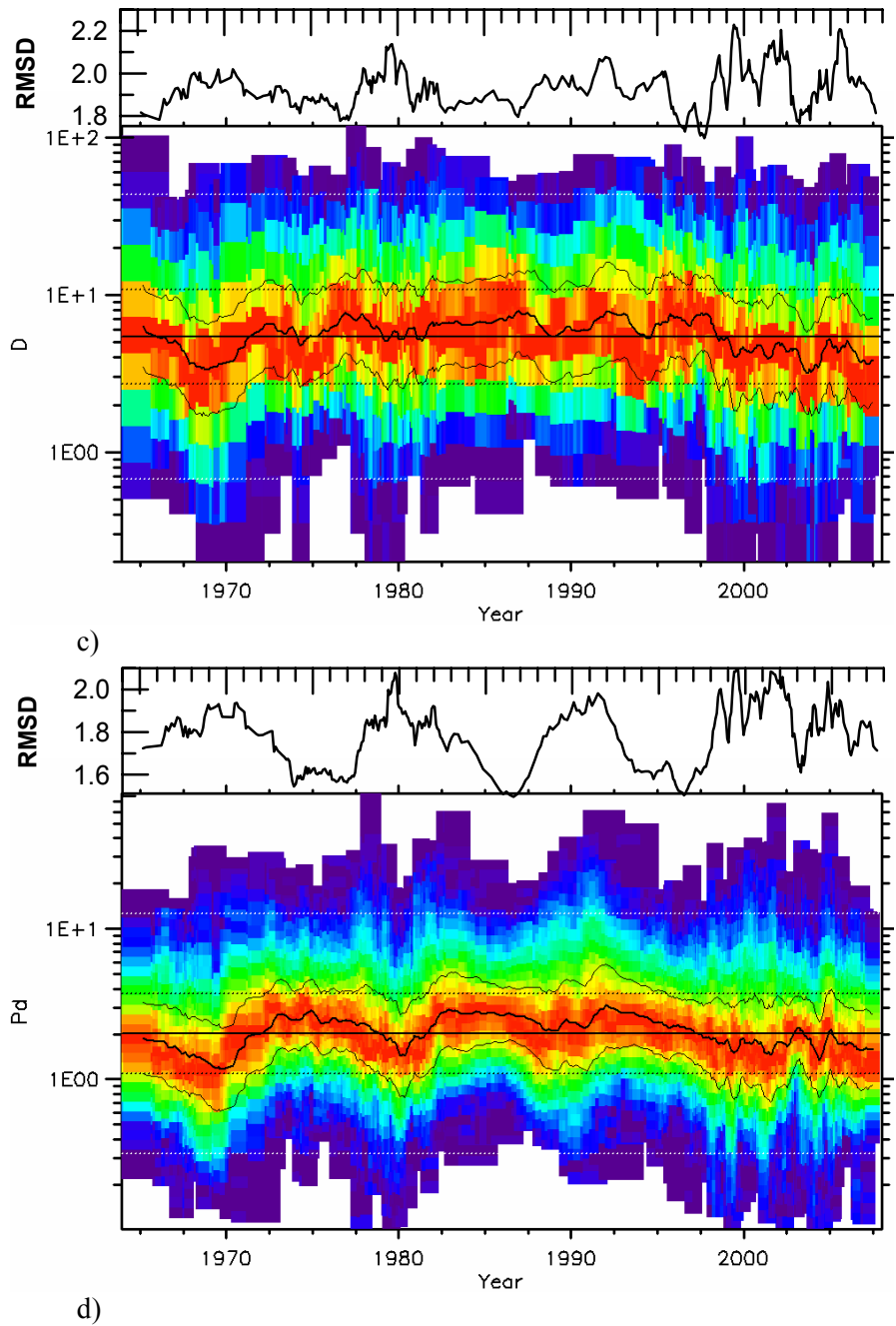

Figure 15. Running histograms of solar cycle variations in solar wind plasma parameters:
(a) velocity; (b) temperature; (c) density; and (d) dynamic pressure. Occurrence number in the running
histograms is indicated in rainbow palette from violet (minimum) to red (maximum). Running
dispersion is expressed in *RMSD* and shown in the top panels. Running mean and running 1-$\sigma$ deviation
are indicated, respectively, by thick and thin black curves. The average is indicated by horizontal black
solid line. The 1-$\sigma$ and 3-$\sigma$ corridors are restricted, respectively,
by horizontal black dotted lines and white dotted lines.

The $E_y$ component is characterized mainly by periods of ~14 days, 28 days, 1 year 1.2 years, and 1.4 years, which inherit partially from the periodicities in solar wind velocity and IMF $B_z$. The dominant periodicity of $E_y$ is characterized by a broad peak at ~28-day period, which is apparently associated with the solar synodic period. The solar cycle periodicities of >2 years vanish in the $E_y$ variations because of high variability of the $B_z$.

**Periodicities of Dimensionless Parameters**

Periodograms of the plasma $\beta$, sonic and Alfvén Mach numbers are shown in Figure 13. The parameters $\beta$ and $M_s$ have very similar periods: solar synodic period with harmonics (27 days, 13.5 days and 9 days), annual and semiannual variations inherit from the dynamics of solar wind velocity and temperature (see Figure 11a). However the solar cycle variations (>2 years) in plasma $\beta$ and $M_s$ are slightly different. Comparing with Figure 11 we can find that the periodicities in the sonic Mach number are mostly related to the solar wind plasma variations with periods of 5-years and ~9.8 years. The periodicity of 2.5 years might be a harmonic of the 5-year period. The variations in plasma $\beta$ are rather related to the dynamics of IMF intensity $B$ (see Figure 12a) with periods of ~5.1 years and 10.3 years.

It is interesting that the periodicities of Alfvén Mach number $M_a$ (see Figure 13b) are practically same as the characteristic periods of sunspot numbers (see Figure 10), excepting the period of 131 days. From Table 3 we can see that the number $M_a$ is a complex parameter of solar wind velocity, proton density and IMF $B$. Comparing Figures 11a, 11b and 12a, we can see that those parameters have pretty different periodograms. Perhaps interference of different periodicities leads to degeneration of the variations into the basic harmonics of three main periodicities: solar synodic period, annual period of the Earth orbital rotation, and sunspot cycle.

# 4. Solar Cycle Variations

Solar cycle variation of heliospheric parameters is a manifestation of various solar sources of the solar wind and IMF operating at different phases of the solar cycle [e.g. *McComas et al.*, 2003]. It is widely accepted that the solar coronal holes is a main source of the solar wind during declining phases of solar cycle [e.g. *Tsurutani et al.*, 1995]. Transient events form the solar wind at solar maximum [e.g. *Cane and Richardson*, 2003]. Slow and dense solar wind streams of heliospheric current sheet dominant in the ecliptic plane during solar minimum [e.g. *Smith*, 2001]. As a result, statistical properties of the heliospheric parameters vary within the cycle such that not only the average and dispersion but also the shape of statistical distribution can change [e.g. *Dmitriev et al.*, 2000; 2005c].

We study solar-cycle variations using a method of running histogram [*Dmitriev et al.*, 2002a; 2005c]. In this method a histogram of statistical distribution is accumulated within a relatively narrow time window, which is consequently shifted with a short time step. As result, a long time interval of measurements is represented by a sequence of running statistical distributions of the measured parameter. For each distribution we calculate running mode, mean and dispersion (*RMSD*). The choice of running time window and step depends mainly on data sampling. The variations of statistical distribution can be represented more brightly when we take into account the characteristic periodicities of analyzing parameter. In the previous section we find that practically all heliospheric parameters have prominent variation with solar synodic period of 27.3 and annual variation. We choose the step of 27 days and time window of 189 days. The time interval of 189 days is approximately equal to 7 solar synodic periods and also corresponds to the half-year period. By this way we eliminate short-time variations associated with solar rotation, but the annual and longer variations can be analyzed.

Figure 14 shows the running histogram of sunspot number. This parameter demonstrates very high variability during the four last solar cycles (Figure 14). The running mean varies in a wide range and during solar maxima and minima exceeds 1-$\sigma$ deviation from the 40-year average of ~50. Large statistically significant variations during the cycle have been also found for the radio emission flux of the Sun at the wave length 10.7 cm and for the total solar irradiance [*Dmitriev et al.*, 2005c]. The dispersion of sunspot number is lowest during solar maximum. A short-time enhancement of the dispersion coincides with decrease of sunspot number in the Gnevyshev gap, which corresponds to polarity reversal of the solar magnetic field in 1969, 1980, 1990 and 2001. The running dispersion is highest during rising and declining phases, while the solar cycle minima are characterized by moderate dispersion.

## Solar Cycle in Plasma Parameters

Solar-cycle dynamics of the statistical distributions of SW plasma parameters is presented in Figure 15. All cyclic variations in the running mean are not very large and lay inside 1-$\sigma$ corridor around the average of ~420 km/s. Specific cyclic patterns of the variations can be discernible for all plasma parameters. *Dmitriev et al.* [2002b] demonstrate that dynamics of solar wind plasma during rising and declining phases are systematically different such that the cycle variation in solar wind parameters represents hysteresis behavior in dependence on the sunspot number.

Most prominent variations with pretty well organized pattern are revealed in the SW velocity (Figure 15a). Note that the statistical distribution of velocity is deviated from lognormal PDF and hence the running averages can be different from running modes. One can clearly see that the variations in running average represent solar-cycle very roughly, because of very wide statistical distribution with long tails. Moreover, the running averages lose informative meaning because the running histogram often has two or even three peaks.

The running mode of solar wind velocity is more representative. In Figure 15a we can distinguish a clear pattern of the both 11-year and 22-year cycles. The 22[nd] cycle in velocity resembles the 20[th] cycle as reported before [e.g. *Cliver et al.*, 1996]. Declining phases of both cycles contain long-duration intervals of fast solar wind with very high dispersion. Detailed comparison reveals that the cycles 21[st] and 23[rd] do not have so prominent velocity enhancements as the cycles 20[th] and 22[nd]. However, the common pattern of velocity variations repeats from cycle to cycle.

The slow solar wind streams with velocity <400 km/s are dominant during the minimum and beginning of rising phase of the solar cycle in 1965~1967, 1976~1977; 1985~1987, 1995~1998, and 2007. Relatively fast solar wind with velocity of 400 to 500 km/s appears in the late stage of rising phase (years 1967~1968, 1978~1979, 1988, and 1999). Solar maximum and beginning of declining phase are represented by pretty wide distribution with mode of ~400 km/s (years 1969~1972, 1989~1992, and 2000~2002). The maximum of 21$^{st}$ cycle demonstrates a little bit different behavior of the velocity with deep decrease to ~350 km/s in 1980 and strong enhancement to ~500 km/s in 1982. Note that in 1980 the solar wind velocity drops down to extremely low values of <200 km, while the other solar maxima are not accompanied by so slow solar wind. It is difficult to decide weather this effect real or experimental artefact.

Most significant variations in the solar wind velocity and substantial enhancements of the velocity dispersion are observed during the second half of declining phase in 1973~1976, 1983~1985, 1992~1995, and 2003~2006. During those time intervals we reveal two or three peaks in the velocity distribution at 600~700 km/s, at ~500 km/s and at ~350~400 km/s corresponding, respectively, to fast, intermediate and slow solar wind streams [*Veselovsky et al.*, 1998c; *Dmitriev et al.*, 2000]. Note that extremely fast solar wind with velocities of >1000 km/s occurs during declining phases. The fast and intermediate solar wind streams disappear abruptly after onset of solar minima.

In Figure 15b we find that solar wind temperature has very similar variations with the velocity, as indicated in many previous studies [e.g. *Luhmann et al.*, 1993]. The statistical distribution of temperature is close to lognormal. Hence the running average is very close to running mode in most cases, excepting strong temperature enhancements during declining phase, when the mode exceeds 1-$\sigma$ deviation from the average of 85000 K, similarly to the solar wind velocity. The most prominent temperature enhancements are revealed in the 20$^{th}$ and 22$^{nd}$ cycles, which have very similar patterns in the temperature variations.

Low temperatures of ~60000 K accompany solar minimum. During the rising phase and in solar maximum the temperature gradually grows up to ~100000 K. Short-time enhancements of the temperature in the late stage of rising phase in 1978 and 1999 correspond to fast solar wind streams appearing at that time. The declining phase is characterized by highest temperatures of about 2~3•10$^{5}$ K. At that time we can find statistical distributions with two peaks (see years 1975, 1994, and 2006), which correspond to very low temperatures of ~50000 K and very high temperatures of ~200000 K. There are no distributions with prominent 3-peak structure. Similarly to the solar wind velocity, the temperature drops down abruptly at the beginning of solar minimum.

From dynamics of running *RMSD* in Figure 15b we find that the rising and especially declining phases are characterized by largest variations in the temperature. Extremely high and extremely low temperatures are observed mainly around solar maximum. However, the highest temperatures in the cycle may occur in the beginning of rising phase and during late declining phase.

Solar cycle variations in the SW proton density are different from those in velocity and temperature as we can see in Figure 15d. The statistical distribution of density is very close to lognormal PDF. We do not reveal any multi-peak structure in the running histograms. However, very often the running distribution has large negative skewness, especially during declining phase, when the average becomes higher than mode. At that time the running mode can be smaller than 1-$\sigma$ lower deviation from the average. In contrast, the running average always lies within 1-$\sigma$ corridor around the average of 5.3 cm$^{-3}$. We do not find similarity between the 20th and 22nd cycles in the dynamics of average proton density. However, 22-year cycle can be revealed in the variations in running dispersion. For the running mean we can find only a pattern of variations inside the 11-year cycle.

The running mode of solar wind density is highest (~7 cm$^{-3}$) in the solar minimum. During rising phase, the density decreases and reaches minimum of <5 cm$^{-3}$ in the solar maximum. Note that dispersion of density is maximal around the solar maximum. The density has a tendency to enhance in the beginning of declining phase and then decrease substantially to values of <5 cm$^{-3}$ during the second half of the declining phase. The density grows fast with onset of the solar minimum. It is interesting that such pattern is very close to that in running dispersion of the sunspot number. In Figure 15c we can also find that the variation in running dispersion of the density anticorrelates with the running mean. This dynamic pattern with two minima preceding the solar maximum and minimum, has a clear 5-year periodicity. That periodicity is revealed very prominent in the periodogram of solar wind density (see Figure 11b).

It seems that the 23rd cycle in solar wind density is much different from the previous ones. Main distinguishing feature is significant negative trend in the density variations starting in 1998. The declining phase in 2002~2007 is enriched by very and extremely low densities. This fact is out of keeping with previous findings that the vast majority of tenuous solar wind streams are observed during rising phase and in maximum of solar cycle [*Crooker et al.*, 2000; *Richardson et al.*, 2000]. In Figure 15c we can clearly see this feature in the solar cycles 20th to 22nd.

Since 1998 the OMNI database is mostly replenished by the ACE data. Comparing the ACE proton density with data measured by the Wind and IMP-8 satellites we can find that the density measured by ACE is systematically lower [*Dmitriev et al.*, 2002a]. Moreover this difference is growing with time probably because of ageing effect of about several percent per year [*Dmitriev et al.*, 2005a]. Accurate accounting of this effect is a subject of future work. Currently it is difficult to interpret the data on solar wind density during the 23rd solar cycle.

We should also note that the occurrence of extremely high densities is distributed in solar cycle pretty randomly. They have a tendency to group around solar maximum. Though one can find such extreme events during other phases, including solar minimum. This is in agreement with a fact that the highest densities are generated in trailing edges of fast interplanetary transients and inside eruptive filaments [*Burlaga et al.*, 1998; *Crooker et al.*, 2000]. The former events group around the solar maximum, which is characterized by very high variability of the density [*Cane and Richardson*, 2003].

Figure 15d demonstrate solar cycle dynamics of the SW dynamic pressure. The running histograms of the dynamic pressure are fitted very well by lognormal PDF. Hence the running mean practically coincides with running mode. Variation in the running average is not very high and restricted by 1-$\sigma$ deviation from the average of 2 nPa. The running dispersion of dynamic pressure correlates very well with the sunspot number: it is lowest in the solar minima and highest around solar maxima. Both extremely high and very low dynamic pressures occur around solar maximum.

The solar cycle variation in the dynamic pressure is remarkably regular and anticorrelate with sunspot number [*Crooker and Gringauz*, 1993; *Luhmann et al.*, 1993]. Note that the dynamic pressure accounts the helium contribution, which correlates with sunspot number, i.e. it is highest (lowest) in solar maximum (minimum) [*Feldman et al.*, 1978; *Aellig et al.*, 2001]. However, the dynamic pressure variations are mainly contributed by the variations in proton density and velocity. As a result, the pressure is smallest in the solar maximum. In the beginning of declining phase the pressure enhances substantially up to ~3 times. This feature is reported by *Richardson et al.* [2001] for the 22nd solar cycle and by *McComas et al.* [2003] for the 23rd solar cycle. Apparently, that enhancement is due to growth of the solar wind density and velocity. The pressure remains high and decreasing slowly during whole declining phase and in solar minimum. The rising phase is accompanied by relatively fast decrease of the dynamic pressure to the minimum value in solar maximum.

Note that this pattern is not well suitable for the 23rd solar cycle. The steep rising of the dynamic pressure observed in 2002 by *McComas et al.* [2003] is followed by pretty fast decrease to the values comparable with the pressure minimum in 2001, as we find in Figure 12d. Comparing with dynamics of the solar wind velocity and density, we find that the enhancement of dynamic pressure in 2002 is caused rather by strong jump of the velocity, while the density drops down at that time. In the previous cycles the dynamic pressure enhancements are produced by growth of the solar wind density.

**Solar Cycle in IMF**

Figure 16 represents running histograms of IMF statistical distribution. Similarly to the SW plasma parameters the solar cycle variation in IMF running mean is restricted by 1-$\sigma$ corridor around the average.

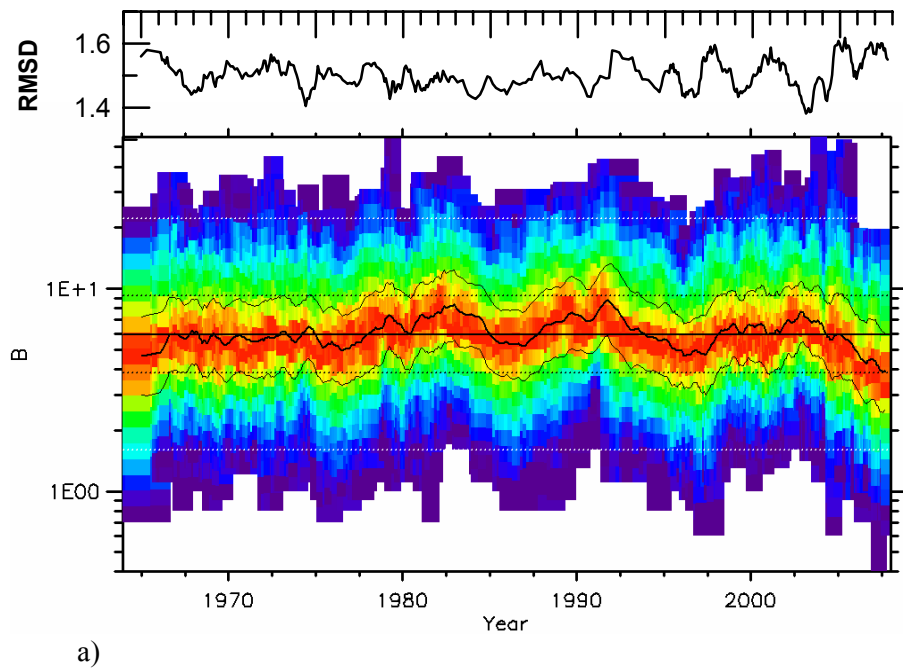

a)

Figure 16. (Continued)

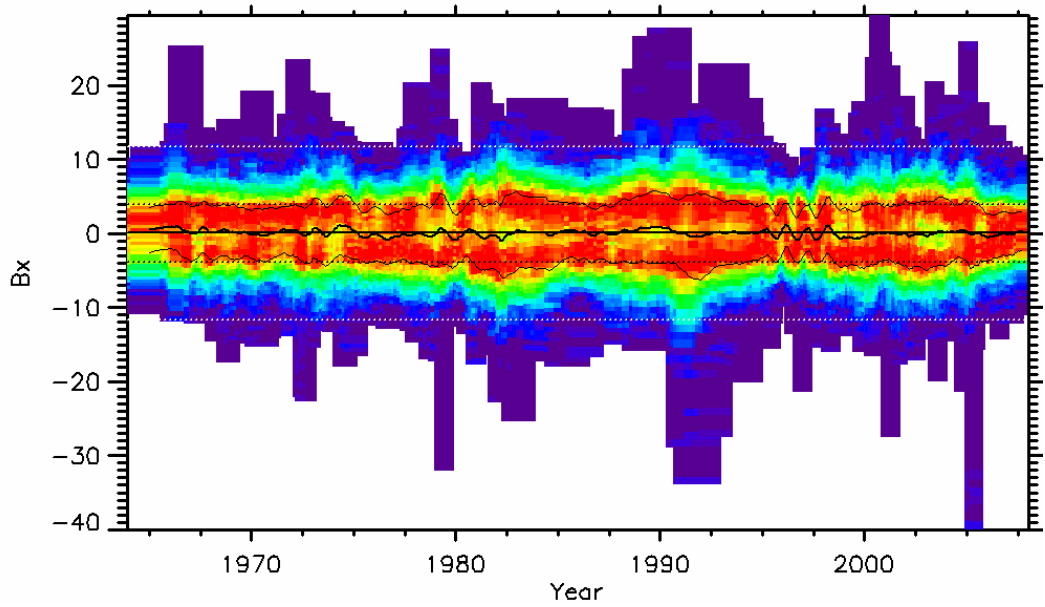

b)

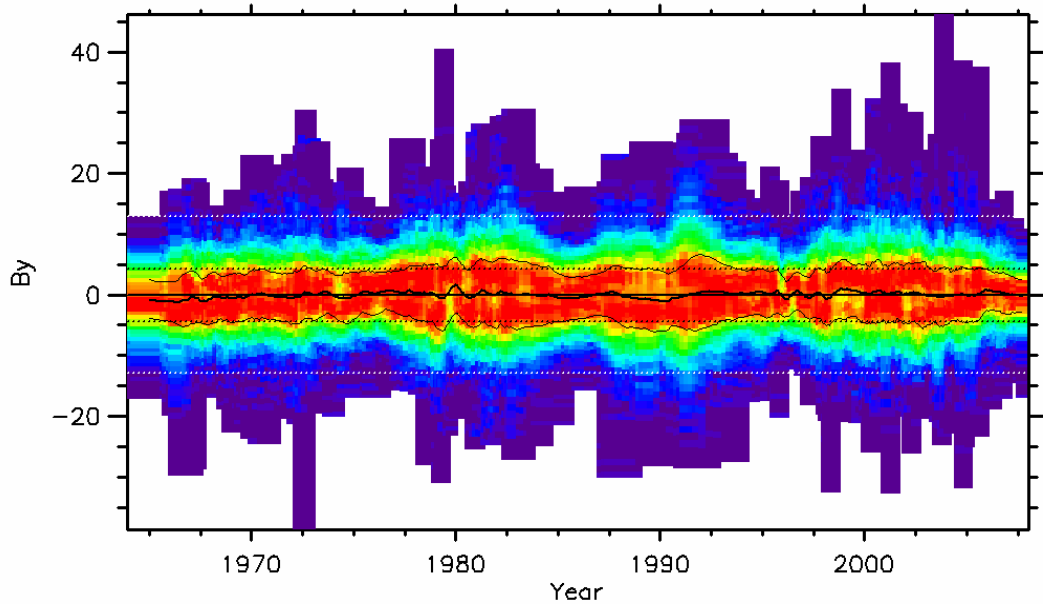

c)

Figure 16. (Continued)

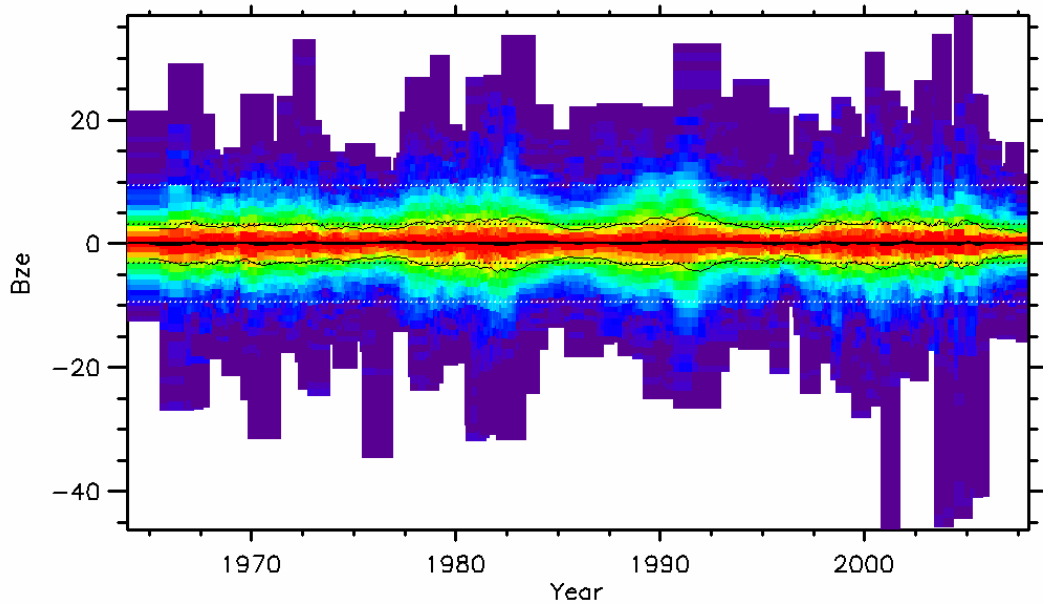

d)

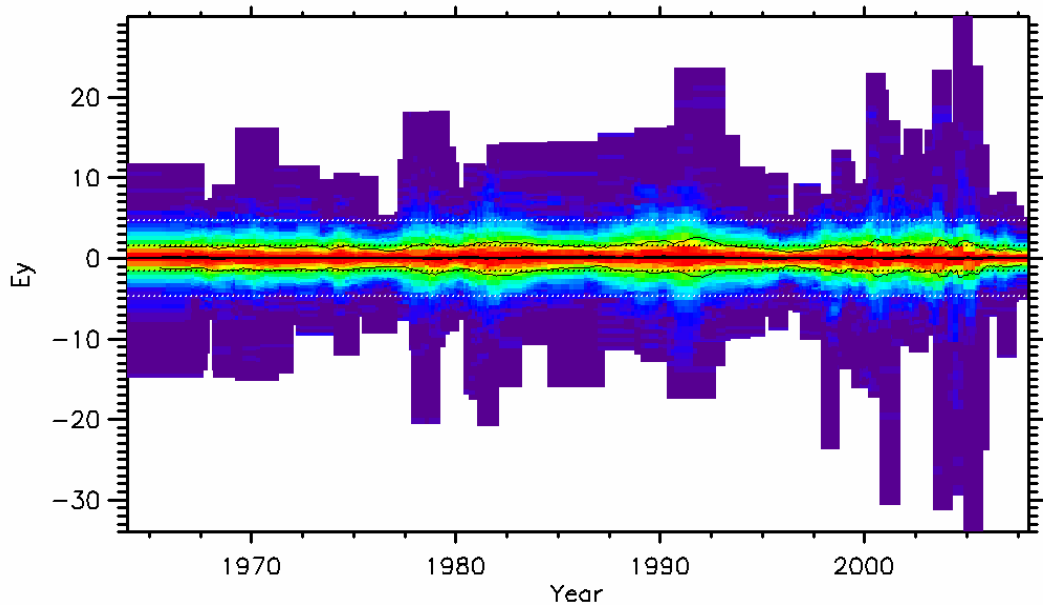

e)

Figure 16. Running histograms of solar cycle variations in IMF parameters: (a) total intensity $B$; (b) $B_x$; (c) $B_y$; (d) $B_z$; and (e) induced interplanetary electric field $E_y$. Occurrence number in the running histograms is indicated in rainbow palette from violet (minimum) to red (maximum). Running dispersion of the total intensity $B$ is expressed in RMSD and shown in the top panel of plot (a). Running mean and running 1-$\sigma$ deviation are indicated, respectively, by thick and thin black curves. The average is indicated by horizontal black solid line. The 1-$\sigma$ and 3-$\sigma$ corridors are restricted, respectively, by horizontal black dotted lines and white dotted lines.

The IMF also demonstrates hysteresis behavior during solar cycle consisting in the phase shifts between the variations in IMF parameters and sunspot number [*Dmitriev et al.*, 2002b]. *Slavin et al.* [1986] mention about poor correlation of the IMF parameters with the sunspot number and find better correlation with the total solar magnetic flux. However a phase shift still persists between the IMF and solar magnetic flux variations.

Solar cycle dynamics of running histograms of the IMF intensity *B* is presented in Figure 16a. Running statistical distributions of the intensity can be well fitted by lognormal PDF and hence running mode and mean are very close. The IMF intensity is lowest in the solar minimum. During rising phase the intensity grows gradually. Solar maximum is characterized by a small local decrease of the IMF intensity. That decrease corresponds to the Gnevyshev gap and polarity reversal of the solar magnetic field when the sunspot number also has local minimum (see Figure 14). After solar maximum, the IMF intensity reaches highest values above the average of 6 nT and then gradually decreases during declining phase. Comparing Figure 16a with Figure 14 we can find that the running mean of IMF intensity correlate in general with running mean of sunspot number, excepting IMF enhancements during the declining phase.

Similar pattern of the IMF dynamics is reported by *Slavin et al.* [1986]. They attribute the IMF enhancements at declining phase to the fast solar wind streams. Comparing Figure 16a with Figure 15a we can find that the maximum of IMF intensity indeed coincides with peak in the solar wind velocity in the 20th, 21st and 23rd solar cycle, while in 1992 (22nd cycle) only moderate solar wind velocities of ~400 km/s correspond to the IMF maximum. Furthermore, we can find that the fast solar wind streams are not necessarily accompanied with the intense IMF.

Solar cycle variation in the IMF dispersion is intensively discussed in the context of IMF multifractal properties [*Burlaga and Ness*, 1998; *Burlaga*, 2001]. It is shown that the relative error of IMF strength is essentially invariant over 15-year period from 1979 to 1994. The *RMSD* in logarithmic scale has exactly the same meaning of relative error (see Equation 9). In Figure 16a (top panel) we can see that variation in running *RMSD* is not very high (~20%) and pretty noisy. However, detailed analysis permits revealing some regular variations. Namely, the *RMSD* has local maximum during polarity reversal simultaneously with the maximum in running dispersion of sunspot number [see Figure 14]. The rising and declining phases of solar cycle are characterized by more-or less gradual decrease preceded by fast growth of the *RMSD*. The onset of solar minimum is accompanied with short-time enhancement of the *RMSD*. Such dynamics of the running dispersion is different from the solar cycle variation in running averages.

Occurrence of extreme IMF intensities is more ordered in solar cycle. The extremely weak IMF occurs mainly during solar minimum and beginning of rising phase and declining phases. Extremely strong IMF is observed usually around solar maximum. It is interesting that such dynamics of extremes resembles the dynamics of running average.

Another interesting feature is absence of 22-year cycle in the IMF intensity. In the periodogram of IMF B (see Figure 12a) the ~22-year periodicity amplitude is diminishing. The method of running histograms also demonstrates clearly that the IMF intensity variations in the 20th (21st) solar cycle does not much those in the cycle 22nd (23rd). Furthermore, the rising phase and maximum of 23rd solar cycle resembles mostly those in the 20th solar cycle [*Dmitriev et al.*, 2002b; 2005c]. However during declining phase, the 23rd cycle rather resembles the previous 22nd solar cycle, as we can see in Figure 16a. Note that the total solar magnetic flux SF also does not demonstrate the 22-year periodicity [*Dmitriev et al.*, 2005c]. However, this result is very preliminary because of low quality of the SF data during the 20[th] solar cycle.

Solar cycle variations in the IMF components $B_x$ and $B_y$ in GSE coordinate system are presented in Figures 16b and 16c, respectively. The running histograms are characterized by 2-peak structure. The absolute value of peaks is situated at ~3 nT. The $B_x$ and $B_y$ demonstrate practically identical solar cycle variations. Dynamics of their running dispersion is very close to the solar cycle variation in IMF intensity. Namely, the dispersion is smallest during solar minimum. During rising phase the dispersion grows gradually. Short-time decease in the solar maximum corresponds to the time interval of polarity reversal. During declining phase the running dispersion reaches the maximum and the gap between peaks at running histograms becomes most prominent. Then the dispersion decreases to the lowest values in solar minimum. Note that extremely large intensities of the $B_x$ and $B_y$ are characterized by the same dynamic pattern.

Interesting feature in dynamics of the $B_x$ and $B_y$ components is wave-like variations in running mean. They are characterized by anticorrelation between the $B_x$ and $B_y$ that is proper for the IMF orientation along the Archimedean spiral (see Figure 7). The variations are associated with consequent weakening of occurrence number peak at positive or negative branches of the running histogram. The period of variations varies from 1 to ~5 years. 1-year variations are originated from annual variation of the Earth's heliographic latitude within a range of ±7°. They occur often in the minimum and at beginning of rising phase. That might indicate to very thing heliospheric current sheet separating magnetic field lines with opposite polarities. Longer periods are usually observed in the end of rising phase and during solar maximum. Those variations are associated with a global north-south asymmetry of the solar magnetic field such that plasma streams from north or south coronal hole prevail in the ecliptic plane for several years. We can indicate two intervals of most prominent solar asymmetry associated with dominant magnetic field from south coronal hole: from 1988 to 1991, when long-lasting positive $B_x$ prevails, and from 1998 to 1999 with prevailing negative $B_x$.

Solar cycle variations in the IMF $B_z$ component is shown in Figures 16d. The variations are very similar to those for the $B_x$ and $B_y$ components. The running dispersion is smallest in the solar minimum. The dispersion increases gradually during rising phase. The time-interval of polarity reversal in the solar maximum is accompanied with a short-time decease of the dispersion. During declining phase the running dispersion reaches the maximum and then decreases to the lowest values in solar minimum. The extreme enhancements of IMF $B_z$ group also around solar maximum. A good correlation of the components with IMF strength was reported by *Slavin et al.* [1986]. They attribute the IMF enhancements to fast solar wind streams. However this relationship is not obvious.

This fact can be clearly seen in dynamics of the induced electric field $E_y$ presented in Figure 16e. Note that $E_y$ is a direct multiplication of the $B_z$ component with the SW velocity. We can see that variations in the electric field are slightly different from the dynamics of $B_z$. Several peaks corresponding to strong $B_z$ component vanish in the $E_y$ because of suppression by relatively small solar wind velocity. On the other hand, in the $E_y$ variations we can find several enhancements, which are not coincident with the strong $B_z$ intensities. Apparently those enhancements are caused by very fast solar wind. Different solar cycle dynamics of the velocity and IMF is pointed out by *Luhmann et al.* [1993]. They conclude that the solar cycle in IMF is rather a variation of solar surface field than the changing contributions of coronal transients and stream interfaces.

## Solar Cycle in Dimensionless Parameters

Solar cycle variations in Alfvén Mach number $M_a$, plasma $\beta$, and sonic Mach number $M_s$ are presented in Figure 17. The running histograms of those parameters are close to lognormal PDF. So the running mean is pretty close to running mode, thought they can be slightly different due to non-zero skewness of the running statistical distributions. Similarly to the SW plasma and IMF parameters, the variations in running mode are not very strong and lie within 1-$\sigma$ corridor around the average.

The Alfvén Mach number and plasma $\beta$ demonstrate very similar solar cycle variations, which are close to those in solar wind dynamic pressure (see Figure 15d). Namely they anticorrelate with the sunspot number. *Luhmann et al.*, [1993] also reported about the anticorrelation of the $M_a$ with sunspot number.

Variations in running dispersion of the Alfvén Mach number are different from those for the plasma $\beta$ and more complex. That complexity is coming from the variations in IMF. Numerically the dispersions of $M_a$ and IMF $B$ are comparable and vary in the same dynamic range. The dispersion of $\beta$ varies in wider range. Despite of those differences, we can find common features in the variations. The running dispersion of both $M_a$ and $\beta$ is highest around solar maximum and small in the solar minimum, especially for the plasma $\beta$. Such pattern is apparently inherited from the solar cycle variation in proton density (see Figure 15c).

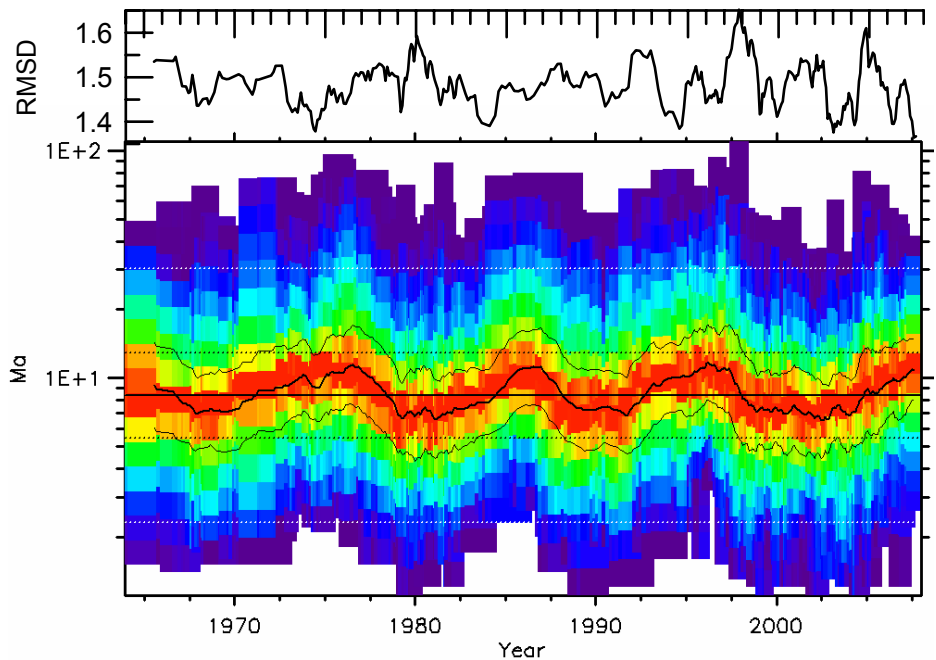

a)

Figure 17. (Continued)

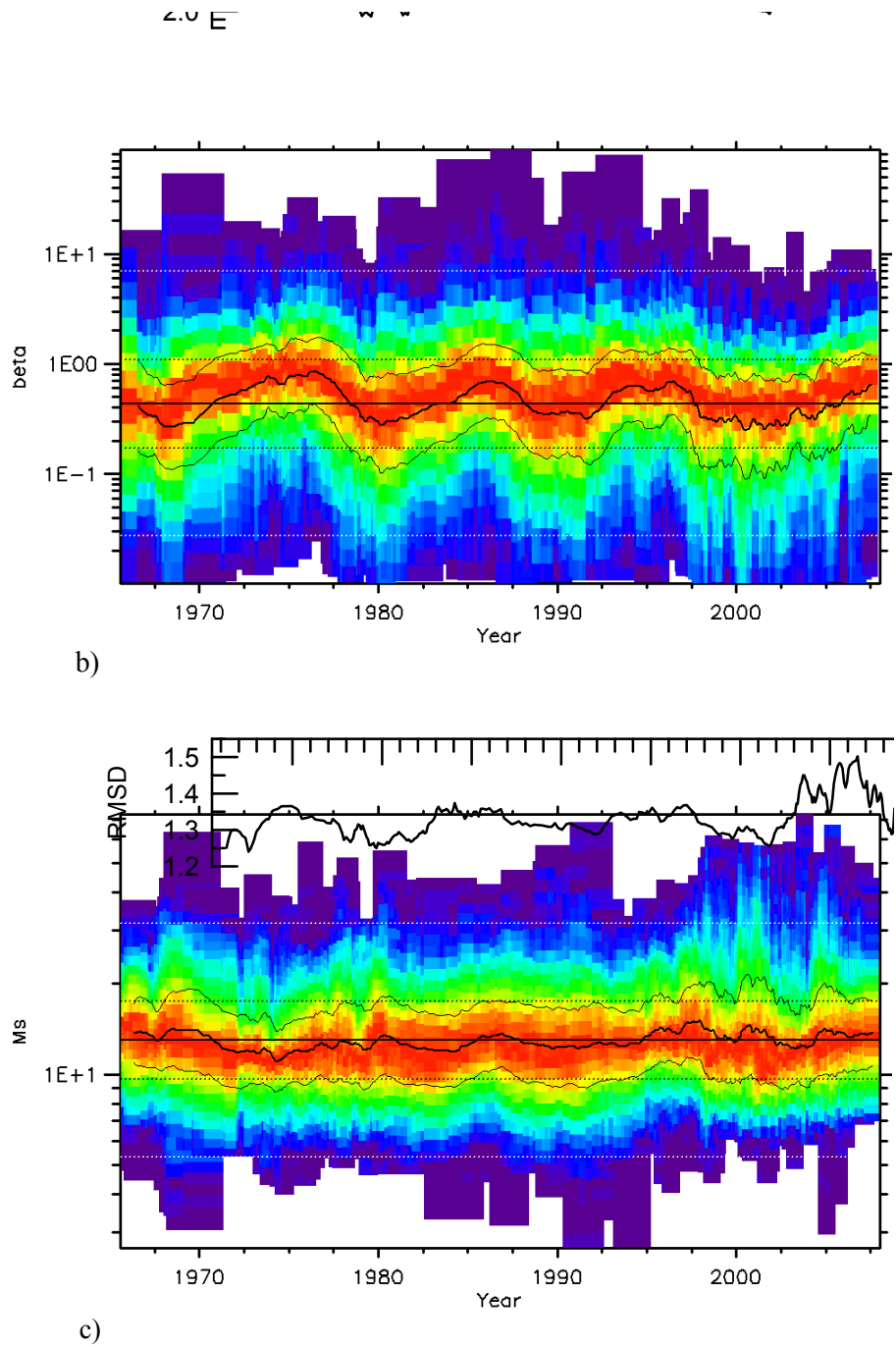

b)

c)

Figure 17. Running histograms of solar cycle variations in dimensionless heliospheric numbers: (a) Alfvén Much number $M_a$; (b) proton plasma $\beta$; and (c) Sonic Mach number $M_s$. Occurrence number in the running histograms is indicated in rainbow palette from violet (minimum) to red (maximum). Running dispersion is expressed in *RMSD* and shown in the top panels. Running mean and running 1-$\sigma$ deviation are indicated, respectively, by thick and thin black curves. The average is indicated by horizontal black solid line. The 1-$\sigma$ and 3-$\sigma$ corridors are restricted, respectively, by horizontal black dotted lines and white dotted lines.

Extremely high magnitudes of Alfvén Mach number are distributed rather uniformly along the solar cycle, which reflects high variability of the IMF intensity. In contrast, extremely low $M_a$ and plasma $\beta$ occur around solar maximum that corresponds to high occurrence rate of very low proton density. Extremely high values of plasma $\beta$ are observed during rising and declining phases that might be attributed to superposition of events with extremely high temperatures and weak IMF strength.

Solar cycle variations in the sonic Mach number are extremely weak, pretty gradual and totally non-regular (see Figure 17c). This degeneration indicates to very close relationship between the solar cycle variations in solar wind velocity and temperature. Indeed, we have found very close dynamics in running modes of those parameters (see Figure 15a and 15b). It is interesting that in contrast to solar wind velocity and temperature, the running dispersion of $M_s$ demonstrates pretty well correlation with sunspot number while the extreme values of $M_s$ are distributed practically uniformly along the solar cycle.

## 5. Summary and Conclusions

Based on the review of statistical properties we can indicate some general features in the dynamics of heliospheric parameters.

1. The IMF intensity $B$ demonstrates pretty unique dynamics with solar cycles. The intensity correlates with sunspot numbers. The periodogram of IMF B contains the periodicities, which are very close to the characteristic periods of sunspot number variations. Note that we don't find any indications on the 22-year period in variability of this parameter in the data set under consideration. While the 21[st] and 22[nd] cycles in IMF intensity are similar, the 23[rd] solar cycle is different because of deeper decrease of the IMF intensity during the declining phase.

2. Solar wind proton density is also characterized by a specific dynamics. It seems that the running mean density correlates with running dispersion of the sunspot number. The periodogram of density is different from those for the sunspot number and for the IMF strength $B$. We can not find 22-periodicity in the solar wind density. Moreover, a 5-year periodicity is very prominent in the periodogram of density. This periodicity prevails in the solar cycle variation of running mean, which has two-wave structure with minima just before the solar maximum and minimum. That is not a case in the 23[rd] solar cycle, which demonstrates very deep gradual decrease of the SW density during recovery phase that is significantly different from the three previous cycles.

3. Statistical distributions of both IMF intensity and SW density are very close to lognormal in temporal scale of ~0.5 year and longer. Though dispersions of those parameters are slightly different. Running logarithmic dispersion of IMF $B$ varies with solar cycle in the range from 1.4 to 1.6. The logarithmic dispersion of density is higher and varies from ~1.7 to ~2.2.

4. The solar wind velocity and temperature are characterized by common dynamics, which is different from the dynamics of IMF strength and SW density. The velocity and temperature have very similar periodograms and close solar cycle variations in running histograms. Their very similar dynamics leads to disappearance of the solar cycle variation in mean sonic Mach number. They demonstrate clearly the ~22-year variation. Statistical distributions of the velocity and density are close to lognormal PDF only in long time scale of tens of years. Within the solar cycle in short temporal scales of ~0.5 year, the distributions are skewed and can acquire two or three peaks.

The 3-peak structure of the velocity statistical distribution lends an additional support to the concept of a three-component solar wind model [*Schwenn*, 1990]. Slow, fast and intermediate solar wind flows are statistically distinct especially during declining solar cycle phases because of the geometry conditions for the tilted rotator [*Veselovsky et al.*, 1998c; *Dmitriev, et al.*, 2000]. The slow solar wind of ~400 km/s, which occupies the heliospheric plasma sheet, originates somewhere inside coronal streamers. Coronal holes with their mostly open unipolar magnetic configurations are known to be the sources of the fastest solar wind streams with speeds of >600 km/s. The transition region between slow and fast streams corresponds to the intermediate average flow velocities of ~500 km/s. The dominant magnetic topology here is not clear, but seems to be mixed and intermittent [e.g. *Posner, et al.*, 2001].

Different dynamics of the IMF intensity, solar wind density at one side and solar wind velocity and temperature at other side indicates to different origin of those heliospheric parameters. Perhaps the velocity and temperature of the three-component solar wind are rather driven by magnetic topology of the source regions and geometry of the global magnetic field of the Sun, which varies substantially with solar cycle due to solar magnetic field reversal. The slow streams in the heliospheric plasma sheet and intermediate flows from extended transition region prevail during solar minimum and rising phase when the solar dipole is tilted only slightly. Large-scale coronal holes diminish during solar maximum, which is enriched by active regions with close configuration of magnetic field lines. At that time slow solar wind dominates. After the reversal, the solar magnetic dipole is substantially tilted and, thus, during declining phase the fast streams from coronal holes concurrent with intermediate and slow solar wind streams.

Origin of the lognormally distributed IMF intensity is still a subject of investigation. Considering the statistical distribution of $B$xy we have found that variations of the magnetic field in the ecliptic plane can not be attributed to the variations of solar wind velocity only. Solar cycle variations and periodicities in the IMF intensity are different from those in solar wind velocity. Namely, the IMF enhancements are not necessarily accompanied with fast solar wind streams enriched by strong Alfvén waves and compressed leading edges with enhanced magnetic field. And vice versa, the fast solar wind is characterized often by moderate and even week IMF intensity.

Hence a significant portion of the IMF variations should be originated from the solar source. Solar eruptive events such as ICME and interplanetary sheath regions contribute mainly to excess of statistics at very large IMF intensities. Perhaps they might cause the increase of the average magnetic field observed around the solar maximum. However that does not explain the lognormal shape of the variations in IMF strength as well as in solar wind density.

The lognormal shape of statistical distribution has a deep physical meaning. A parameter $x$ with lognormal distribution can be originated from a specific multiplication generator. The coefficient of multiplication $k=x/X_0$ of that generator can be represented as $k=\exp(\nu)$, where $\nu$ is distributed normally with the mode $\nu_0=0$ and dispersion equal to the dispersion of parameter *x*. In other words, the variable with lognormal distribution is generated as a result of numerous multiplications of random variables.

The lognormal statistical distributions of the solar wind parameters and IMF might indirectly indicate to multiplicative transformation of local characteristics under alternating random amplification and weakening of waves, compression and rarefaction of inhomogeneities in turbulent processes of transfer of plasma mass, energy and momentum on the Sun and in the heliosphere. For example, in the downstream region of fast interplanetary shock, the values of plasma density and magnetic field are multiplied by compression. We can assume that the solar wind plasma density and IMF intensity might be generated and/or modulated in the regions of alternate compression/decompression in the solar atmosphere.

On the other hand the lognormal distribution itself is not sufficient for conclusions about predominantly random and irregular nature of a multiplicative process because regular processes with very high levels of complexity and multi-dimensions are rather difficult to distinguish from the random process. Both of those possibilities do not seem excluding each other in interpretation of observations. They rather supplement the interpretation from different points of view.

The fact that statistical properties of heliospheric parameters do not satisfy to normal distribution is not surprising. Normal statistics is proper for equilibrium and stationary processes. In contrast, the solar wind plasma and IMF are non-equilibrium and non-stationary. They change with distance and with heliographic latitude and longitude, i.e. they are characterized by various spatial gradients. Moreover, the heliosphere is populated by various kinds of large-scale transient events, interaction regions and waves. We have demonstrated that the key heliospheric parameters are transitional. They are characterized by meaningful temporal variations within solar cycle and from cycle to cycle. However, those variations are not totally random but reveal specific spatial and temporal patterns. Our results clearly show that the lognormal statistics is more-or-less proper for such non-equilibrium and non-stationary random processes with characteristic spatial and temporal scales.

Finally, we should comment that we investigated here only individual one-point, one-time and one-parameter statistical representations. Correlations between different parameters and higher order multi-point and multi-time structure functions were not considered, but remained mostly beyond the scope of our consideration. It is a next step for future analysis enabling visualization of physical nonlinearities, memory effects and non-local links on the Sun and in the heliosphere. Existing knowledge in this regard is also in its infant stage in spite of increasing information about turbulent processes in solar wind and interplanetary magnetic field [*Bruno and Carbone*, 2005].

We conclude that summary of statistical studies of individual SW and IMF parameters was presented. Growing amount of direct in-situ measurements near the Earth orbit during space era allowed a robust empirical modeling of average SW and IMF characteristics and variations. The resulting probability distribution functions together with their reliability and variability estimates given in formulae, Tables and simple graphs can be used for scientific and technical applications. The statistics up to now encompasses only four recent solar cycles. It is not sufficient for characterization of typical and a-typical solar cycle behavior of SW and IMF parameters contrary to premature claims in this respect found sometimes in literature. Nevertheless, the fundamental quantitative knowledge accumulates and fits rather well modern views and ideas based on MHD and kinetic plasma electromagnetic theories.

## Acknowledgements

We thank Joe King and Natalia Papitashvili from NASA/NSSDC and QSS Group, Inc. for providing OMNI database. This study was supported by the RFBR grants 07-02-00147, 06-05-64500, INTAS 03-51-6202, MSU Interdisciplinary Scientific Project and grant NSC96-2923-M-008-001MY3/07-02-92004HHC_a from the National Science Council of Taiwan. It is also fulfilled as a part of the Programs of the Russian Academy of Sciences: "Origin and evolution of stars and galaxies" (P-04), "Solar activity and physical processes in the Sun-Earth system" (P-16, Part 3) and "Plasma processes in the Solar system (OFN-16).

## References

Aellig, M., A. Lazarus, and J. Steinberg (2001), The Solar Wind Helium Abundance: Variation with Wind Speed and the Solar Cycle, *Geophys. Res. Lett.*, 28(14), 2767-2770.

Akasofu, S.-I. (1979), Interplanetary energy flux associated with magnetospheric substorms, *Planet. Space. Sci.*, *27*, 425.

Baker, D., W. Feldman, S. Gary, D. McComas, and J. Middleditch (1986), Plasma Fluctuations and Large-Scale Mixing Near Comet Giacobini-Zinner, *Geophys. Res. Lett.*, 13(3), 271-274.

Belcher, J., and L. Davis Jr. (1971), Large-Amplitude Alfvèn Waves in the Interplanetary Medium, 2, *J. Geophys. Res*., 76(16), 3534-3563.

Bieber, J., J. Chen, W. Matthaeus, C. Smith, and M. Pomerantz (1993), Long-Term Variations of Interplanetary Magnetic Field Spectra with Implications for Cosmic Ray Modulation, *J. Geophys. Res.,* 98(A3), 3585-3603.

Bruno, R., and V. Carbone (2005), The Solar Wind as a Turbulence Laboratory, *Living Rev. Solar Phys.*, 2, 4. http://www.livingreviews.org/lrsp-2005-4

Bolzan, M. J. A., R. R. Rosa, F. M. Ramos, P. R. Fagundes, Y. Sahai (2005), Generalized thermostatistics and wavelet analysis of solar wind and proton density variability, *Journal of Atmospheric and Solar-Terrestrial Physics,* 67, 1843-1851.

Borrini, G., J. Gosling, S. Bame, W. Feldman, and J. Wilcox (1981), Solar Wind Helium and Hydrogen Structure Near the Heliospheric Current Sheet: A Signal of Coronal Streamers at 1 AU, *J. Geophys. Res*., 86(A6), 4565-4573.

Borrini, G., J. Gosling, S. Bame, and W. Feldman (1982a), An Analysis of Shock Wave Disturbances Observed at 1 AU from 1971 Through 1978, *J. Geophys. Res.,* 87(A6), 4365-4373.

Borrini, G., J. Gosling, S. Bame, and W. Feldman (1982b), Helium Abundance Enhancements in the Solar Wind, *J. Geophys. Res*., 87(A9), 7370-7378.

Bothmer, V. and R. Schwenn (1995), The interplanetary and solar causes of major geomagnetic storms, *J. Geomag. Geoelectr.*, 47, 1127-1132.

Burlaga, L. F. (1992), Multifractal Structure of the Magnetic Field and Plasma in Recurrent Streams at 1 AU, *J. Geophys. Res*., 97(A4), 4283-4293.

Burlaga, L. F. (2001), Lognormal and multifractal distributions of the heliospheric magnetic field, *J. Geophys. Res*., 106(A8), 15,917-15,927.

Burlaga, L. F. (2005), Interplanetary magnetohydrodynamics, New York : Oxford University Press, 1995.

Burlaga, L. F., and M. A. Forman (2002), Large-scale speed fluctuations at 1 AU on scales from 1 hour to ~1 year: 1999 and 1995, *J. Geophys. Res.*, 107(A11), 1403, doi:10.1029/2002JA009271.

Burlaga, L. F., and J. King (1979), Intense Interplanetary Magnetic Fields Observed by Geocentric Spacecraft During 1963-1975, J. Geophys. Res., 84(A11), 6633-6640.

Burlaga, L. F., and A. Lazarus (2000), Lognormal distributions and spectra of solar wind plasma fluctuations: Wind 1995-1998, *J. Geophys. Res.*, 105(A2), 2357-2364.

Burlaga, L. F., and N. Ness (1998), Magnetic field strength distributions and spectra in the heliosphere and their significance for cosmic ray modulation: Voyager 1, 1980–1994, *J. Geophys. Res.*, 103(A12), 29719-29732.

Burlaga, L. F., and A. Szabo (1999), Fast and slow flows in the solar wind near the ecliptic at 1 AU?, *Space Science Reviews*, 87, 137-140.

Burlaga, L. F., K. Behannon, and L. Klein (1987), Compound Streams, Magnetic Clouds, and Major Geomagnetic Storms, J. Geophys. Res., 92(A6), 5725-5734.

Burlaga, L. F., R. Fitzenreiter, R. Lepping, K. Ogilvie, A. Szabo, A. Lazarus, J. Steinberg, et al. (1998), A magnetic cloud containing prominence material: January 1997, *J. Geophys. Res.,* 103(A1), 277-285.

Burlaga, L. F., R. Skoug, C. Smith, D. Webb, T. Zurbuchen, and A. Reinard (2001), Fast ejecta during the ascending phase of solar cycle 23: ACE observations, 1998-1999, *J. Geophys. Res.*, 106(A10), 20957-20977.

Burton, R.K., R.L. McPherron, and C.T. Russell (1975), An empirical relationship between interplanetary conditions and *Dst*, *J. Geophys. Res., 80*, 4204.

Cane H. V., and I. G. Richardson (2003), Interplanetary coronal mass ejections in the near-Earth solar wind during 1996-2002, *J. Geophys. Res.*, 108 (A4), 1156, doi:10.1029/2002JA009817.

Charvátová, I. (2007), The prominent 1.6-year periodicity in solar motion due to the inner planets, *Ann. Geophys.*, 25, 1227-1232.

Cliver, E., V. Boriakoff, and K. Bounar (1996), The 22-year cycle of geomagnetic and solar wind activity, J. Geophys. Res., 101(A12), 27091-27109.

Couzens, D. A., and J. H. King (1986), *Interplanetary Medium Data Book, Suppl,* 3, 1977-1985, Rep. NSSDC/WDC-A-R&S 86-04, NASA, 1986.

Crooker, N.U. (1983), Solar cycle variations of the solar wind, in *Solar Wind Five* edited by M. Neugebauer, NASA, CP-2280, 303-313.

Crooker, N., and K. Gringauz (1993), On the Low Correlation Between Long-Term Averages of Solar Wind Speed and Geomagnetic Activity After 1976, *J. Geophys. Res.,* 98(A1), 59-62.

Crooker, N., S. Shodhan, J. Gosling, J. Simmerer, R. Lepping, J. Steinberg, and S. Kahler (2000), Density Extremes in the Solar Wind, *Geophys. Res. Lett.*, 27(23), 3769-3772.

Dal Lago, A., W. D. Gonzalez, A. L. C. de Gonzalez, L.E. A. Vieira (2001), Compression of magnetic clouds in interplanetary space and increase in their geoeffectiveness, J. Atmosph. *Solar Terrest Phys.*, 63, 451-455

Deeming, T. J. (1975), Fourier analysis with unequally-spaced data, Astrophysics and Space Science, 36, 137-158.

Dmitriev A. V., A. V. Suvorova, I. S. Veselovsky (2000), Solar Wind and Interplanetary Magnetic Field Parameters at the Earth's Orbit During Three Solar Cycles, *Phys. Chem. of the Earth*, Part C, 25(1-2), 125-128.

Dmitriev A. V., J. K. Chao, Y. H. Yang, C.-H. Lin, and D. J. Wu (2002a), Possible Sources of the Difference between a Model Prediction and Observations of Bow Shock Crossings, *Terrestrial, Atmospheric and Oceanic Sciences* (TAO), 13(4), 499-521.

Dmitriev, A. V., Suvorova, A. V., Veselovsky, I. S. (2002b), Expected hysteresis of the 23-rd solar cycle in the heliosphere, *Adv. Space Res.*, 29(3), 475-479.

Dmitriev A., J.-K. Chao, D.-J. Wu (2003), Comparative study of bow shock models using Wind and Geotail observations, *J. Geophys. Res.*, 108(A12), 1464, doi:10.1029/2003JA010027.

Dmitriev A. V., A. V. Suvorova, J.-K. Chao, Y.-H. Yang (2004), Dawn-dusk asymmetry of geosynchronous magnetopause crossings, *J. Geophys. Res.*, 109, A05203 doi: 10.1029/2003JA010171.

Dmitriev, A. V., I. S. Veselovsky, O.S. Yakovchouk, (2005a), Problems of consistency in solar wind parameters between data sets OMNI and OMNI-2, "*Solar activity as space weather factor*". *Proceedings of the 9th International Conference on the Physics of the Sun*, St Petersburg, July 4-9, 2005, VVM Publishers, pp. 51-56 (in Russian).

Dmitriev A. V., J.-K. Chao, A. V. Suvorova, K. Ackerson, K. Ishisaka, Y. Kasaba, H. Kojima, H. Matsumoto (2005b), Indirect estimation of the solar wind conditions in 29-31 October 2003, *J. Geophys. Res.*, 110, A09S02, doi:10.1029/2004JA010806.

Dmitriev, A. V., I. S. Veselovsky, A. V. Suvorova, (2005c), Comparison of heliospheric conditions near the Earth during four recent solar maxima, *Adv. Space Res.*, 36, 2339-2344.

El-Borie, M. A., Duldig, M.L., and Humble (1997), J.E., Interplanetary plasma and magnetic field observations at 1AU, *25th International Cosmic Ray Conference*, *Contributed Papers ,1,* SH1-3, 317-320, 1997.

El-Borie, M. A. (2002), On long-term periodicities n the solar-wind ion density and speed measurements during the period 1973-2000, *Solar Physics*, 208(2), pp. 345-358.

El-Borie, M. A., and S. S. Al-Thoyaib (2002), Power spectrum of cosmic-ray fluctuations during consecutive solar minimum and maximum periods, *Solar Physics*, 209, 397-407.

Feldman, W., J. Asbridge, S. Bame, and J. Gosling (1978), Long-Term Variations of Selected Solar Wind Properties: Imp 6, 7, and 8 Results, *J. Geophys. Res.*, 83(A5), 2177-2189.

Feynman, J., and A. Ruzmaikin (1994), Distributions of the Interplanetary Magnetic Field Revisited, *J. Geophys. Res.*, 99(A9), 17645-17651.

Freeman, J., and R. Lopez (1985), The cold solar wind, *J. Geophys. Res.*, 90(A10), 9885-9887.

Freeman, J., T. Tollen, and A. Arya (1993), Comparison between Helios data normalized to 1 AU and OMNI data, in Solar-Terrestrial Predictions-IV, Hruska, H., Shea, M. A., Smart, D. F., and Heckman, G., Eds., Boulder: *NOAA/ERL*, 2, 540-549.

Gazis, P.R. (1996), Solar cycle variation in the heliosphere, *Rev. Geophys.,34*, 379-402.

Gazis, P. R., J. D. Richardson, and K. I. Paularena (1995), Long Term Periodicity in Solar Wind Velocity During the Last Three Solar Cycles, *Geophys. Res. Lett.*, 22(10), 1165-1168.

Gnevyshev, M. N., and A. I. Ohl (1948), 22-year solar activity cycle, *Astronomicheskii Zhurnal*, 25, 18-20 (in Russian).

Hartlep, T., W. Matthaeus, N. Padhye, and C. Smith (2000), Magnetic field strength distribution in interplanetary turbulence, *J. Geophys. Res.*, 105(A3), 5135-5139.

Iijima, T., and T. A. Potemra (1982), The relationship between interplanetary quantities and Birkeland current densities, *Geophys. Res. Lett., 9*, 442.

Ipavich, F., A. B. Galvin, S. E. Lasley, J. A. Paquette, S. Hefti, K.-U. Reiche, M. A. Coplan, et al. (1998), Solar wind measurements with SOHO: The CELIAS/MTOF proton monitor, *J. Geophys. Res*., 103(A8), 17205-17213.

Jain, R, B. Chandel, M. Aggarwal1, R. Devi, and A. K. Gwal1 (2008), Periodicities Inferred from SOXS Mission in Soft and Hard X-ray Emission from the Solar Corona, Abs. on *International Workshop Solar Variability, Earth's Climate & the Space Environmen*t, Bozeman, Montana, June 1-6, 2008, 30.

Janardhan P., K. Fujiki, H. S. Sawant, M. Kojima, K. Hakamada, R. Krishnan (2008), Source regions of solar wind disappearance events, *J. Geophys. Res*., 113, A03102, doi:10.1029/2007JA012608.

Juckett, D. A. (2000), Solar activity cycles, north/south asymmetries, and differential rotation associated with solar spin-orbit variations, *Solar Physics*, 191, 201-226.

Kane, R. P. (2005), Differences in the quasi-biennial oscillation and quasi-triennial oscillation characteristics of the solar, interplanetary, and terrestrial parameters, *J. Geophys. Res.,* 110, A01108, doi:10.1029/2004JA010606.

King, J. H. (1977), Ed. *Interplanetary Medium Data Book*, NSSDC/WDC-A-R/S 77-04.

King, J. H. (1981), On the Enhancement of the IMF Magnitude During 1978-1979, *J. Geophys. Res.*, 86(A6), 4828-4830.

King, J. H., N. E. Papitashvili (2005), Solar wind spatial scales in and comparisons of hourly Wind and ACE plasma and magnetic field data, *J. Geophys. Res*., 110, A02104, doi:10.1029/2004JA010649.

Lazarus, A. J. (2000), The day the solar wind almost disappeared, *Science, 287*(5461), 2172-2173.

Lazarus, A. J., and K. I. Paularena (1993), Solar wind parameters from the MIT plasma experiment of IMP-8, *Solar-Terrestrial Predictions-IV,* Hruska, H., Shea, M. A., Smart, D. F., and Heckman, G., Eds., Boulder: NOAA/ERL, 2, 720.

Lopez, R. (1987), Solar Cycle Invariance in Solar Wind Proton Temperature Relationships, *J. Geophys. Res*., 92(A10), 11189-11194.

Lopez, R., and J. Freeman (1986), Solar Wind Proton Temperature-Velocity Relationship, *J. Geophys. Res.,* 91(A2), 1701-1705.

Luhmann, J., T.-L. Zhang, S. Petrinec, C. Russell, P. Gazis, and A. Barnes (1993), Solar Cycle 21 Effects on the Interplanetary Magnetic Field and Related Parameters at 0.7 and 1.0 AU, *J. Geophys. Res.,* 98(A4), 5559-5572.

Mavromichalaki, H., B. Petropoulos, C. Plainaki, O. Dionatos, I. Zouganeils (2005), Coronal index as a solar activity index applied to space weather, *Adv. Space Res.*, 35, 410-415.

McComas D. J., H. A. Elliott, N. A. Schwadron, J. T. Gosling, R. M. Skoug, B. E. Goldstein (2003), The three-dimensional solar wind around solar maximum, *Geophys. Res. Lett*., 30 (10), 1517, doi:10.1029/2003GL017136.

McIntosh, P. S., R. J. Thompson, E. C. Willock (1992), A 600-day periodicity in solar coronal holes, *Nature,* 360, 322 - 324.

Mood, A. M., F. A. Graybill, D. C. Boes (1974), Introduction to the theory of statistics, McGraw-Hill, Singapore.

Mullan, D. J. , C. W. Smith (2006),  Solar Wind Statistics at 1 AU: Alfven Speed and Plasma Beta, *Solar Phys. 234* (2), 325-338.

Mursula, K., and J. H. Vilppola (2004), Fluctuations of the solar dynamo observed in the solar wind and interplanetary magnetic field at 1 AU and in the outer heliosphere, *Solar Physics*, 221, 337-349.

Mursula, K., and B. Zieger (1996), The 13.5-day periodicity in the Sun, solar wind, and geomagnetic activity: The last three solar cycles, *J. Geophys. Res.,* 101(A12), 27077-27090.

Neugebauer, M., E. J. Smith, A. Ruzmaikin, J. Feynman, and A. H. Vaughan (2000), The solar magnetic field and the solar wind: Existence of preferred longitudes, *J. Geophys. Res.*, 105(A2), 2315-2324.

Newbury, J. A., Russell, C. T., Phillips, J. L., Gary, S. P. (1998), "Electron temperature in the ambient solar wind: Typical properties and a lower bound at 1 AU", *J. Geophys. Res.,* 103, 9553-9566.

Ogilvie, K., M. Coplan, and R. Zwickl (1982), Helium, Hydrogen, and Oxygen Velocities Observed on ISEE 3, *J. Geophys. Res.*, 87(A9), 7363-7369.

Owens, M. J., and P. J. Cargill (2002), Correlation of magnetic field intensities and solar wind speeds of events observed by ACE, *J. Geophys. Res.*, 107(A5), 1050, doi:10.1029/2001JA000238.

Owens M. J., P. J. Cargill, C. Pagel, G. L. Siscoe, N. U. Crooker (2005), Characteristic magnetic field and speed properties of interplanetary coronal mass ejections and their sheath regions, *J. Geophys. Res.*, 110, A01105, doi:10.1029/2004JA010814.

Oyama, K., K. Hirao, T. Hirano, K. Yumoto, and T. Saito (1986), Was the solar wind decelerated by comet Halley? *Nature*, 321, 310-313.

Paularena, K., A. Szabo, and J. Richardson (1995), Coincident 1.3-Year Periodicities in the ap Geomagnetic Index and the Solar Wind, *Geophys. Res. Lett.*, 22(21), 3001-3004.

Posner, A., T. Zurbuchen, N. Schwadron, L. Fisk, G. Gloeckler, J. Linker, Z. Mikić;, and P. Riley (2001), Nature of the boundary between open and closed magnetic field line regions at the Sun revealed by composition data and numerical models, *J. Geophys. Res.*, 106(A8), 15869-15879.

Rangarajan, G. K., and L. M. Barreto (2000), Long term variability in solar wind velocity and IMF intensity and the relationship between solar wind parameters & geomagnetic activity, *Earth Planets Space*, 52, 121-132.

Richardson, I. G., and H. V. Cane (1995), Regions of Abnormally Low Proton Temperature in the Solar Wind (1965-1991) and their Association with Ejecta, *J. Geophys. Res.*, *100*(A12), 23,397–23,412.

Richardson, J., K. Paularena, J. Belcher, and A. Lazarus (1994), Solar Wind Oscillations with a 1.3 Year Period, *Geophys. Res. Lett.*, 21(14), 1559-1560.

Richardson, I., D. Berdichevsky, M. Desch, and C. Farrugia (2000), Solar-Cycle Variation of Low Density Solar Wind during more than Three Solar Cycles, *Geophys. Res. Lett.,* 27(23), 3761-3764.

Richardson, J. D., C. Wang, and K. I. Paularena (2001), The solar wind: from solar minimum to solar maximum, *Advances in Space Research*, 27(3), 471-479.

Rouillard, A., and M. Lockwood (2004), Oscillations in the open solar magnetic flux with a period of 1.68 years: imprint on galactic cosmic rays and implications for heliospheric shielding, *Annales Geophysicae*, 22, 4381-4395.

Russell, C. T., and S. M. Petrinec (1993), On the relative intercalibration of solar wind detectors, in Solar-Terrestrial Predictions-IV, Hruska, H., Shea, M. A., Smart, D. F., and Heckman, G., Eds., Boulder: NOAA/ERL, 2, 663-670.

Schwenn, R. (1990), Large-scale structure of the interplanetary medium, in *Physics of the Inner Helioshpere, 2,* edited by R. Schwenn and E. Marsch, Berlin-Heidelberg, Springer-Verlag, 99-133.

Schwenn, R. and Marsch E. (Eds.) (1990), Physics of the Inner Heliosphere I, Berlin, Germany, Springer-Verlag.

Skoug R. M., J. T. Gosling, J. T. Steinberg, D. J. McComas, C. W. Smith, N. F. Ness, Q. Hu, L. F. Burlaga (2004), Extremely high speed solar wind: 29-30 October 2003, *J. Geophys. Res.*, 109, A09102, doi:10.1029/2004JA010494.

Slavin, J., G. Jungman, and E. Smith (1986), The Interplanetary Magnetic Field During Solar Cycle 21: ISEE-3/ICE Observations, *Geophys. Res. Lett.*, 13(6), 513-516.

Smith, E. (2001), The heliospheric current sheet, *J. Geophys. Res.,* 106(A8), 15819-15831.

Spreiter, J.R., A.L. Summers, and A.Y. Alksne (1966), Hydromagnetic flow around the magnetosphere, *Planet. Space Sci., 14*, 223.

Szabo, A., R. Lepping, and J. King (1995), Magnetic Field Observations of the 1.3-Year Solar Wind Oscillation, *Geophys. Res. Lett.*, 22(14), 1845-1848.

Thomson, D. J. C. G. Maclennan , L. J. Lanzerotti (2002), Propagation of solar oscillations through the interplanetary medium, *Nature,* 376, 139 - 144; doi:10.1038/376139a0.

Tsyganenko, N. A. (2000a), Solar wind control of the tail lobe magnetic field as deduced from Geotail, AMPTE/IRM, and ISEE 2 data, *J. Geophys. Res.*, 105, 5517.

Tsyganenko, N. A. (2000b), Modeling the inner magnetosphere: The asymmetric ring current and Region 2 Birkeland currents revisited, *J. Geophys. Res.*, 105, 27,739.

Tsurutani, B., and W. Gonzalez (1987), The cause of high-intensity long-duration continuous AE activity (HILDCAAS): interplanetary Alfvén wave trains, *Planet. Space Sci.*, 35(4), 405-412.

Tsurutani, B., W. Gonzalez, A. Gonzalez, F. Tang, J. Arballo, and M. Okada (1995), Interplanetary Origin of Geomagnetic Activity in the Declining Phase of the Solar Cycle, *J. Geophys. Res.*, 100(A11), 21717-21733.

Tsurutani, B. T., Mannucci, A. J., Iijima, B., Abdu, M. A., Sobral, J. H. A., Gonzalez, W., Guarnieri, F., et al. (2004): Global dayside ionospheric uplift and enhancement associated with interplanetary electric fields, *J. Geophys. Res.*, 109, A08302, doi:10.1029/2003JA010342.

Usmanov A. V., M. L. Goldstein, K. W. Ogilvie, W. M. Farrell, G. R. Lawrence (2005), Low-density anomalies and sub-Alfvènic solar wind, *J. Geophys. Res.*, 110, A01106, doi:10.1029/2004JA010699.

Valdés-Galicia, J. F., A. Lara, B. Mendoza (2005), The solar magnetic .ux mid-term periodicities, and the solar dynamo, Journal of Atmospheric and Solar-Terrestrial *Physics,* 67, 1697-1701.

Veselovsky, I.S., Physics of the interplanetary plasma (1984), *Itogi Nauki i Tekhniki, "Issledovania Kosmicheskogo Prostranstva"*, 22, Moscow, VINITI, (in Russian).

Veselovsky, I. S. (2004), Electric currents and magnetic fields in the solar corona and heliopshere, *Solar-terrestrial physics*, 6, 119 (in Russian).

Veselovsky, I. S., and M. V. Tarsina (2001),  Angular Distribution of the Interplanetary Magnetic Field Vector , *Geomag. and Aeron.* 41(4), 452-456.

Veselovsky, I. S., and M. V. Tarsina (2002a), Intrinsic nonlinearity of the solar cycles, *Adv. Space Res.*, 29(3), 417-420.

Veselovsky, I. S., and M. V. Tarsina (2002b), Rhythmic and a-rhythmic variations in the heliosphere near the Earth, *Atlas of Time Variations of Natural, Anthrop and Social Processes*, V.3, 457-464, Moscow, Yanus Publishing Co., (in Russian).

Veselovsky, I.S., A.V. Dmitriev, and A.V. Suvorova (1998a), Average parameters of the solar wind and interplanetary magnetic field at the Earth's orbit for the last three solar cycles, *Solar System Research*, 32, 4, 310-315.

Veselovsky I.S., Dmitirev A.V., Suvorova A.V., Panassenko O.A. (1998b), Plasma and magnetic field parameters in the heliosphere at the Earth's orbit, Preprint of the Institute of Nuclear Physics, Moscow State University #98-18/519, Moscow, 17pp.

Veselovsky I.S., Dmitirev A.V., Suvorova A.V., Panassenko O.A. (1998c), Statistical and spectral properties of the heliospheric plasma and magnetic fields at the Earth's orbit, Preprint of the Institute of Nuclear Physics, Moscow State University #98-18/519, Moscow, 20pp.

Veselovsky I. S., A. V. Dmitriev, O. A. Panassenko, and A. V. Suvorova 1999, Solar cycles in the energy and mass outputs of the heliospheric plasma, *Astronomy Reports*, 43, 7, 485-486.

Veselovsky, I. S., Dmitriev, A. V., Suvorova, A. V., Minaeva, Yu. S. (2000a), Structure of Long-Term Variations of the Plasma Parameters and Magnetic Field in the Near-Earth Heliosphere, *Solar System Res. 34*(1), 75-85.

Veselovsky I.S., Dmitriev A.V., Suvorova A.V., Tarsina M.V. (2000b), Solar wind variation with the cycle, *J. Astrophys. Astr.*, 21, p. 423-429.

Veselovsky I.S., Dmitriev A.V., Suvorova A.V. (2001), Plasma and magnetic fields in the heliosphere at the growth phase of solar cycle 23: comparison with previous solar cycles, *Solar System Res.*, 35, 3, 262-266.

Veselovsky, I. S., I. G. Persiantsev, A. Yu. Ryazanov, Yu. S. Shugai (2006), One-parameter representation of the daily averaged solar-wind velocity, *Solar System Research*, 40(5), 427-431, DOI: 10.1134/S0038094606050078.

Vrsnak, B., M. Temmer, A. M. Veronig (2007), Coronal holes and solar wind high-speed streams: I. forecasting the solar wind parameters, *Solar Phys.,* 240, 315-330, DOI 10.1007/s11207-007-0285-8

Zhang, G.-L. and Xu, Y.-F., An interplanetary view of solar cycle variations, in *Solar-Terrestrial Predictions-IV*, 2, edited by H. Hruska , M.A. Shea, D.F. Smart, and G. Heckman, NOAA/ERL, Boulder, 396-401, 1993.

Zurbuchen, T. H., S. Hefti, L. A. Fisk, G. Gloeckler, N. A. Schwadron, C. W. Smith, N. F. Ness, R. M. Skoug, D. J. McComas, and L. F. Burlaga (2001), On the origin of microscale magnetic holes in the solar wind, *J. Geophys. Res.*, 106(A8), 16,001-16,010.

Zwickl, R. (1993), Comparison of Los Alamos plasma ion sensors: IMP-8 and ISEE-3, in Solar-Terrestrial Predictions-IV, Hruska, H., Shea, M. A., Smart, D. F., and Heckman, G., Eds., Boulder: NOAA/ERL, 2, 723.